\documentclass{article}

\usepackage{arxiv}

\usepackage[utf8]{inputenc} % allow utf-8 input
\usepackage[T1]{fontenc}    % use 8-bit T1 fonts
\usepackage{hyperref}       % hyperlinks
\usepackage{url}            % simple URL typesetting
\usepackage{booktabs}       % professional-quality tables
\usepackage{amsfonts}       % blackboard math symbols
\usepackage{nicefrac}       % compact symbols for 1/2, etc.
\usepackage{microtype}      % microtypography
\usepackage{lipsum}
\usepackage{graphicx}
\graphicspath{ {./images/} }

\usepackage{amssymb}
\usepackage{amsthm}
\usepackage{mathrsfs}
\usepackage{amsmath}
\usepackage{float}
\usepackage[linewidth=1pt]{mdframed}
\usepackage{tabularx}
\DeclareSymbolFont{bbold}{U}{bbold}{m}{n}
\DeclareSymbolFontAlphabet{\mathbbold}{bbold}
\usepackage{subcaption}

\usepackage{graphicx} % Required for inserting images
\usepackage{subcaption}
\usepackage{amsmath}
\usepackage{listings}
\usepackage{tabularx}
\usepackage{ amssymb }
\usepackage{cancel}
\usepackage{physics}
\usepackage{cancel}
\usepackage[linguistics]{forest}

\def\e0{\varepsilon_0}

\def\s0{\sigma_0}

\def\bk0{\mathbb{0}}

\title{The effect of fiber plasticity on domain formation in soft biological composites---Part II: An imperfection analysis}

\author{
 Dimitris Iordanidis \\
  Sibley School of Mechanical and Aerospace Engineering\\
  Cornell University, Ithaca, NY, USA \\
   \And
 Fernanda F. Fontenele \\
  Sibley School of Mechanical and Aerospace Engineering\\
  Cornell University, Ithaca, NY, USA \\ 
  \And
 Konstantinos Poulios \\
  Department of Civil and Mechanical Engineering \\
  Technical University of Denmark, Kgs. Lyngby, Denmark \\ 
  \And
 Michalis Agoras \\
  Department of Mechanical Engineering\\
  University of Thessaly, Volos, Greece \\
  \texttt{agoras@uth.gr} \\
  \And
 Nikolaos Bouklas \\
  Sibley School of Mechanical and Aerospace Engineering\\
  Cornell University, Ithaca, NY, USA \\ 
  \& Pasteur Labs, Brooklyn, NY\\
  \texttt{nbouklas@cornell.edu} \\
}

\begin{document}
\maketitle\footnote{DI and FFF have an equal contribution to the manuscript.}
\begin{abstract}

The main objective of this work is to numerically investigate the effect of geometric imperfections on the macroscopic response and domain formation in soft biological composites that exhibit plasticity in the stiff (fiber)
phase. 
This work builds on the corresponding bifurcation analysis in Part I of this study \cite{agoras2025effect} for simple laminates with perfectly flat layers under plane-strain, non-monotonic loading conditions, aligned with the layer direction. 
The post-bifurcation solution obtained in Part I for these materials corresponds physically to the formation of twin lamellar domains perpendicular to the loading axis,
which is consistent with the chevron-like deformation patterns that develop in tendons under cyclic loading. 
As biological materials are highly imperfect, and specifically tendons exhibit a high degree of so-called “crimp” in the collagen fibers, in this study the effect of imperfections to the response is explored. 
For all composites with small initial imperfections that have been considered, the results of this work have been found to be in complete agreement with the corresponding analytical results of Part I, and, domains have been found to emerge at a macroscopically compressive state.
However, as the imperfection amplitude is increased and becomes of the order to the layer width, or greater, domains begin to develop at macroscopically tensile stresses, which is in agreement with the fact that the loading of soft biological materials such as tendons and ligaments is tensile in nature. 
Thus, the findings of this work suggest strongly that plasticity and geometric imperfections of collagen fibers may play a key role on the onset and evolution of domains in actual soft biological composites.
\end{abstract}

% keywords can be removed
%\keywords{First keyword \and Second keyword \and More}

\section{Introduction}\label{Intro}

Unraveling the damage cascade in soft and fibrous tissues such as tendon and ligaments necessitates first understanding the mechanisms that control their response and the deformation patterns that emerge due to their load bearing function. Interestingly, these highly hydrated biological materials flip the traditional template for engineering design of composite materials --where ductility is introduced in the matrix phase--, as collagen fibers commonly exhibit ductile behavior \cite{tang2010deformation,veres2013repeated,fontenele2023understanding} whereas deformations of the matrix phase (composed of elastin, small leucin-rich repeat
proteoglycans, glycosaminoglycans and tenocytes) are mostly recoverable. Additionally, they exhibit phenomenal resistance to failure while undergoing millions of loading cycles and large deformations. As such, a thorough understanding of their microstructural characteristics and multiscale response is of interest not just to understand the progression of diseases like tendinopathy, but also to guide the design of the next generation of composites for advanced applications. One intriguing phenomenon that is central to this study, is that in cyclically loaded tendon, collagen fibers collectively form localized kinks and repeating chevron-like patterns \cite{veres2014mechanically,herod2016collagen} significantly disrupting the tenocyte microenvironment and are treated as early markers of tendinopathy \cite{fung2009subrupture,andarawis2011tendon}. These patterns are reminiscent of chevron folding in layered rock formations, but also of the patterns of {\it{Lopha cristagalli}} seashell as extensively discussed in \cite{bigoni2016folding}. Improved understanding and predictive capabilities for this phenomenon is also one of the targets of  Part I \cite{agoras2025effect} of this study\footnote{The work of Part I in \cite{agoras2025effect} for brevity will just be referred as Part I further in the manuscript.}.

\begin{figure}[ht!bp]
  \centering
    \includegraphics[width=0.15\linewidth]{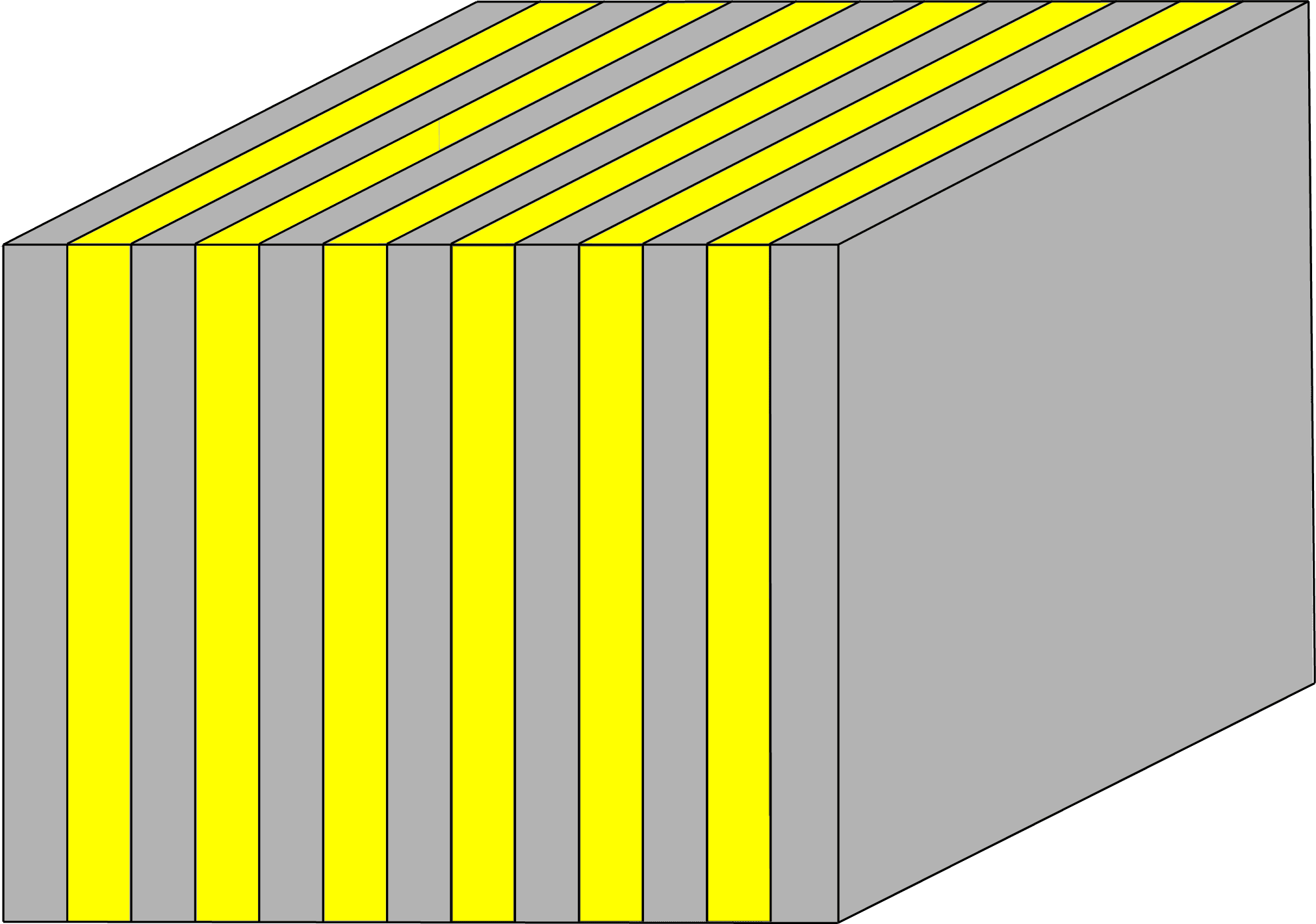}
    \caption{\label{fig:RVE} Schematic of the two-phase laminate composite.} 
\end{figure}

In Part I of this work, a comprehensive analysis of the effect of fiber plasticity on the macroscopic response and
domain formation is presented. 
%%MA
%This approach extended the idea presented in the work of \cite{fontenele2022fiber} in which the microstructure of the laminate approximation was implicitly incorporated into the model by considering the directionality of the fibers in their Voigt approximation. 
%
The effect of fiber plasticity on the critical conditions for the emergence of these domains has also been studied by \cite{fontenele2022fiber} in the context of a Voigt-type homogenization model, incorporating transversely isotropic fiber behavior.
For simplicity, the study in Part I focused on the homogenized behavior of two-phase laminates, comprised of purely elastic (matrix phase) and elastoplastic (fiber phase) layers. First, a bifurcation analysis examining the loss of ellipticity of the homogenized composite in monotonic and nonmonotonic aligned loading under plane strain conditions was presented, and consequently explored the long-wavelength post-bifurcation response and corresponding formation of a  rank-2 laminate with a twinned micrustructure at the mesoscopic level assuming constant per-phase fields..
It was shown that the domain formation for the rank-2 laminate introduced a softer response ,compared to the principal path, which acts as a stress-relaxation mechanism, resulting  primarily through microlayer rotation combined with increased longitudinal shear of the soft matrix phase. These findings, and especially the emerging deformation patterns are consistent with those observed in studies of tendon \cite{veres2014mechanically,herod2016collagen}. Additionally, while considering an elastoplastic fiber phase, the results of Part I are in qualitative agreement with respect to both domain formation and macroscopic response with the corresponding results of \cite{furer2018macroscopic} for laminates with purely Neo-Hookean phases; remarking also on the consistency of the response with \cite{d2016localization} even though a rank-2 laminate was not specifically targeted in their study. Domain formation was predicted during the compressive part of a load cycle, subsequent to tensile loading, accumulation of plastic deformation and elastic unloading. 

%%MA
%The aforementioned study presented in Part I, required taking into account the complexity of the underlying fibrous microstructure as well as the material constitutive nonlinearities. 
In Part I of this work, we employed simple lamellar microstructures as two-dimensional approximations of the actual fibrous microstructures of interest and we accounted for both constitutive and geometric nonlinearities at the level of the constituent layers.
A key restriction of the corresponding bifurcation and post-bifurcation analysis of Part I lies in the assumption of a geometrically perfect sample microstructure, which is fundamental towards obtaining simple analytical expressions that can be solved without resorting to a numerical partial differential equation solver. To obtain higher fidelity approximations of the deformation cascade, it is important to note that the hierarchical nature of biological composites (see \cite{baer1991hierarchical}, \cite{fratzl2003cellulose} and the references therein) exhibits imperfections at a variety of scales and through various mechanisms; from branching of collagen fibrils that contradict the description as a unidirectional long-fiber composite, to variations in the thickness of collagen fibers (assemblies of multiple fibrils), one of the more prominent imperfections is that of the existence of so--called ``crimp'' in the collagen fiber alignment. Crimp refers to some waviness in the geometry of the fibrillar microstructure in the unloaded, stress-free configuration, and is postulated to be crucial in the homeostasis and function of the tissue \cite{franchi2010tendon,jarvinen2004collagen}. As such, it is crucial to explore the effect of these imperfections in the response of the materials that are examined in this study.

Regarding experiments and modeling involving instabilities in laminate and fiber-reinforced composites, the reader is directed to our more in-depth literature review of the literature in Part I . To briefly summarize relevant experimental studies for engineering composites we highlight the work in \cite{vogler1997initiation} and \cite{kyriakides1995compressive} where in the event of an overload, these materials exhibit a sudden and catastrophic failure that is accompanied by highly localized (kink-type) zones of deformation. The underlying theoretical understanding had already been presented in \cite{budiansky1983micromechanics} --and later expanded in \cite{budiansky1993compressive}--. The mathematical idea that can indicate the emergence of localization is the loss of ellipticity in the governing equations, which allows for discontinuities to emerge. For a homogeneous elastoplastic material, \cite{rice1976localization} outlined the methodology for predicting localization by treating it as a bifurcation problem. 
%In a similar direction, 
\cite{triantafyllidis1985comparison} considered the microstructural bifurcation problem for periodic, layered composites and 
%%MA
%realized that potential unstable solutions to the incremental elastostatic equations were possible, thus further solidifying micro-buckling as the onset of macroscopic failure. It wasn't until recently, in the work of \cite{d2016localization} that it was shown that loss of ellipticity is a sufficient but not necessary condition for localization and that the post bifurcation response can be stable in a laminate composite. Furthermore, as it was mentioned previously, the work of \cite{furer2018macroscopic} made clear that the long-wavelength post-bifurcation behavior, for  laminates with purely Neo-Hookean phases, manifests itself with the formation of a rank-2 laminate.
%
showed that the onset of bifurcations of the long-wavelength type coincides with the loss of ellipticity of the incremental equations of equilibrium for the homogenized behavior of the laminate.
\cite{d2016localization} investigated the long-wavelength, post-bifurcation behavior of a class of hyperelastic laminates and found that the macroscopic response of these materials is stable for a wide range of local material behaviors, including the special case of Neo-Hookean phases, and that it becomes unstable only in situations where the incremental strength of the soft layers deteriorates significantly with increasing strain.
\cite{furer2018macroscopic} computed the relaxation of the effective stored-energy function for incompressible Neo-Hookean laminates and showed that the long-wavelength post-bifurcation solution described by the relaxed energy corresponds physically to the formation of a rank-2 laminate in the post-bifurcation regime.

This work concerns itself with the numerical analysis of the effect of geometrical imperfections in fiber-reinforced composites with fiber plasticity, while at the the same time aiming to validate and build on the results presented in Part I. Utilizing the local constitutive laws presented in Part I, the study examines the effect of perturbations to the initial geometry, while focusing on modeling a periodic representative volume element (RVE) with a lengthscale relevant to the laminate microsctructure and the imperfection wavelength, under plane strain conditions.  The paper is organized in the following manner: an outline of the relevant constitutive theories and corresponding finite element implementation for the RVE are presented in Section \ref{sec::methodology}. Subsequently, in Section \ref{sec::lastic} a laminate with purely Neo-Hookean phases, is examined under monotonic aligned compression utilizing a variety of perturbations to the geometry of the RVE under consideration. In Section \ref{sec::plastic} the more rich scenario of an elastoplastic laminate under non-monotonic, aligned loading is investigated. The concluding remarks are presented in Section \ref{sec::conclusion}.

\section{Theory and methodology}\label{sec::methodology}
The material under consideration is a two-phase laminate with periodically alternating layers (see Fig. (\ref{fig:RVE})). In the reference configuration the region occupied by the body is  $\Omega_0=\Omega^{(f)}_0\cup\Omega^{(m)}_0$, where $\Omega^{(r)}_0$  is the region of a homogeneous phase, with the superscripts (f),(m) denoting the fiber (stiff) and matrix (soft) phases respectively.

\subsection{Kinematics}\label{subsec::kinematics}
The positional vector of a material point in $\Omega_0$ is $\boldsymbol{X}$, while the current placement of the same point is given by the bijective function $\boldsymbol{y(\boldsymbol{X},t)}\mapsto \boldsymbol{x}$ and so the displacement field is $\boldsymbol{u}=\boldsymbol{x}-\boldsymbol{X}$. The deformation gradient is defined as $\boldsymbol{F}=\nabla_{\boldsymbol{X}}\boldsymbol{y}= \boldsymbol{I}+\nabla_{\boldsymbol{X}}\boldsymbol{u}$, where $\boldsymbol{I}$ is second order identity tensor and the subscript on the gradient operator will be implied from now on. No material interpenetration is allowed which requires that $J=\text{det}(\boldsymbol{F})>0$.

In this work, for the reasons explained in the introduction, the family of composites that is studied consists of, in general, elastoplastic phases (the case of pure elasticity is pertained within the elastoplastic framework). Our formulation assumes the usual multiplicative decomposition of the deformation gradient (\cite{lee1969elastic}), per phase, as:
\begin{equation}
\boldsymbol{F}^{(r)} = \boldsymbol{F}^{(r,e)}\boldsymbol{F}^{(r,p)}\label{MultComp}
\end{equation}
which introduces a fictitious intermediate configuration, that is produced by taking each material point, with its neighborhood, in the deformed state and elastically unloading it to the zero stress state.

\subsection{Constitutive theory}\label{subsec::constitutive}
Our general framework is that of J2-flow  plasticity theory at finite strains and under isothermal conditions, for isotropic materials. The yield criterion used, which specifies the elastic domain of a phase, is that of Von Mises with isotropic hardening defined over the intermediate configuration through the Mandel stress tensor $\boldsymbol{\Sigma}^{(r)}$ as :
\begin{equation}
\Phi\left(\boldsymbol{\Sigma}^{(r)},\Bar{\epsilon}^{(r,p)}\right) = \sqrt{\frac{3}{2}}\norm{\boldsymbol{\Sigma}^{d^{(r)}}} -\Sigma_y^{(r)} + \beta(\Bar{\epsilon}^{(r,p)})\leq0, \quad \boldsymbol{\Sigma}^{(r)}=\boldsymbol{F}^{(r,e)^T}\boldsymbol{\tau}^{(r)}\boldsymbol{F}^{(r,e)^{-T}}\label{YieldCriterion}
\end{equation}
with $\boldsymbol{\Sigma}^{d^{(r)}}$, $\Sigma_y^{(r)}$ being the deviatoric Mandel stress and initial yield limit respectively. The Kirchhoff stress tensor $\boldsymbol{\tau}^{(r)}$ is also introduced, along with the, in general, non-linear hardening function $\beta(\Bar{\epsilon}^{(r,p)})$  of the accumulated plastic strain $\Bar{\epsilon}^{(r,p)}\geq0$.

We assume that each phase poses a free energy function of the following uncoupled form:

\begin{equation}
\psi^{(r)}(\boldsymbol{b}^{(r,e)},\Bar{\epsilon}^{(r,p)}) = \psi^{(r,e)}(\boldsymbol{b}^{(r,e)}) + \psi^{(r,p)}(\Bar{\epsilon}^{(r,p)})
\end{equation}
where $\psi^{(r,e)}(\boldsymbol{b}^{(r,e)})$ is an objective, isotropic potential, depending on the elastic left Cauchy-Green tensor $\boldsymbol{b}^{(r,e)} = \boldsymbol{F}^{(r,e)}\boldsymbol{F}^{(r,e)^T}$, that accounts for the stored elastic energy per unit of undeformed volume. We also have the plastic potential $\psi^{(r,p)}(\Bar{\epsilon}^{(r,p)})$ which is taken to be a convex function describing the isotropic hardening mechanism via the relation $\beta = -\partial_{\Bar{\epsilon}^{(r,p)}}\psi^{(r,p)}$. 

By exploiting the Coleman-Noll produce, along with the maximum dissipation principle one obtains the elastic constitutive law (\ref{CollNollPros}), the associative flow rule (\ref{FlowRule}) that dictates plastic deformation, and the evolution equation of our plastic internal variable (\ref{EvolutionEqu}), written as:
\begin{align}
&\boldsymbol{\tau}^{(r)} = 2 \boldsymbol{b}^{(r,e)}\dfrac{\partial \psi^{(r)}}{\partial \boldsymbol{b}^{(r,e)}} \Rightarrow \boldsymbol{P}^{(r)}=\boldsymbol{\tau}^{(r)}\boldsymbol{F}^{(r)^{-T}} \label{CollNollPros} \\ 
&\dot{\boldsymbol{F}}^{(r,p)} = \dot{\Lambda}^{(r)}\boldsymbol{M}^{(r)}\boldsymbol{F}^{(r,p)}, \quad \boldsymbol{M}^{(r)} = \frac{\partial \Phi}{\partial \boldsymbol{\Sigma}}=\frac{\sqrt{3}}{\sqrt{2}\norm{\boldsymbol{\Sigma}^{d^{(r)}}}}\boldsymbol{\Sigma}^{d^{(r)}}\label{FlowRule} \\ 
&\dot{\Bar{\epsilon}}^{(r,p)}=\dot{\Lambda}^{(r)} \label{EvolutionEqu}
\end{align}
where $\dot{\Lambda}^{(r)}\geq0$ is the Lagrange/plastic multiplier introduced for the maximization problem and $\boldsymbol{P}^{(r)}$ is the 1st Piola-Kirchhoff stress tensor. In the associative framework of plasticity, the direction of the plastic flow $\boldsymbol{M}^{(r)}$ is normal to the yield surface.

The plastic multiplier and the yield function need to satisfy the following condition:
\begin{equation}
\dot{\Lambda}^{(r)} \cdot \Phi\left(\boldsymbol{\Sigma}^{(r)},\Bar{\epsilon}^{(r,p)}\right) = 0 \label{KKT}
\end{equation}
which conveys  that during elastic deformation when $\Phi<0 \Rightarrow \dot{\Lambda}^{(r)}=0$, and for continuous plastic loading when $\dot{\Lambda}^{(r)}>0$, we have the consistency condition $\dot{\Phi} = 0$. For a detailed analysis of isotropic plasticity the interested reader is referred to the works of \cite{simo1988framework}, \cite{miehe1994associative}, \cite{simo1992associative}.

In alignment with the analytical work in part I, the free energy function and the derived elastic constitutive law are specialized to:
\begin{align}
&\psi^{(r)}(\boldsymbol{b}^{(r,e)},\Bar{\epsilon}^{(r,p)}) =\frac{\mu^{(r)}}{2}\left(\text{Tr}(\boldsymbol{b}^{(r,e)})-3-2\ln{J^{(r)}}\right) +\frac{\beta^{(r)}}{2}(J^{(r)}-1)^2  +\frac{1}{2}h^{(r)}\bar{\epsilon}^{(r,p)^2}\label{ConstLaw} \\[10pt]
&\boldsymbol{\tau}^{(r)} = \mu^{(r)}\boldsymbol{b}^{(r,e)} +\left( \beta^{(r)} (J^{(r)}-1)J^{(r)} - \mu^{(r)}\right) \boldsymbol{I}
\end{align}
where $\mu^{(r)},\kappa^{(r)}$, are the small strain shear and bulk moduli respectively, with $\beta^{(r)}=\kappa^{(r)}-2\mu^{(r)}/3>0$. Due to the plastic incompressibility ($J^{(r,p)}=1$), which follows from the pressure insensitivity of our yield criterion, we use the same symbol for the determinant of the deformation gradient ($J^{(r)}\equiv J^{(r,e)}$). The current form of the plastic potential gives rise to a linearly hardening material.

\subsection{Flow rule integration scheme}\label{subsec::flow}
The flow theory of plasticity employed necessitates the incremental treatment of the problem. For that we consider the time-discrete solutions of our primary variables at steps $t_0=0, t_1,\dots, t_n,t_{n+1},\dots,T$. The time integration of the flow rule is tackled using the scheme presented in \cite{pereda1993finite}, which in a discrete step from $t_n$ to $t_{n+1}$ yielding the following approximate result:
\begin{equation}
\boldsymbol{F}_{}^{(r,e)} = \boldsymbol{F}^{(r)} \cdot \boldsymbol{F}_{n}^{(r,p)^{-1}}\left[\boldsymbol{I}-\Delta\Lambda^{(r)}\boldsymbol{M}^{(r)}(\boldsymbol{F}_{n}^{(r,e)}) +  \frac{1}{2}\left(\Delta\Lambda^{(r)}\boldsymbol{M}^{(r)}(\boldsymbol{F}_{n}^{(r,e)})\right)^2 \right] \label{TimeInteg}
\end{equation}
where quantities with the subscript "n" refer to the state of the body at the previous time instant -- that is considered as the reference configuration within the current step and which for simplicity we write as $\Omega_0^{(r)}$ instead of $\Omega_{0,n}^{(r)}$ -- and $\Delta \Lambda^{(r)}=\bar{\epsilon}^{(r,p)}-\bar{\epsilon}^{(r,p)}_n$ is the magnitude of the plastic strain increment. We emphasize that equation (\ref{TimeInteg}) entails both the unknowns $\Delta \Lambda^{(r)}$, $\boldsymbol{F}^{(r)}$ and the known plastic $ \boldsymbol{F}_{n}^{(r,p)^{-1}}$ and elastic $ \boldsymbol{F}_{n}^{(r,e)}$ states from the previous loading step. This implies that information needs to be stored after each successful application of a loading increment.

\subsection{Finite element implementation}\label{subsec::FE}
The strong forms of the boundary value problems for each phase, neglecting body and inertia forces, can now be written, on the reference configuration, as:
\begin{equation}
\nabla \cdot \boldsymbol{P}(\boldsymbol{F}^{(r,e)}) = \boldsymbol{0} \quad \text{in} \quad \Omega_0^{(r)}, \qquad
\boldsymbol{u}(\boldsymbol{X}) = \hat{\boldsymbol{u}}^{(r)} \quad \text{on} \quad \partial_u\Omega_0^{(r)}, \qquad
 \boldsymbol{P}^{(r)}\cdot \boldsymbol{N} = \hat{\boldsymbol{T}}^{(r)} \quad \text{on} \quad \partial_T\Omega_0^{(r)}
\end{equation}
where $\partial_T\Omega_0^{(r)},\partial_u\Omega_0^{(r)}$ - with $\partial_T\Omega_0^{(r)}\cap \partial_u\Omega_0^{(r)}=\emptyset$ and $\partial_T\Omega_0^{(r)}\cup \partial_u\Omega_0^{(r)}=\partial\Omega_0^{(r)}$ - are the portions of the boundary of a phase on which traction and displacement fields are prescribed respectively and are denoted with the ``\ \textasciicircum\ '' superscript. An admissible solution to these governing equations, which hold point-wise, is one that also respects the inequality of the yield criterion in equation (\ref{YieldCriterion}).

Our ultimate goal is to use the finite element method to discretize the problem and satisfy the governing equations in a weak sense. In order to do so, the principal of virtual work is used, which is appropriate for the non conservative system in hand. In this physical framework, given the reference configuration of the body we assume that a virtual displacement field $\delta \boldsymbol{u}(\boldsymbol{X})$, which satisfies the displacement boundary conditions in their homogeneous form, is prescribed. Requiring the virtual internal work produced by the body to equal the virtual work expended by the prescribed traction, we write the following expression to hold:
\begin{equation}
\int_{\Omega_0^{(r)}}\boldsymbol{P}^{(r)}: \delta \boldsymbol{F}dV - \cancel{\int_{\partial_T\Omega_0^{(r)}}\hat{\boldsymbol{T}}^{(r)}\cdot \delta \boldsymbol{u}dS} = 0 \label{VirtualWork}
\end{equation}
where the second term has been crossed off as further we will solely consider Dirichlet boundary conditions. 

Further, we specialize the  implementation to 2D boundary value problems without altering our notation.  The plane problems analyzed in this work are in plane strain conditions for which $F_{13}^{(r)}=F_{23}^{(r)}=F_{31}^{(r)}=F_{32}^{(r)} = 0$ and $F_{33}^{(r)}=1$. Since  $\delta \boldsymbol{F} = \nabla \otimes \delta \boldsymbol{u}$ only the in-plane components of $\boldsymbol{P}^{(r)}$ are involved in the equilibrium equations. The procedures outlined further in this section are carried out using the finite element package GetFEM, developed by \cite{GetFem}.

\subsection{Enforcement of periodic boundary conditions}\label{subsec::periodic}
\begin{figure}[ht!bp]
  \centering
    \includegraphics[width=0.40\linewidth]{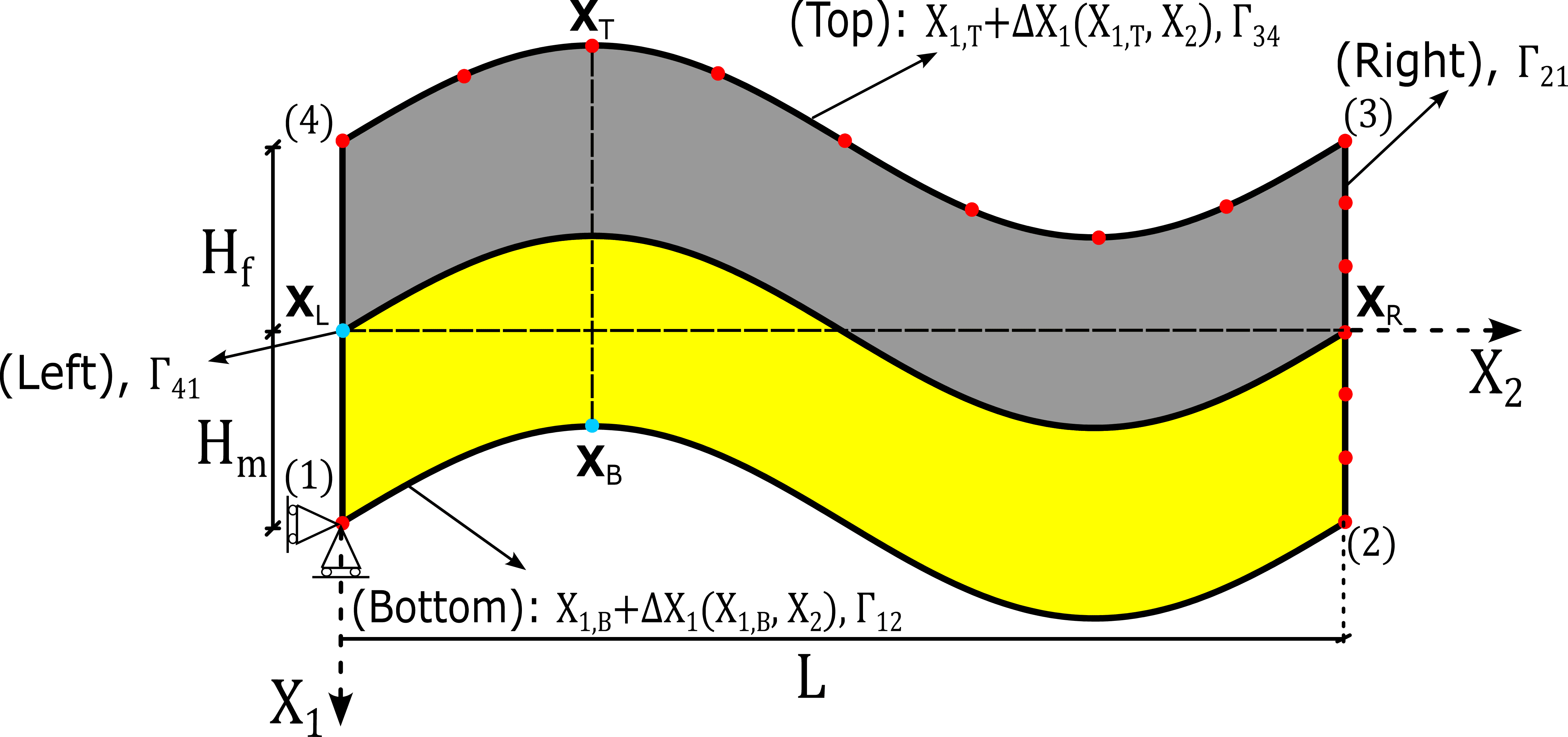}
    \caption{\label{fig:FEMsetup} Schematic for the perturbed geometry of the unit cell.} \label{FEMunitCell}
\end{figure}
The assumption of periodicity in material and geometric properties --as the latter relate to mesoscopic long-wavelength perturbations--,  allows the restriction of the domain to that of a unit cell/RVE, as explained in \cite{michel1999effective} and depicted in Fig. \ref{FEMunitCell}. The vertical faces of the undeformed unit cell do not need to be straight, but they are defined that way for simplicity. For any calculation performed on this computational domain aiming to simulate the behavior of a unit cell, periodic boundary conditions need to be prescribed (\cite{kouznetsova2001approach}). 

For the displacement controlled simulations performed in the present work, these conditions are comprised of the tying equations between the displacements of homologous boundary material points. Following, in order to retain the periodic shape and orientation of opposite boundary faces, we enforce the following restrictions:
\begin{align}
\boldsymbol{u}(\boldsymbol{X}_{R})-\boldsymbol{u}(\boldsymbol{X}_{L})-\Bar{\boldsymbol{H}}\cdot (\boldsymbol{X}_{R}-\boldsymbol{X}_{L})=\boldsymbol{0}\quad \text{and} \quad \boldsymbol{u}(\boldsymbol{X}_{T})-\boldsymbol{u}(\boldsymbol{X}_{B})-\Bar{\boldsymbol{H}}\cdot (\boldsymbol{X}_{T}-\boldsymbol{X}_{B}) =\boldsymbol{0}\label{PBC}
\end{align} 
The subscripts $(L)\rightarrow \text{Left}$, $(R)\rightarrow \text{Right}$, denote the vertical boundary regions $\Gamma_{41}, \Gamma_{23}$ at $X_2=0, X_2=L$ respectively, while the subscripts $(T)\rightarrow \text{Top}$, $(B)\rightarrow \text{Bottom}$, denote the curved boundary paths between the points (3),(4) and (1),(2) labeled by $\Gamma_{34}, \Gamma_{12}$ respectively. Given the unit cell of Fig. \ref{FEMunitCell}, if the positional vector of material point located on the right face is denoted as $\boldsymbol{X}_R$, then its homologous point located on the left face, denoted as $\boldsymbol{X}_L$, is acquired by traveling parallel to $X_2$-axis until the left face is reached. Similarly for points on the top face mapped to the respective points on the bottom face. To unify our notation we define $\Gamma_+=\Gamma_{23}\cup\Gamma_{34}$, $\Gamma_-=\Gamma_{41}\cup\Gamma_{12}$ and we assume the existence of a function $\boldsymbol{\zeta}: \Gamma_+ \mapsto \Gamma_-$ such that $\forall \boldsymbol{X}_+\in \Gamma_+$, $\boldsymbol{\zeta}(\boldsymbol{X}_+)$ returns its homologous point $\boldsymbol{X}_-$ on $\Gamma_-$ . The volume averaged displacement gradient $\bar{\boldsymbol{H}}$ is also introduced, through which a particular loading program is realized.
For the 2D problems treated here, at each simulation increment, all four components of the strain-like second order tensor $\bar{\boldsymbol{H}}$ should be considered as known/prescribed.

It is emphasized that these equations have been written in terms of material points and not using specific nodes of the discretized geometry, as the GetFEM implementation follows a Mortar-type approach. This is done in order to avoid dependencies on particular mesh structures, and subsequently equations (\ref{PBC}) are enforced in a weak sense. To do so, a new Lagrange multiplier field variable, denoted as $\boldsymbol{\lambda}(\boldsymbol{X})$, is introduced. This vector field, which assumes non-trivial values on the boundary face $\Gamma_{+}$, acts as a Lagrange multiplier responsible for prescribing the periodic boundary conditions. 
Consequently, the corresponding weak form for enforcing the intended periodicity conditions is:
\begin{align}
\int_{\Gamma_{+}}\left(\delta\boldsymbol{u}(\boldsymbol{X}_+)-\delta\boldsymbol{u}(\boldsymbol{X}_-) \right)\cdot \boldsymbol{\lambda}\;d\Gamma+\int_{\Gamma_{+}}\left(\boldsymbol{u}(\boldsymbol{X}_+)-\boldsymbol{u}(\boldsymbol{X}_-)-\Bar{\boldsymbol{H}}\cdot (\boldsymbol{X}_{+}-\boldsymbol{X}_{-})\right)\cdot \delta\boldsymbol{\lambda}\;d\Gamma =0  \label{FullWeakForm}
\end{align}
The first integral is meant to be combined with the weak form from equation.~\eqref{VirtualWork}, while the second integral introduces the constraint equations which are necessary in order to solve for the multiplier field $\boldsymbol{\lambda}$.

\subsection{Nonlinear solution scheme}\label{subsec::nlscheme}

Let $\Omega_0^{(r,h)}\subset \mathbb{R}^2$ denote the discretized geometry corresponding to a perturbed lamellar phase in the unit cell. In the same spirit, we denote with $\boldsymbol{u}^{(r,h)}$, $\boldsymbol{\lambda}^{(h)}$ the discretized displacement and Lagrange multiplier fields respectively. Within the $e^{\mathrm{th}}$ element of a discretized unit cell, a field variable $\boldsymbol{\eta}(\boldsymbol{X})$ is interpolated as $\boldsymbol{\eta}^{(h)}= \boldsymbol{N}_e\cdot \boldsymbol{\eta}_e$, where $\boldsymbol{N}_e$ is the local matrix containing the shape functions and $\boldsymbol{\eta}_e$ is the vector containing the nodal degrees of freedom. Similarly we have $\nabla \boldsymbol{\eta}^{(h)}=\boldsymbol{B}_e\cdot \boldsymbol{\eta}_e$, where $\boldsymbol{B}_e$ is the local matrix containing the spatial derivatives of the shape functions (\cite{hughes2003finite}). Specifically, the geometry of the unit cell and the unknown fields $\boldsymbol{u}^{(r,h)}$, $\boldsymbol{\lambda}^{(h)}$ are discretized with 4-node linear elements. 

A solution to the discretized version of the weak equations (\ref{VirtualWork}) and (\ref{FullWeakForm}) is pursued using the Newton-Raphson (NR) method which requires the linearization of the equations about
the k$^{\mathrm{th}}$ step of the iterative process, eventually yielding a global linear system of the following form:
\begin{equation}
\mathbf{K}_{T}\rvert_{\boldsymbol{u}^h_k} \cdot \Delta \mathbf{v}^h_{k+1} = -\mathbf{R}\rvert_{\boldsymbol{u}^h_k} \label{LinearSystem}
\end{equation} 
where both the nodal unknowns $\boldsymbol{u}^h$ and $\boldsymbol{\lambda}^h$ have been lumped into the global vector $\boldsymbol{v}^h$, with $\Delta \boldsymbol{v}_{k+1}^h=\boldsymbol{v}_{k+1}^h-\boldsymbol{v}_{k}^h$ being the increment within the NR algorithm. The last step before solving the linear system in hand is to modify it to account for the supports placed on node (1), necessary to avoid rigid body motions. This is achieved, by using a set of discrete Lagrange multipliers $\lambda_{R_{X_2}}, \lambda_{R_{X_1}}$. 

To respect the yield criterion inequality in our finite element approximation --given the converged outcome of the NR algorithm described above-- the yield function is evaluated on  Gauss points used for the numerical integration of the weak form. If the given state suggests that plasticity should have been developed, on a particular point, the plastic increment is calculated in order to satisfy the consistency condition, else it is set to zero if the state was in the elastic domain. The satisfaction of either case is simultaneously achieved by introducing  the following non-linear equation:   
\begin{equation}
    \max\left\{\sqrt{\frac{3}{2}}\norm{\boldsymbol{\Sigma}^d(\boldsymbol{F}^{(r,e)})} -\Sigma_y^{(r)} -h^{(f)}\cdot(\Bar{\epsilon}_n^{p}+\Delta \Lambda), -\Delta \Lambda\right\} = 0 \label{NumConsist}
 \end{equation}
This is handled, using the NR algorithm  at the integration points level. The now elastoplastically consistent state needs to also satisfy the equilibrium equations and if it does not, the global system is reformulated based on this new state. The process is repeated until both the global equilibrium equations and the local constitutive laws are satisfied.

After a successful increment, in preparation for the next loading step, the following updates are performed, pursuant to the evolution equation (\ref{EvolutionEqu}) and the multiplicative decomposition of the deformation gradient (\ref{MultComp}):
\begin{equation}
\bar{\epsilon}_n^{(p)} \leftarrow \bar{\epsilon}_n^{(p)} + \Delta\Lambda \qquad
\mathbf{F}_{n}^{(f,p)^{-1}} \leftarrow \mathbf{F}^{-1}\cdot\mathbf{F}_{}^{(f,e)^{-1}} \qquad
\mathbf{F}_{n}^{(f,e)} \leftarrow \mathbf{F}_{}^{(f,e)}
\end{equation}
These variables are stored at each Gauss point and for the first load step, the tensorial quantities are initialized with $\boldsymbol{I}$ and the scalar with zero. For all subsequent calculations, 7 Gauss points per element are utilized.
The finite element implementation is concluded by noting that by construction the interface between the 2 materials shares the same set of nodes, meaning that the requirement of displacement continuity on the interphase is strictly enforced. On the other hand, force equilibrium of that interphase is weakly satisfied. 

Additionally, the several line plots of averaged scalar quantities, presented in subsequent sections, have been calculated using the finite element solution, by performing the following integration:
\begin{equation}
\Bar{\xi} =\frac{1}{|\Omega_0|} \int_{\Omega_0} \xi(\boldsymbol{X})dV =\frac{|\Omega_0^{(f)}|}{|\Omega_0|}\frac{1}{|\Omega_0^{(f)}|} \int_{\Omega_0^{(f)}} \xi(\boldsymbol{X})dV + \frac{|\Omega_0^{(m)}|}{|\Omega_0|} \frac{1}{|\Omega_0^{(m)}|}\int_{\Omega_0^{(m)}} \xi(\boldsymbol{X})dV=c_0^{(f)}\Bar{\xi}^{(f)}+c_0^{(m)}\Bar{\xi}^{(m)}
\end{equation}

\section{Numerical investigation of a purely hyperelastic laminate.}\label{sec::lastic}
There are several targets for the investigation in this section. Restricting our attention to a laminate with purely elastic phases, we first wish to compare the effect of perturbations to the response that was predicted analytically for the unperturbed laminate geometry. 
%This can be achieved by focusing on the primary path, loss of ellipticity (LOE) and subsequent post-bifurcation path presented in \cite{furer2018macroscopic}, using a Neo-Hookean model for both phases, or by specializing the results from Part I for the purely elastic case which coincide with the latter. 
First, we test the case of a sinusoidal imperfection, and then proceed to compare to the perturbation proposed in \cite{d2016localization}. Finally, the effect of imperfection wavelength in the response is explored.
\subsection{Baseline case}
We start by considering purely elastic phases with the Neo-Hookean potential described in equation (\ref{ConstLaw}). A unit cell with a sinusoidal perturbation of the lamellar geometry (shown in Fig. \ref{FEMunitCell}) is employed, motivated from crimp imperfections observed in tendon \cite{jarvinen2004collagen}, with 
\begin{equation}
    \Delta X_1= \Delta X_1^{S}(X_2)=-\alpha \sin(w\cdot 2X_2\pi/L)
\end{equation}
%to study the response of the composite;
where $\alpha$ and $w$ are the parameters controlling the amplitude and wavelength of the perturbation, while $L$ denotes the length of the unit cell. For the subsequent numerical results (contours and line plots) we maintain notational consistency with the presentation of the companion article Part I.

%It is important to note that we study perturbations where $L_{\mathrm{macro}}>>L>>H$; which  holds true for the remainder of the study.  
%%MA
%$L_{\mathrm{macro}}$ is the lengthscale describing the size of the macroscopic domain, and $L$ the length of the unit cell under consideration --note the adjusted definition of lengths and size of the unit cell with respect to Part I--.    
Only 2 layers need to be considered for the unit cell, which are taken to have initial dimensions $H_m=H_f=H=L/100$ and thus the corresponding initial volume fractions are $c_0^{(f)}=c_0^{(m)}=0.5$. The parameter values that are selected for the imperfection are $\alpha=10^{-4}L=10^{-2}H$ and $w=1$. To better visualize the pertinent characteristics of the numerical solution, 10 of these unit cells are assembled into a larger scale RVE, but the computation takes place in a single unit cell to minimize computational cost.
%%ΜΑ
%To maintain consistency of this study with that of \cite{furer2018macroscopic} and also Part I, we conduct a convergence analysis with respect to the analytical results, where a unit cell of a decreasing $H/L$ ratio is considered 
%
A convergence analysis of the numerical simulations with respect to the ratio $H/L$ is given in Appendix A.

\begin{figure}[h!tbp]
  \centering
  \subfloat[$\bar{\lambda}_2=0.95$]{\includegraphics[width=0.18\textwidth]{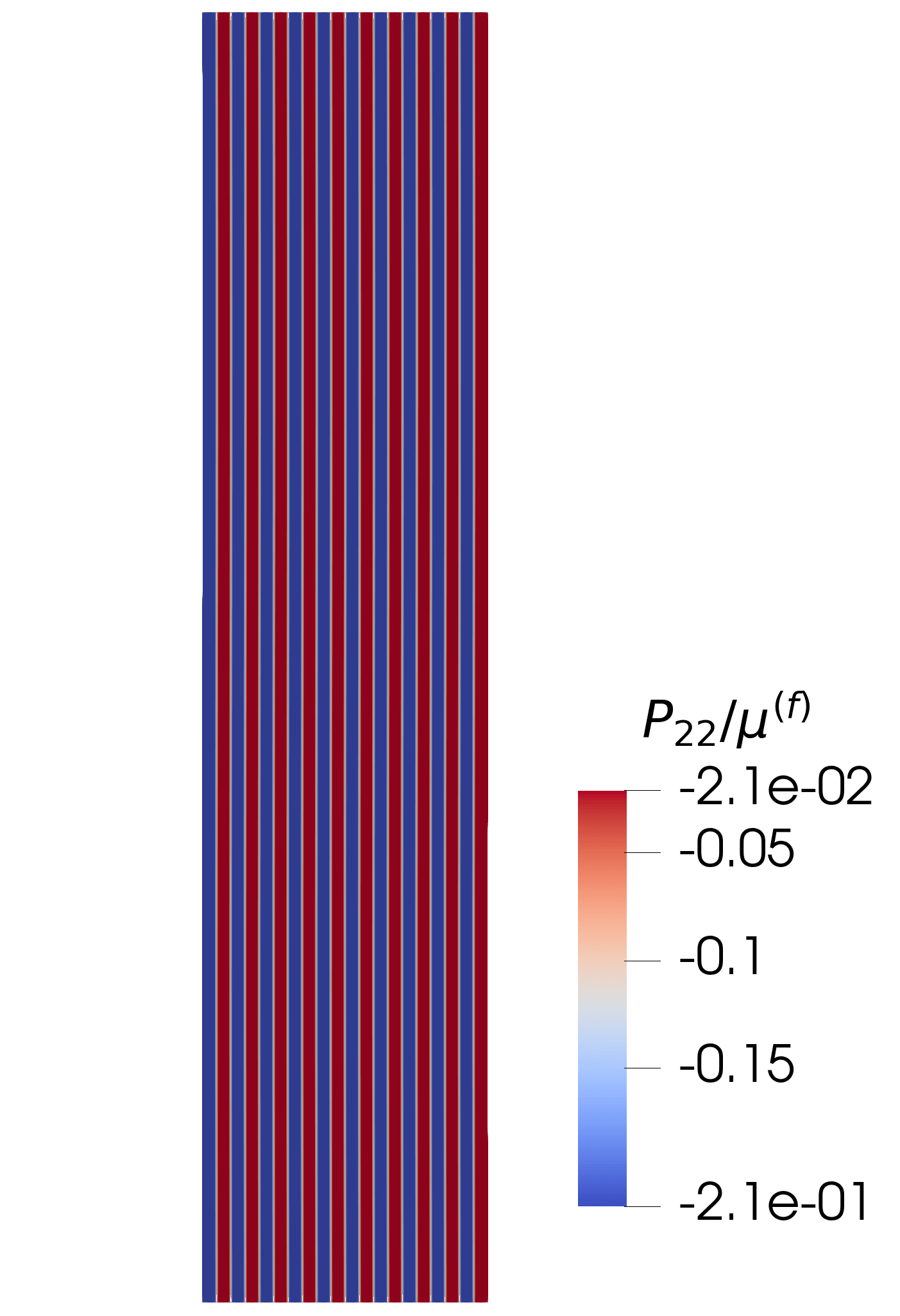}\label{fig:OpeningStep25}}
  \hspace{0mm}
  \subfloat[$\bar{\lambda}_2=0.903$ (LOE)]{\includegraphics[width=0.18\textwidth]{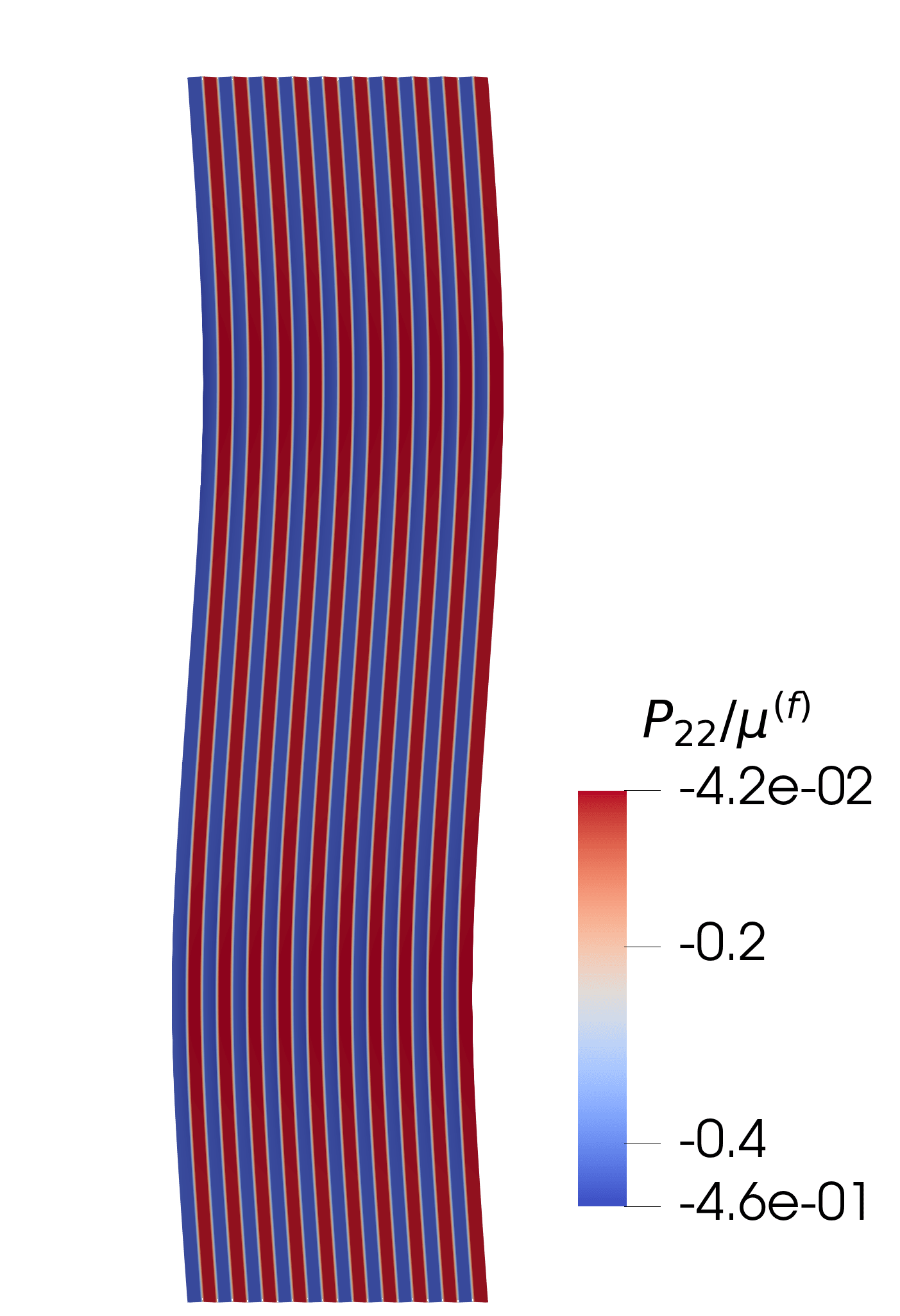}\label{fig:OpeningStepLOE}}

 \vspace{0mm}
  
  \subfloat[$\bar{\lambda}_2=0.85$]{\includegraphics[width=0.18\textwidth]{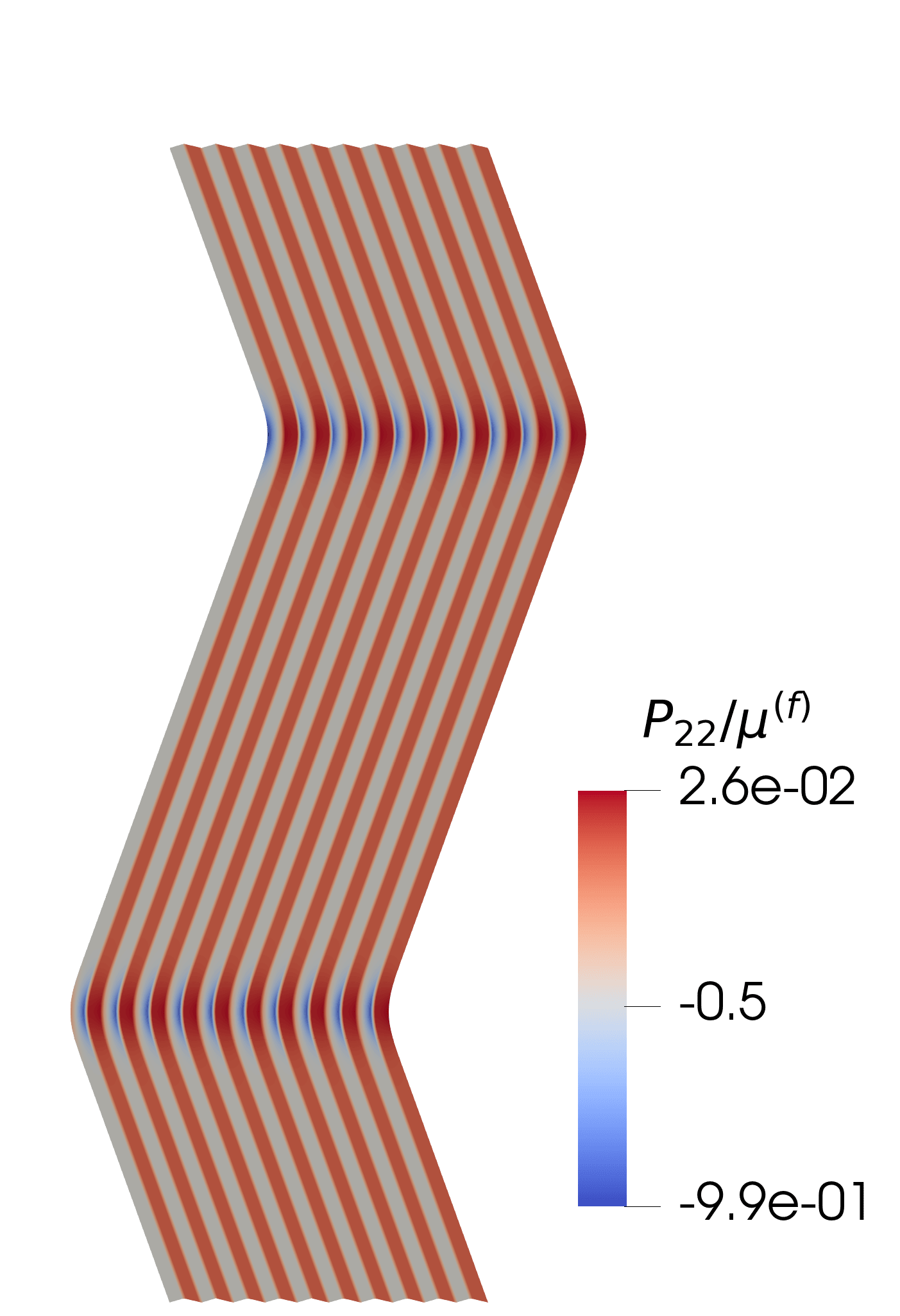}\label{fig:OpeningStep75}}
  \hspace{0mm}
  \subfloat[$\bar{\lambda}_2=0.8$]{\includegraphics[width=0.18\textwidth]{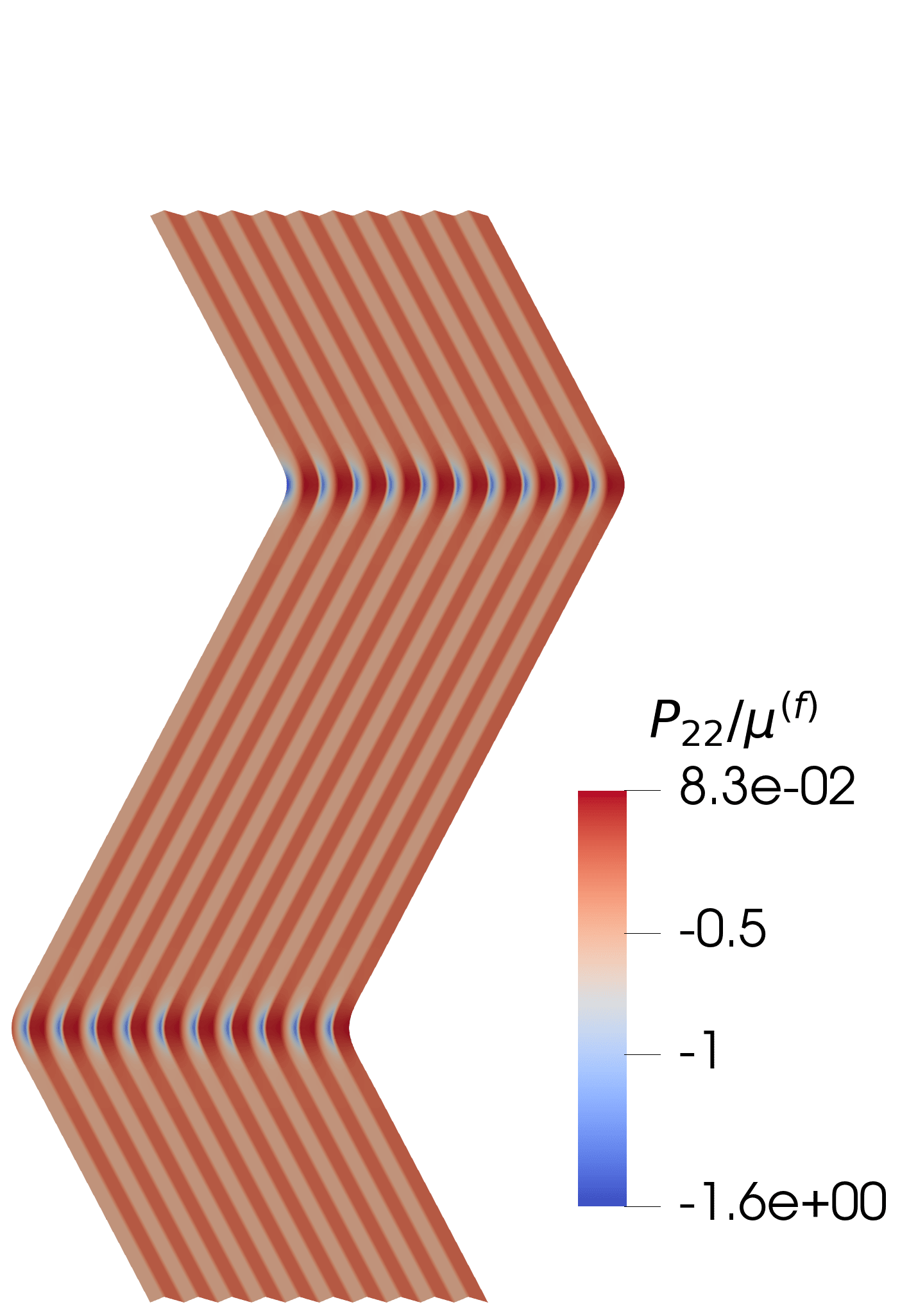}\label{fig:OpeningStep100}}
  \caption{Contour of $P_{22}/\mu^{(f)}$ for purely elastic phases with $\mu^{(f)}/\mu^{(m)}=10$, $\kappa^{(f)}=100\mu^{(f)}$,$\kappa^{(m)}=100\mu^{(m)}$, under monotonic compression.}\label{fig:OpeningFIG}
\end{figure}

In our baseline case, the stiffness contrast is $\mu^{(f)}/\mu^{(m)}=10$ while the phases are modeled as slightly compressible with $\kappa^{(f)}=100\mu^{(f)}$, $\kappa^{(m)}=100\mu^{(m)}$ (see the appendix for a parametric study on the effect of elastic coefficients). The RVE is subjected to monotonic compression, along the fiber direction, which is realized by setting $u_{X_2}^{(2)}=0$ and prescribing $\bar{H}_{22}=\bar{\lambda_2} -1$. The evolution of the deformed configuration is shown in Fig. \ref{fig:OpeningFIG}.

\begin{figure}[h]
  \centering
    \includegraphics[width=0.45\linewidth]{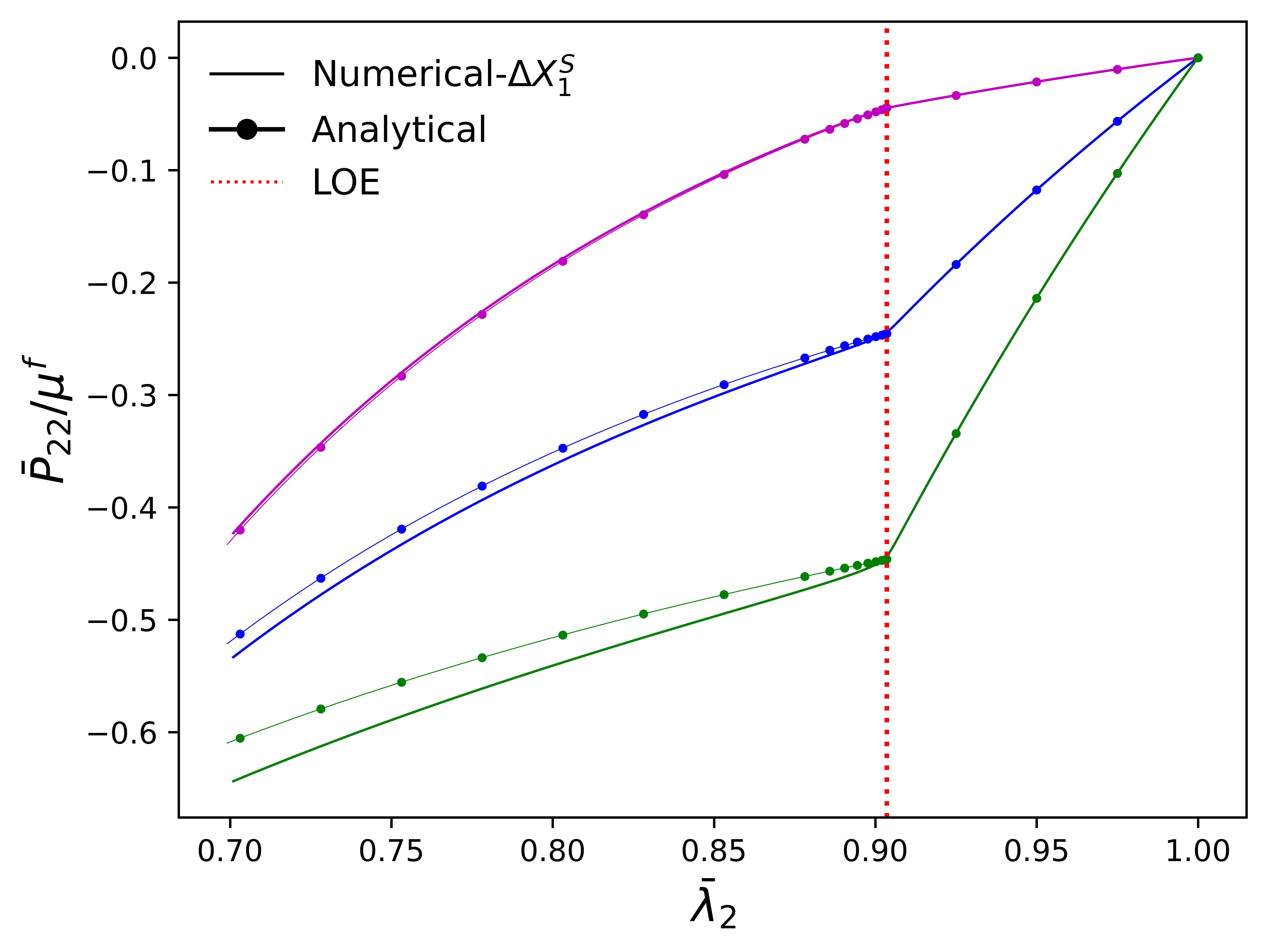}
    \caption{\label{fig:Response_vs_analytical} 
    Comparison between the analytical solution for perfect laminates and the numerical solution for the sinusoidal geometric perturbation. The material properties for the laminates, which are subjected to monotonic compression, are $\mu^{(f)}=10\mu^{(m)}$, $\kappa^{(f)}=100\mu^{(f)}$,$\kappa^{(m)}=100\mu^{(m)}$. The blue color is used for the composite, the green for the fiber phase and the pink for the matrix. The vertical red line indicates the loss of ellipticity in the analytical solution.} 
\end{figure}

\begin{figure}[h!tbp]
  \centering
  \subfloat[$\bar{\lambda}_2=0.8$]{\includegraphics[width=0.35\textwidth]{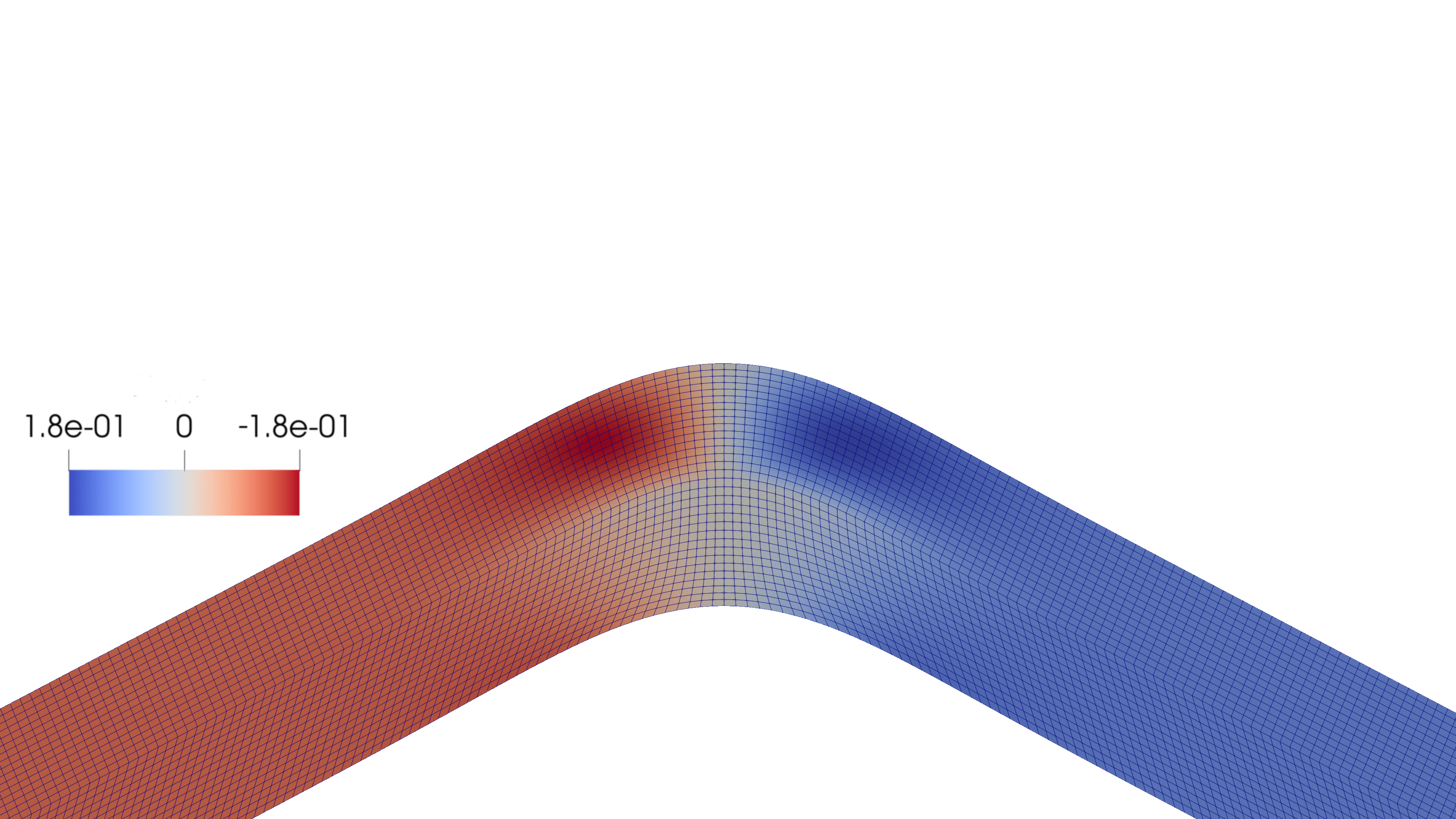}\label{fig:TRZoomStep1}}
  \hspace{0mm}
  \subfloat[$\bar{\lambda}_2=0.45$]{\includegraphics[width=0.35\textwidth]{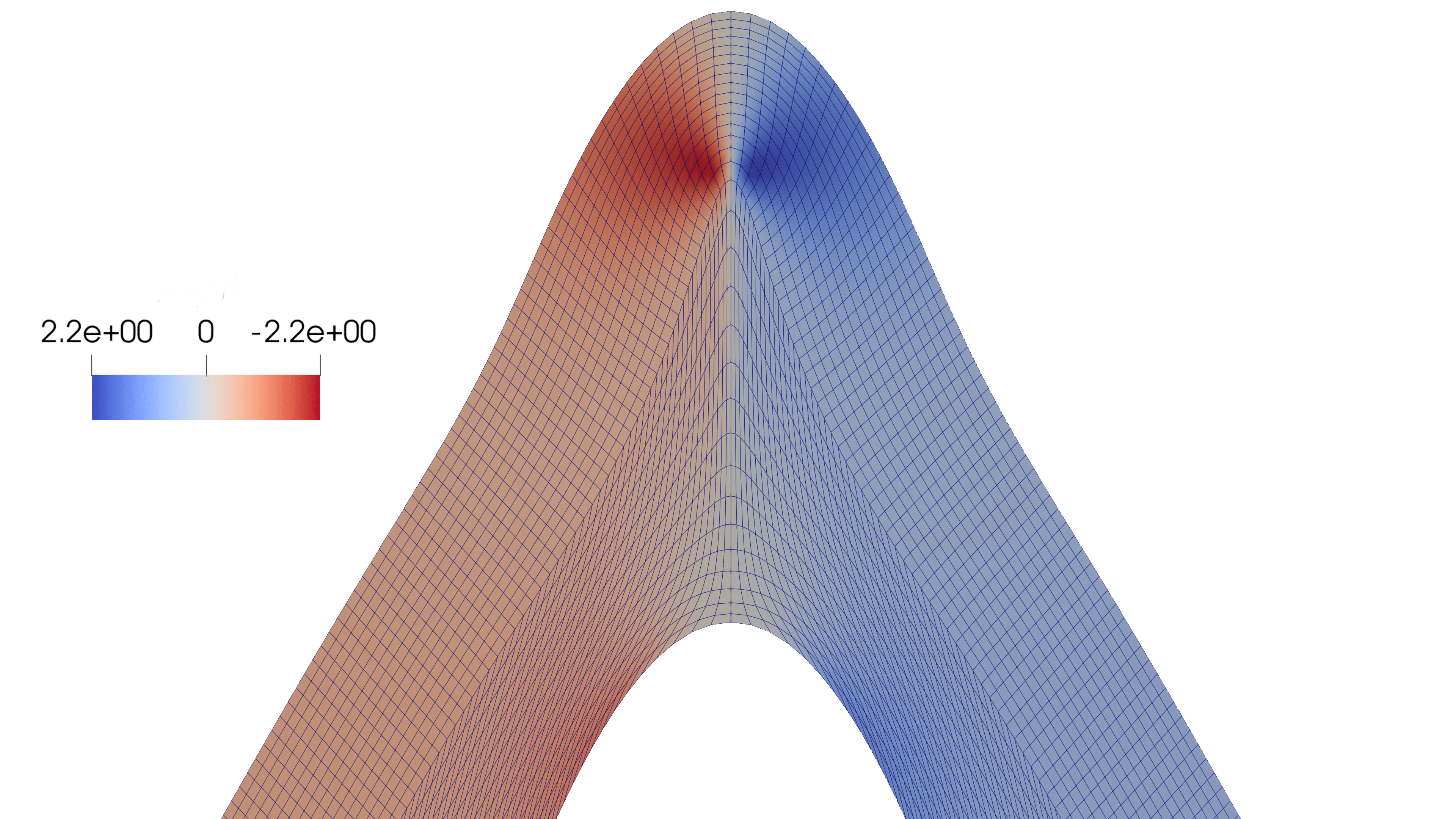}\label{fig:TRZoomStep2}}
  \caption{Contour of $P_{21}/\mu^{(f)}$ for purely elastic phases with $\mu^{(f)}/\mu^{(m)}=10$, $\kappa^{(f)}=100\mu^{(f)}$,$\kappa^{(m)}=100\mu^{(m)}$, under monotonic compression. The laminates are presented rotated. \label{fig:TRZoom}}
\end{figure}

For small applied deformations, the (approximately) homogeneous principal solution is captured (Fig. \ref{fig:OpeningStep25}). Due to the imperfection considered, a gradual transition from the principal to the post bifurcated solution is observed, with a more pronounced deviation from the homogeneous state consistent with the sinusoidal imperfection but with an increased amplitude (Fig. \ref{fig:OpeningStepLOE}). This occurs approximately at the point of loss of macroscopic ellipticity of the elastic analytical solution calculated in Part I for the unperturbed geometry. The numerically calculated response of the imperfect unit cell is presented in Fig. \ref{fig:Response_vs_analytical} where the analytical solution of the perfect unit cell is added for comparison. Further, in Fig. \ref{fig:OpeningStep75}, where the body has shifted into the post bifurcation regime, fields  are approximately constant per-phase, but the emergence of a clear transition region between each phase acts like a diffuse phase boundary for the rank-2 laminate. This formation of finite curvature transition regions is an intrinsic characteristic of the imperfection and can be correlated with the modeled physical system. The evolution of the transition regions is highlighted in Fig. \ref{fig:TRZoom}. Intense bending accompanies these regions and the local stresses get significantly higher, compared to those in the bulk of the material, as is evident by the contour in Fig. \ref{fig:OpeningStep100}. A more detailed exploration of the stress response is presented in the appendix.

The anticipated behavior of the mesolayers being in a combined state of compression and equal magnitude, but different direction, of shear is also adequately captured, as is shown in Fig. \ref{fig:TRZoom}, while the volume fractions of the produced rank-2 laminate, $c^{(-)},c^{(+)}$ are calculated to be 1/2, in agreement with the findings of \cite{furer2018macroscopic} and Part I.

Focusing on the comparison between the averaged numerical solution and the analytical results form Part I --composed of the primary solution, LOE and post-bifurcation solution-- as depicted in Fig. \ref{fig:Response_vs_analytical} we see a good agreement in all parts of the response, including the softer post-bifurcation response. Consistent with the findings of Part I and the results in Fig. \ref{fig:OpeningFIG}, the softening is accommodated by rotation of the mesolayers and corresponding increased shear in the soft phase --note the element deformation per phase in Fig. \ref{fig:TRZoom}--. The localized deformation pattern of the soft phase  in Fig. \ref{fig:TRZoom}(b) at high level of compression is reminiscent of creases observed on free surfaces \cite{hohlfeld2011unfolding,hong2009formation,wu2013swell} and interfaces of soft materials \cite{dortdivanlioglu2021swelling}.

\begin{figure}[h!tbp]
  \centering
  \subfloat[$\bar{\lambda}_2=0.8$, $\bar{\gamma}=0$, $c^{(-)}=0.5$]{\includegraphics[width=0.22\textwidth]{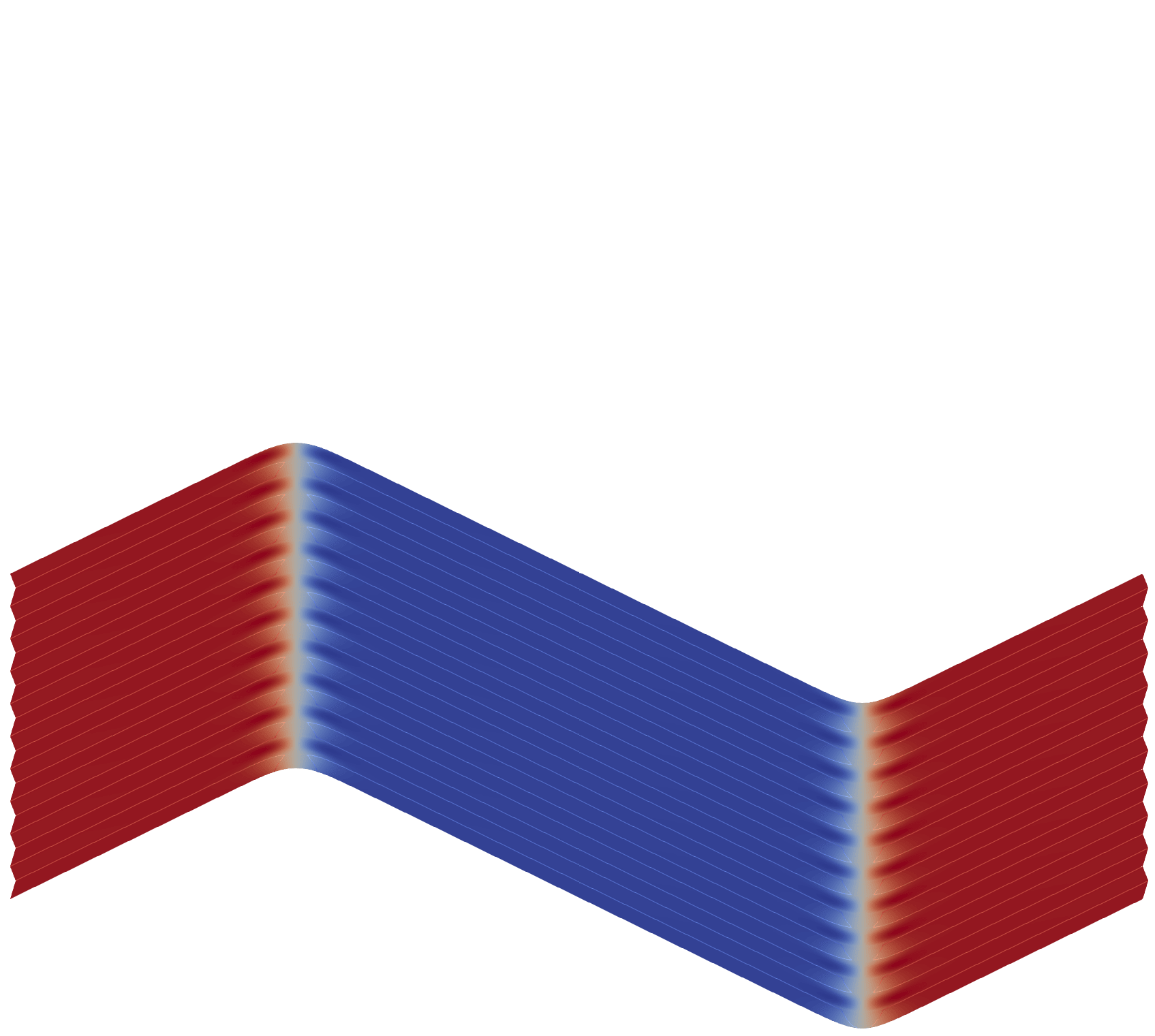}\label{fig:Transformation1}}
  \hspace{0mm}
  \subfloat[$\bar{\gamma}=0.10$, $c^{(-)}=0.62$]{\includegraphics[width=0.22\textwidth]{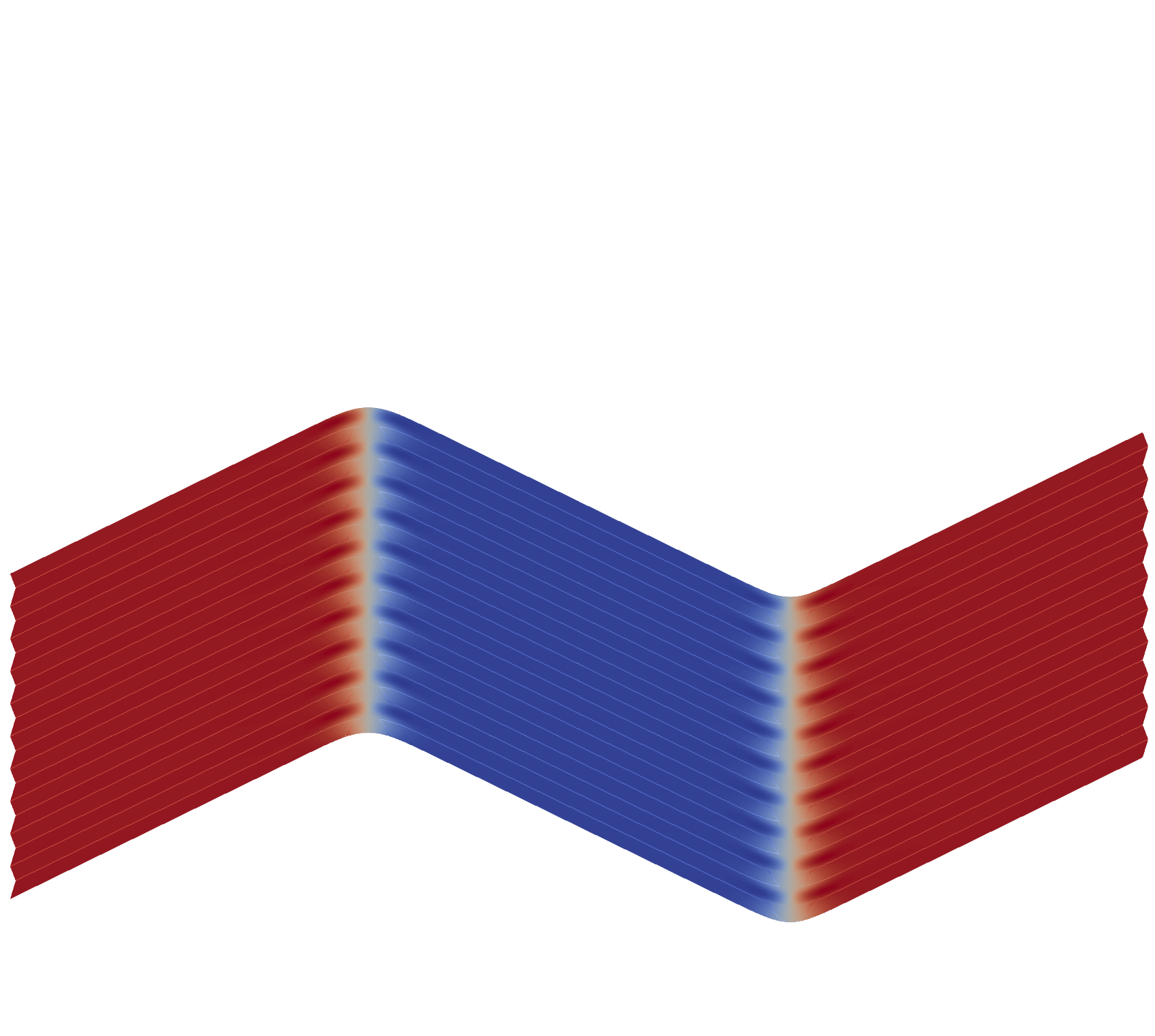}\label{fig:Transformation2}}

 \vspace{3mm}
  
  \subfloat[$\bar{\gamma}=0.20$, $c^{(-)}=0.735$]{\includegraphics[width=0.22\textwidth]{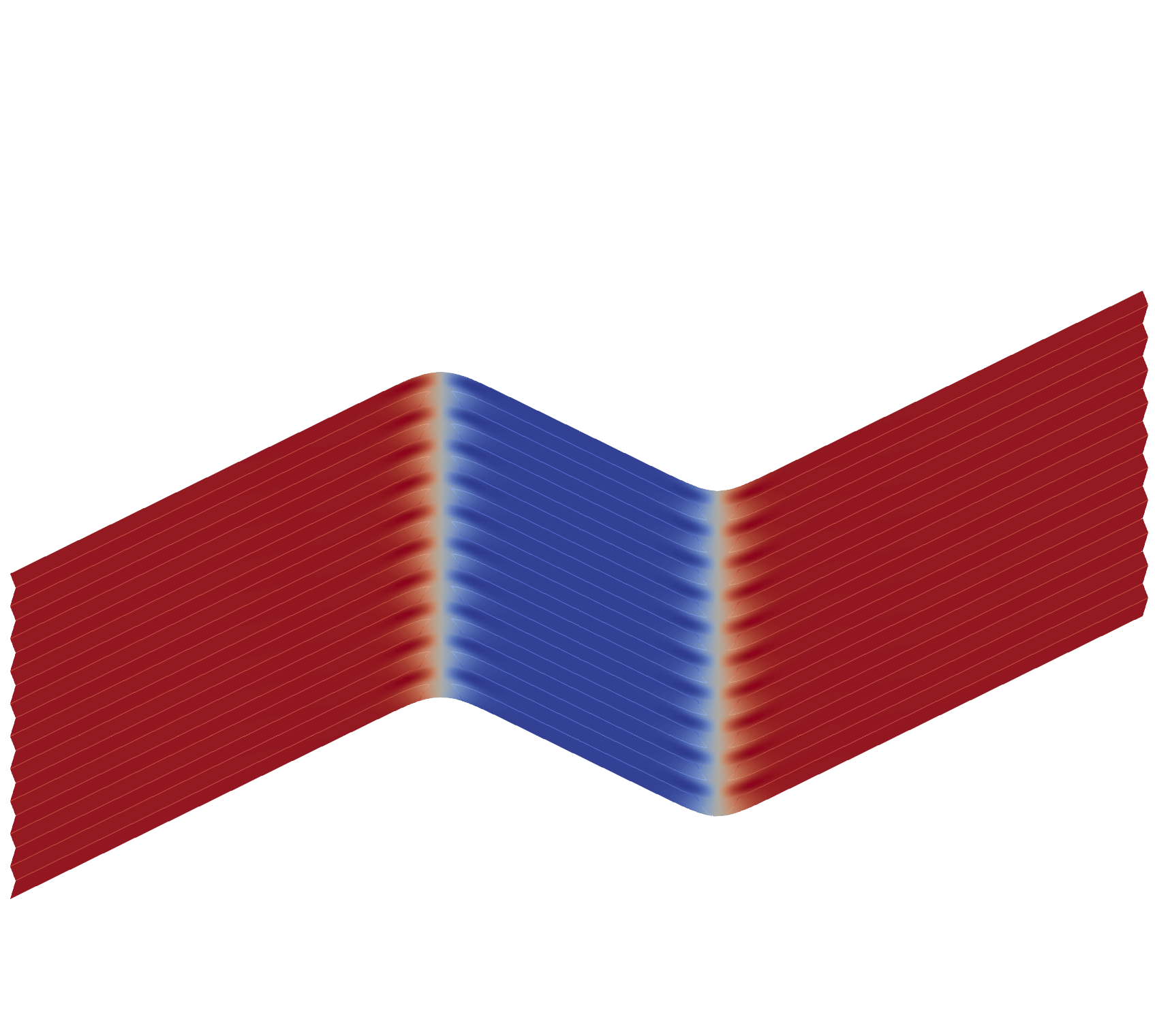}\label{fig:Transformation3}}
  \hspace{0mm}
  \subfloat[$\bar{\gamma}=0.30$, $c^{(-)}=0.855$]{\includegraphics[width=0.22\textwidth]{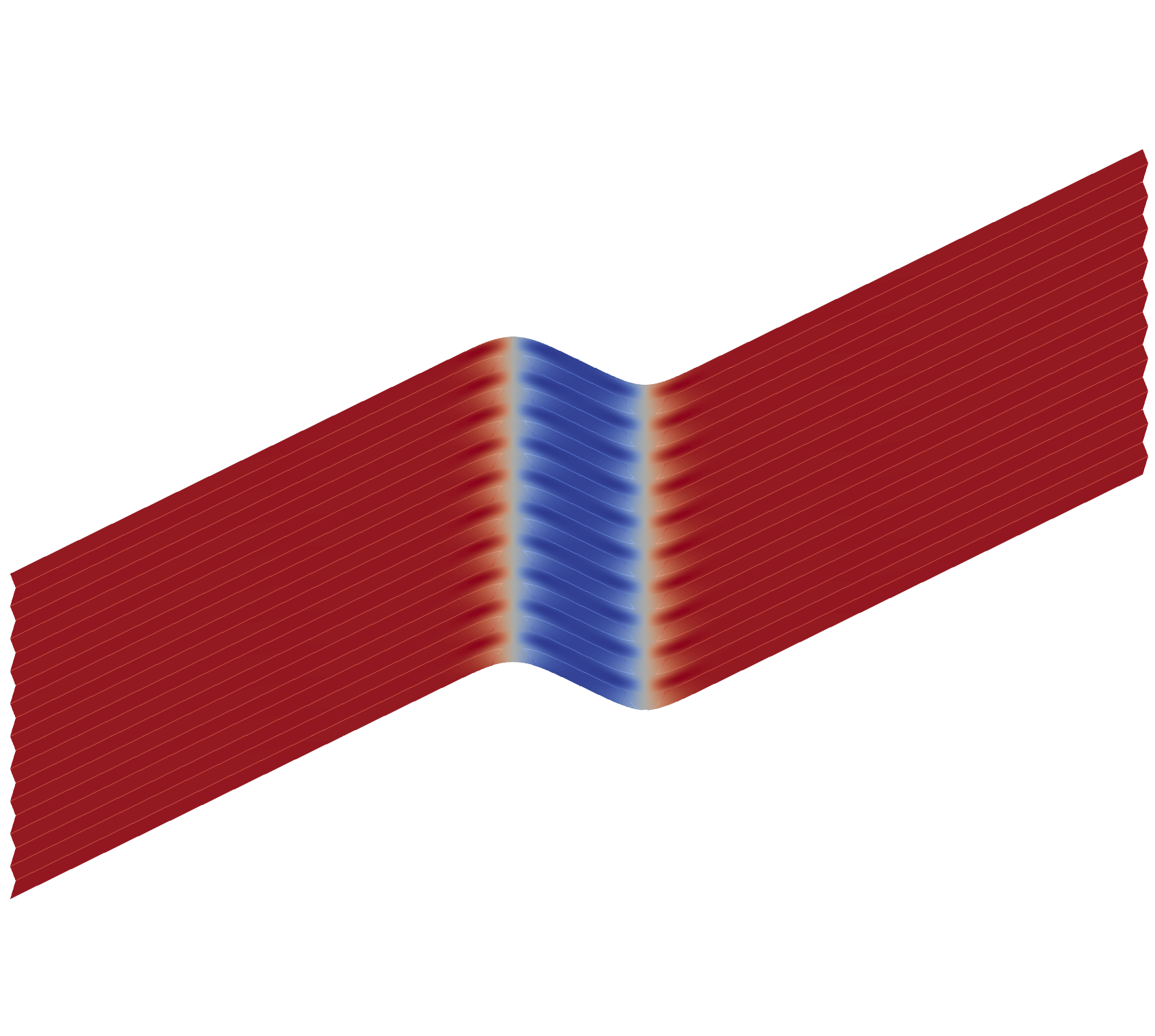}\label{fig:TRZoomStep4}}
  
   \vspace{3mm}
  
  \subfloat[$\bar{\gamma}=0.372$, $c^{(-)}=0.939$]{\includegraphics[width=0.23\textwidth]{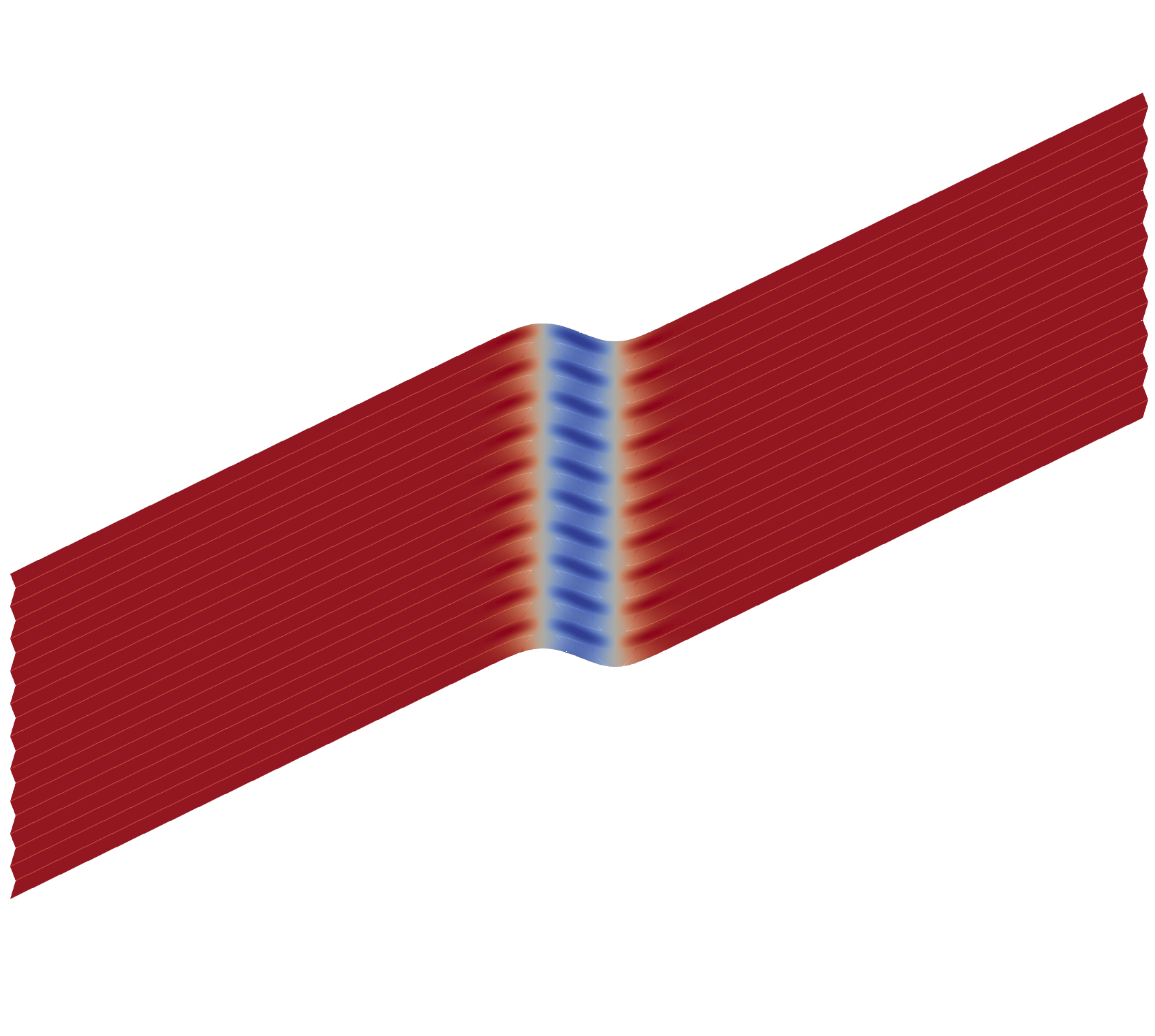}\label{fig:Transformation5}}
  \hspace{0mm}
  \subfloat[$\bar{\gamma}=0.382$, $c^{(-)}=1$]{\includegraphics[width=0.23\textwidth]{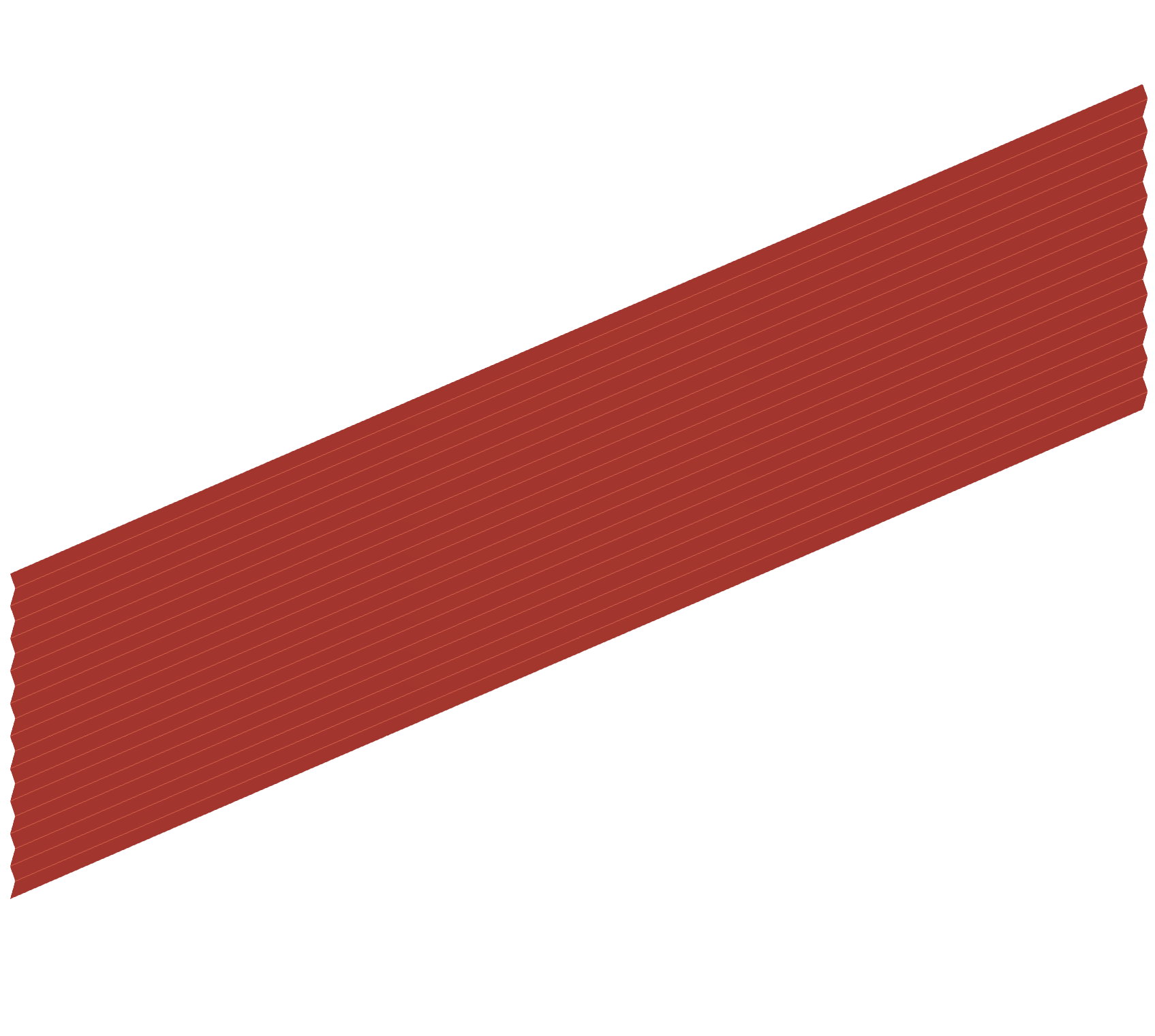}\label{fig:TRZoomStep6}}
  \caption{Evolution of the deformed state under the incremental application of shear, for a purely elastic composite with $\mu^{(f)}=10\mu^{(m)}$, $\kappa^{(f)}=100\mu^{(f)}$,$\kappa^{(m)}=100\mu^{(m)}$. The contours for $P_{21}/\mu^{(f)}$ are plotted. The body has been rotated.}\label{fig:Transformation}
\end{figure}

%%MA
%Motivated by the presentation in 
Making contact with the work by \cite{furer2018macroscopic}, the composite is further subjected in simple shear, after domain-formation has taken place during compression. This is implemented by keeping the $\bar{H}_{22}$ component constant and setting $\bar{H}_{12}=-\bar{\gamma}$. This additional shear loading is captured in Fig. \ref{fig:Transformation}. The incremental application of the shear deformation results in the gradual neutralization of the positive shear field, effectively removing the mesolayer accommodating this mode of deformation and thus transforming the body back into a rank-1 laminate. Application of macroscopic shear has as a result the propagation of the rank-2 phase boundaries --corresponding to the transition regions-- showcasing a phase transformation process. At each successive  step of the loading, part of the positively sheared mesolayer is removed and eventually it ceases to exist, confirming the nature of the post bifurcation solution. The final configuration is one where only the phases of the original rank-1 laminate are present.
%%MA
It is also interesting to observe that the numerical results of Fig. \ref{fig:Transformation} for the evolution of the volume fraction $c^{(-)}$ of the associated mesolayers are in excellent agreement with the following formula
\begin{equation}
	c^{(-)} = \frac{1}{2} \left( 1 + \frac{\bar{\gamma}}{\bar{\gamma}^{rc}} \right)
	\label{cmevol}
\end{equation}
which has been obtained by \cite{furer2018macroscopic} for incompressible Neo-Hookean laminates. In connection with (\ref{cmevol}), it is recalled that 
\begin{equation}
	\bar{\gamma}^{rc}= \bar{\lambda}_{2} \sqrt{ \left( \frac{\bar{\lambda}_{2}^{se}}{\bar{\lambda}_{2}} \right)^2 -1 }, 
	\quad\quad
	\bar{\lambda}_{2}^{se} = \left( 1- \frac{\check{\mu}}{\bar{\mu}} \right)^{1/4},
	\quad\quad
	\check{\mu} = \left( \frac{c^{(m)}}{\mu^{(m)}} + \frac{c^{(f)}}{\mu^{(f)}} \right)^{-1}, 
	\quad\quad
	\bar{\mu} = c^{(m)} \mu^{(m)} + c^{(f)} \mu^{(f)},
	\label{cmevol2}
\end{equation}
where $c^{(m)}$ and $c^{(f)}$ are the volume fractions of the underlying microlayers in the mesolayers. 

\subsection{An alternative type of geometric imperfection}

Loss of ellipticity has traditionally been treated as an indication of  localization of deformation and instabilities, as it was shown that jumps in the gradient of the  deformation gradient become possible in that case. That indication  was often interpreted as a validation of the emergence of localization, but it was shown in \cite{d2016localization} that LOE is a necessary condition but not sufficient. This is a result that does not contradict the formation of domains in the post bifurcation regime. The type of localization they were trying to achieve in that study was of the kink-type, which is commonly accommodated by an unstable response. For that, they introduced an imperfection that was aiming to isolate the --potentially formed-- localized zone of deformation at the center of the specimen. To do so, they assumed an imperfect body with the following geometric perturbation\footnote{Based on Fig. \ref{FEMunitCell}, the origin of the coordinate system for $\Delta X_1^{STW}$ should be placed at points (4).}:
\begin{equation}
\Delta X_1^{STW}(X_1,X_2)= -\xi \sin(\pi X_1/H_{\text{tot}}))\arctan\left[\beta\left(X_2/L -1/2\right)\right]\label{TriantGeom}
\end{equation}
with $\xi=10^{-3}$, $\beta=8$ and $H_{\text{tot}}$ being the total thickness of the RVE. The imperfection described by $\Delta X_1^{STW}$ is similar to the sinusoidal in the sense that both preserve symmetry about the mid point of the domain, with the first one being engineered in such a way that for sufficiently big RVEs, no edge effects occur. 

\begin{figure}[h!tbp]
  \centering
  \subfloat[$\bar{\lambda}_2=0.9035$ (\text{LOE})]{\includegraphics[width=0.45\textwidth]{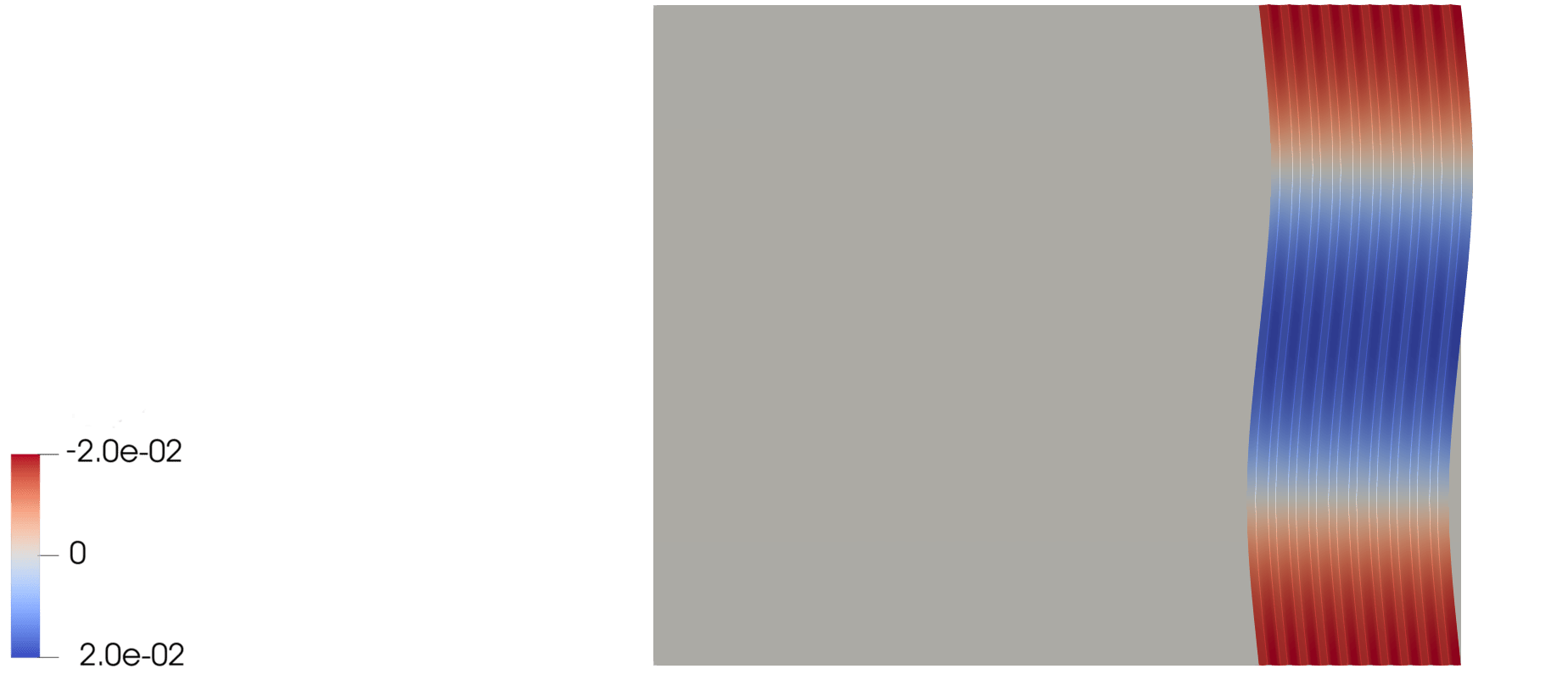}\label{fig:TriantVsMe_LOE}}
  \hspace{0mm}
  \subfloat[$\bar{\lambda}_2=0.9025$]{\includegraphics[width=0.45\textwidth]{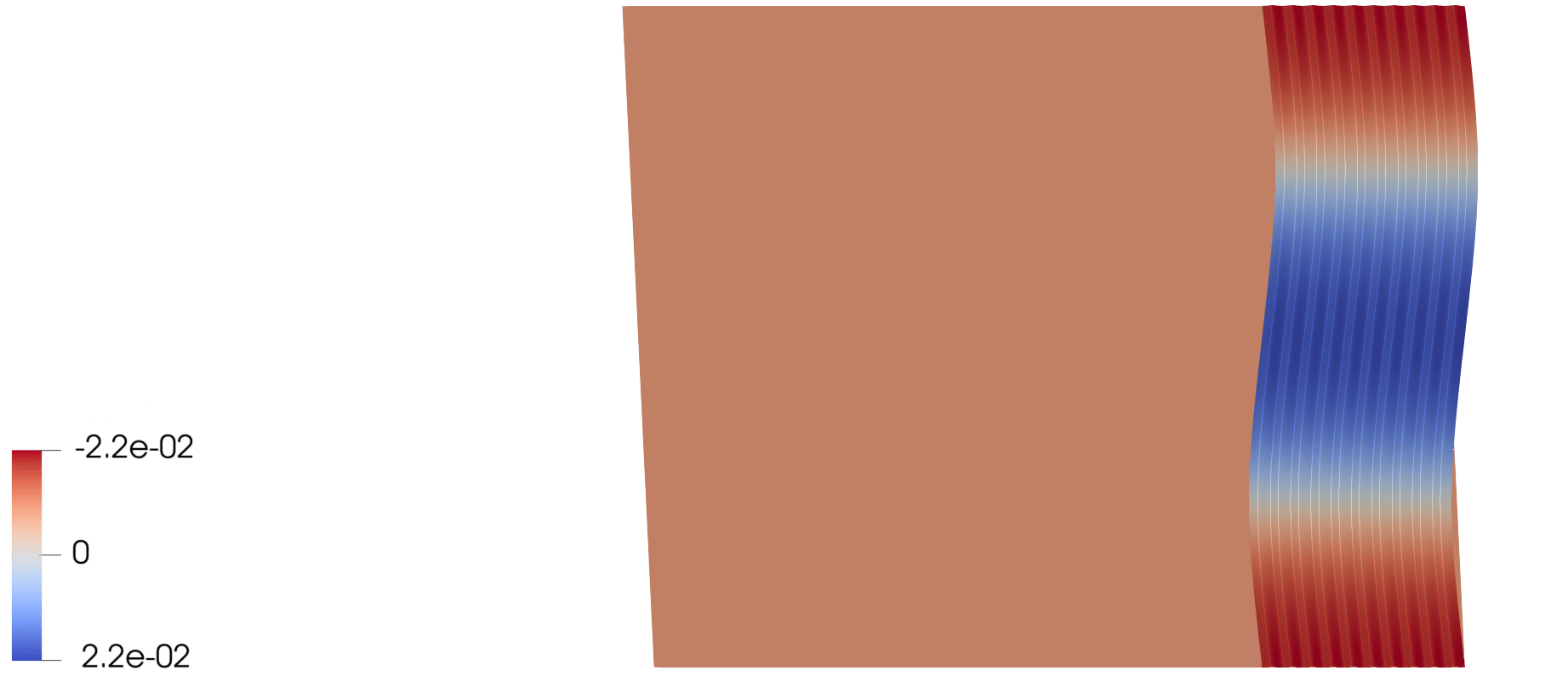}\label{fig:TriantVsMe_L09025}}

 \vspace{5mm}
  
  \subfloat[$\bar{\lambda}_2=0.875$]{\includegraphics[width=0.45\textwidth]{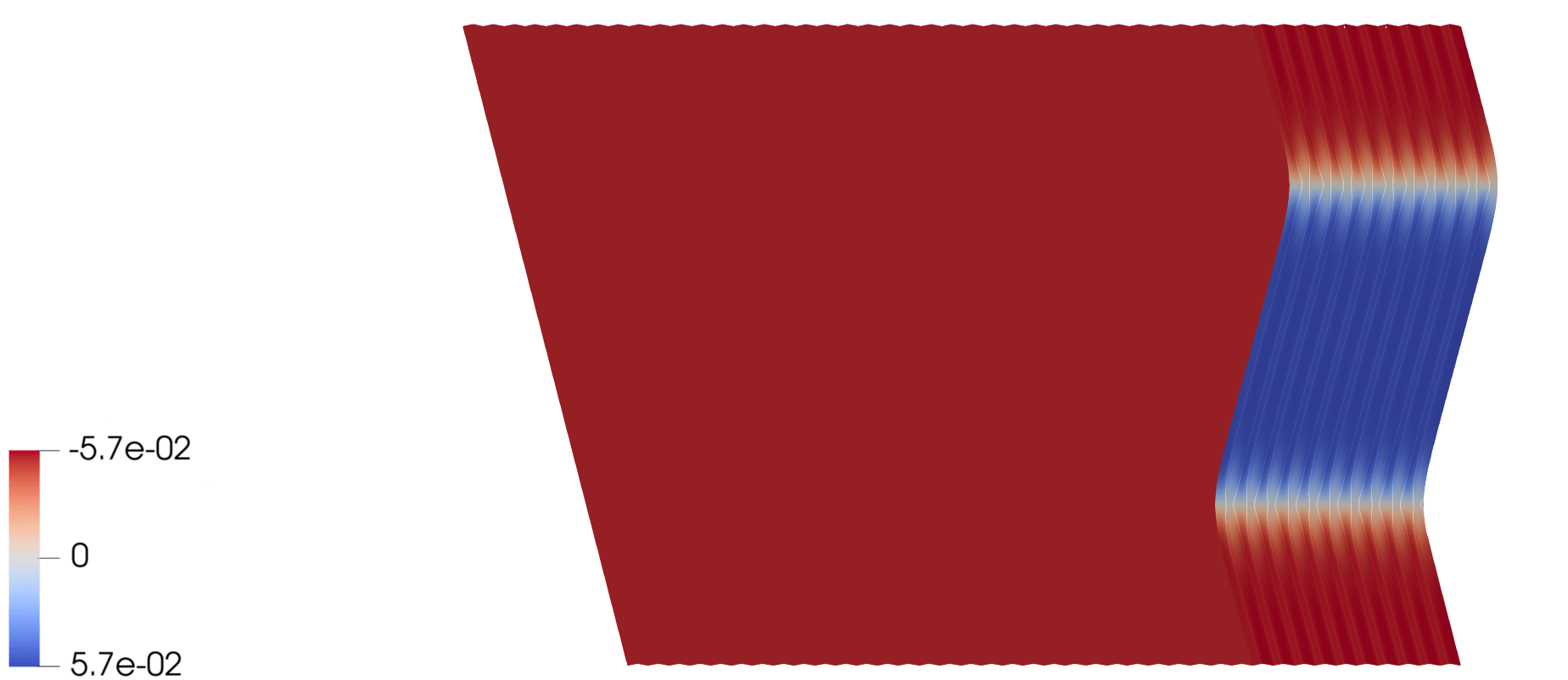}\label{fig:TriantVsMe_L0875}}
  \hspace{0mm}
  \subfloat[$\bar{\lambda}_2=0.7$]{\includegraphics[width=0.45\textwidth]{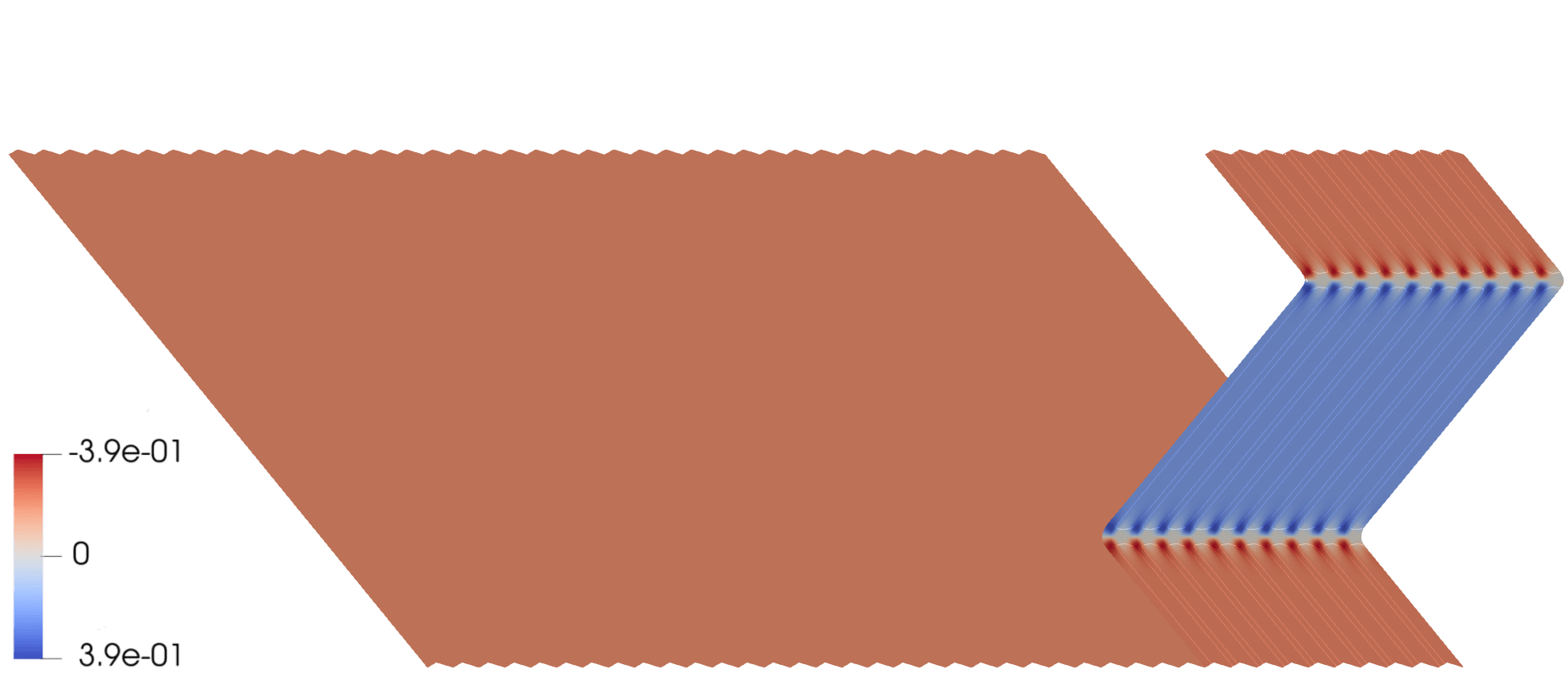}\label{fig:TriantVsMe_L07}}
  \caption{Evolution of the deformed state for the two geometrical perturbations. The square RVE corresponds to $\Delta X_1^{STW}$ while the one in front of it to $\Delta X_1^{S}$. The material properties for the laminates, which are subjected to monotonic compression, are $\mu^{(f)}=10\mu^{(m)}$, $\kappa^{(f)}=100\mu^{(f)}$,$\kappa^{(m)}=100\mu^{(m)}$. The contours in the figure correspond to $P_{21}/\mu^{(f)}$}\label{fig:TriantVsMe}
\end{figure}

Having already established that a sinusoidal imperfection can lead to the formation of domains, a side by side comparison of $\Delta X_1^S$ and $\Delta X_1^{STW}$  is presented in Fig. \ref{fig:TriantVsMe}. Both RVEs are constructed  with $L=1000$, $H_m=H_f=L/80$. The dependency of equation (\ref{TriantGeom}) on the $X_1$ coordinate means that the number of unit cells (as defined for the sinusoidal case, where one unit cell has two layers) used for the final assembly of the RVE will influence the solution. Following \cite{d2016localization}, 40 unit cells are used for consistency with that study meaning that $H_{\text{tot}}=L$. For the sinusoidal perturbation, where there is no $X_1$ dependence of the geometry, 10 unit cells are used for better visualization. The imperfection parameters for $\Delta X_1^{S}$ are $\alpha=10^{-3}L$, $w=1$. The same material parameters as in the baseline case were used for both geometries.

\begin{figure}[h]
  \centering
    \includegraphics[width=0.48\linewidth]{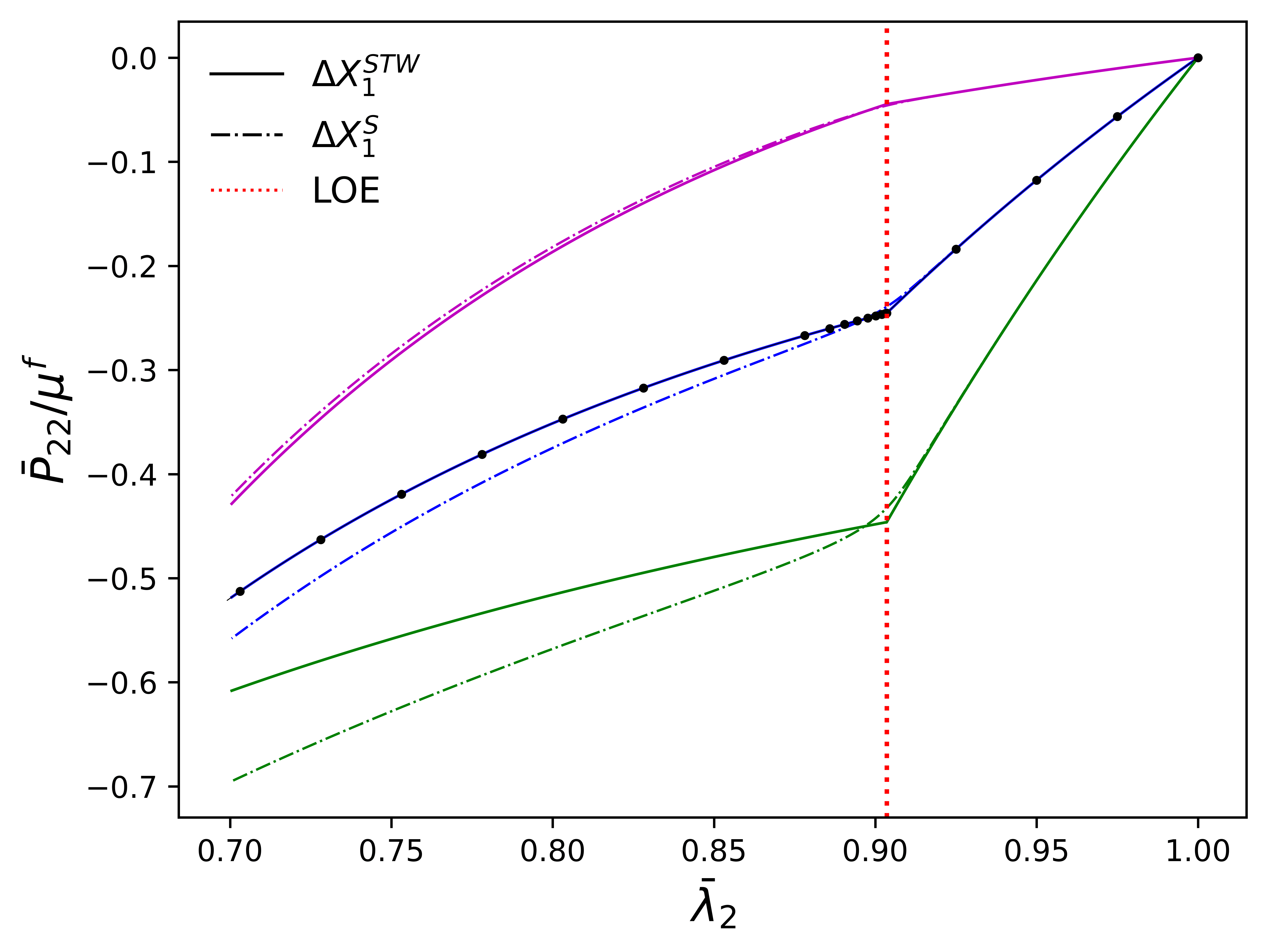}
    \caption{\label{fig:Response_Triant_vs_Me} 
    Comparison between the geometric perturbations. The material properties for the laminates, which are subjected to monotonic compression, are $\mu^{(f)}=10\mu^{(m)}$, $\kappa^{(f)}=100\mu^{(f)}$,$\kappa^{(m)}=100\mu^{(m)}$. The blue color is used for the composite, the green for the fiber phase and the pink for the matrix. The analytical solution for the unperturbed composite has been included with black markers. The vertical red line indicates the loss of ellipticity in the analytical solution.} 
\end{figure}

As seen in Figs. \ref{fig:TriantVsMe_L0875} and \ref{fig:TriantVsMe_L07}, for loads sufficiently past the critical load for LOE, the specimen following the perturbed geometry of $\Delta X_1^{STW}$ does not exhibit any localized deformations, maintaining constant per-phase field, retrieving the solution in \cite{d2016localization} for purely Neo-Hookean layers. Interestingly this deformation mode is indistinguishable (minus the transition regions) with the one attained in one of the two antisymmetric phases of the rank-2 laminate when the $\Delta X_1^{S}$ imperfection is used.  The macroscopic responses for the two types of imperfections are compared in Fig. \ref{fig:Response_Triant_vs_Me}. The essential difference between the two bodies is that the one with the sinusoidal geometry includes in the deformed state the transition regions accompanying domain formation. It is postulated that the additional energy of the phase boundaries of the rank-2 laminate (where intense bending is observed), accounts for the stiffer response of the sinusoidal model. 
 
\subsection{Wave length study}\label{subsec:WaveLength} 
\begin{figure}[h!tbp]
  \centering
  \subfloat[$w=1$]{\includegraphics[width=0.4\textwidth]{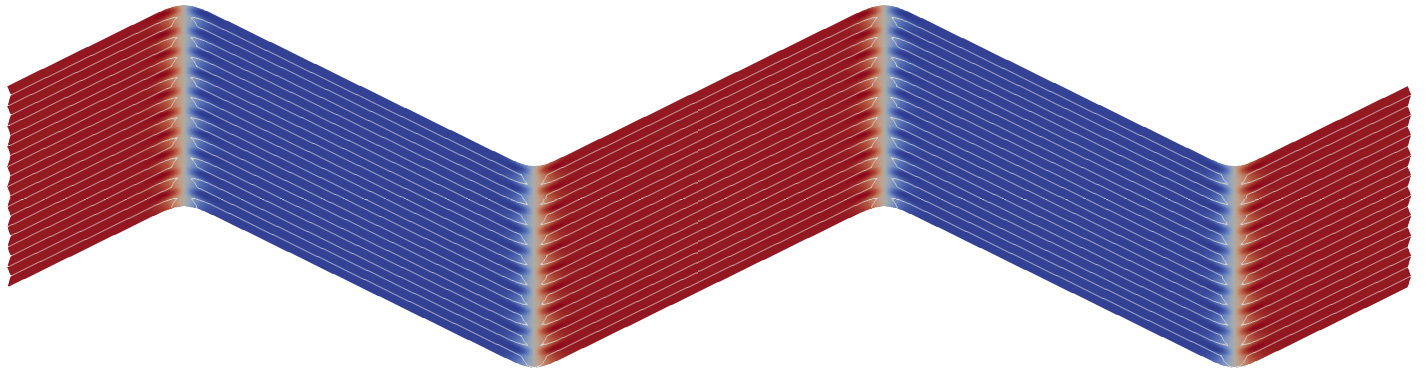}\label{fig:Wave1}}
  
  \vspace{0mm}
  
  \subfloat[$w=2$]{\includegraphics[width=0.4\textwidth]{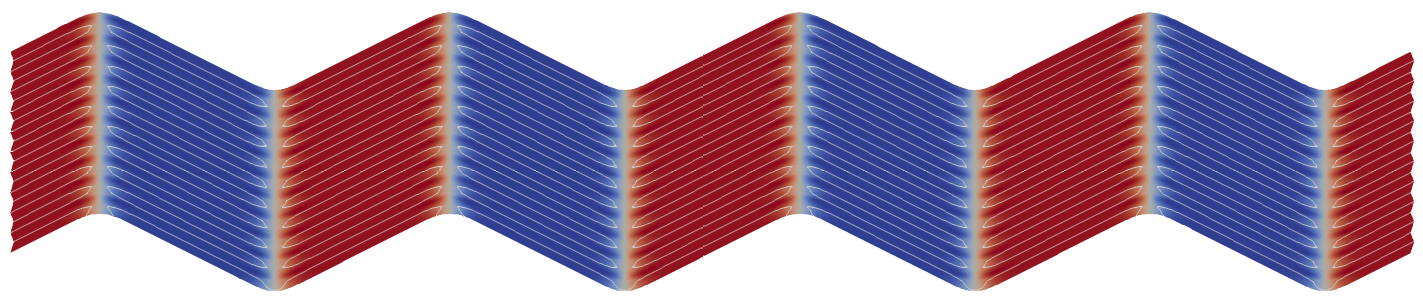}\label{fig:Wave2}}

 \vspace{0mm}
  
  \subfloat[$w=3$]{\includegraphics[width=0.4\textwidth]{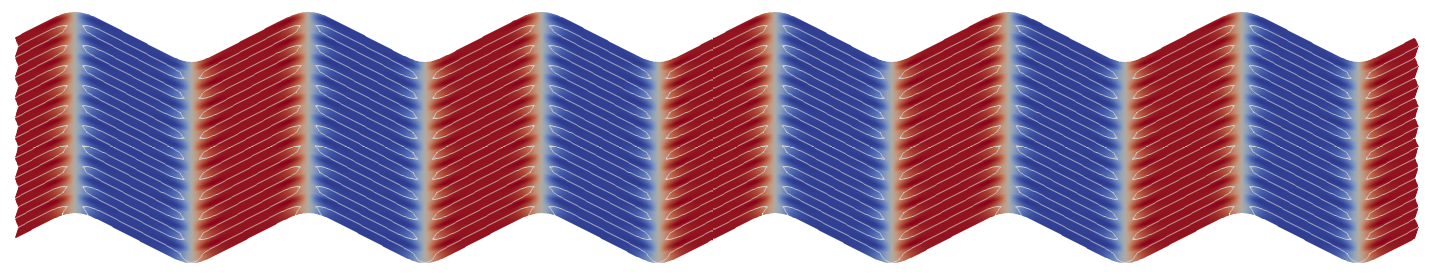}\label{fig:Wave3}}
  \caption{Deformed configurations of composites with different prescribed wavelengths (w) on $\Delta X_1^S$ and the same $\alpha=10^{-4}L$. The material properties for all the geometries are $\mu^{(f)}=10\mu^{(m)}$, $\kappa^{(f)}=100\mu^{(f)}$,$\kappa^{(m)}=100\mu^{(m)}$. The composites were subjected to monotonic compression. The presented state has been rotated.}\label{fig:Wavelenth analysis}
\end{figure}

Motivated by the above result, and the energy associated with the presence of the phase boundaries of the rank-2 laminate, we examine the number of wavelengths included in a unit cell (see also the appendix for studies on imperfection amplitude ($\alpha$) and different geometrical perturbations). By using the baseline case and setting $w=1,2,3$  the results shown in Fig. \ref{fig:Wavelenth analysis} are obtained. Since the domains nucleate around the higher curvature peaks and valleys of the sinusoidal curve, more of them lead to a denser lamellar microstructure for the induced rank-2 composite, with the produced mesolayers retaining their 1/2 volume fraction.

\begin{figure}[h!tbp]
  \centering
\includegraphics[width=0.5\textwidth]{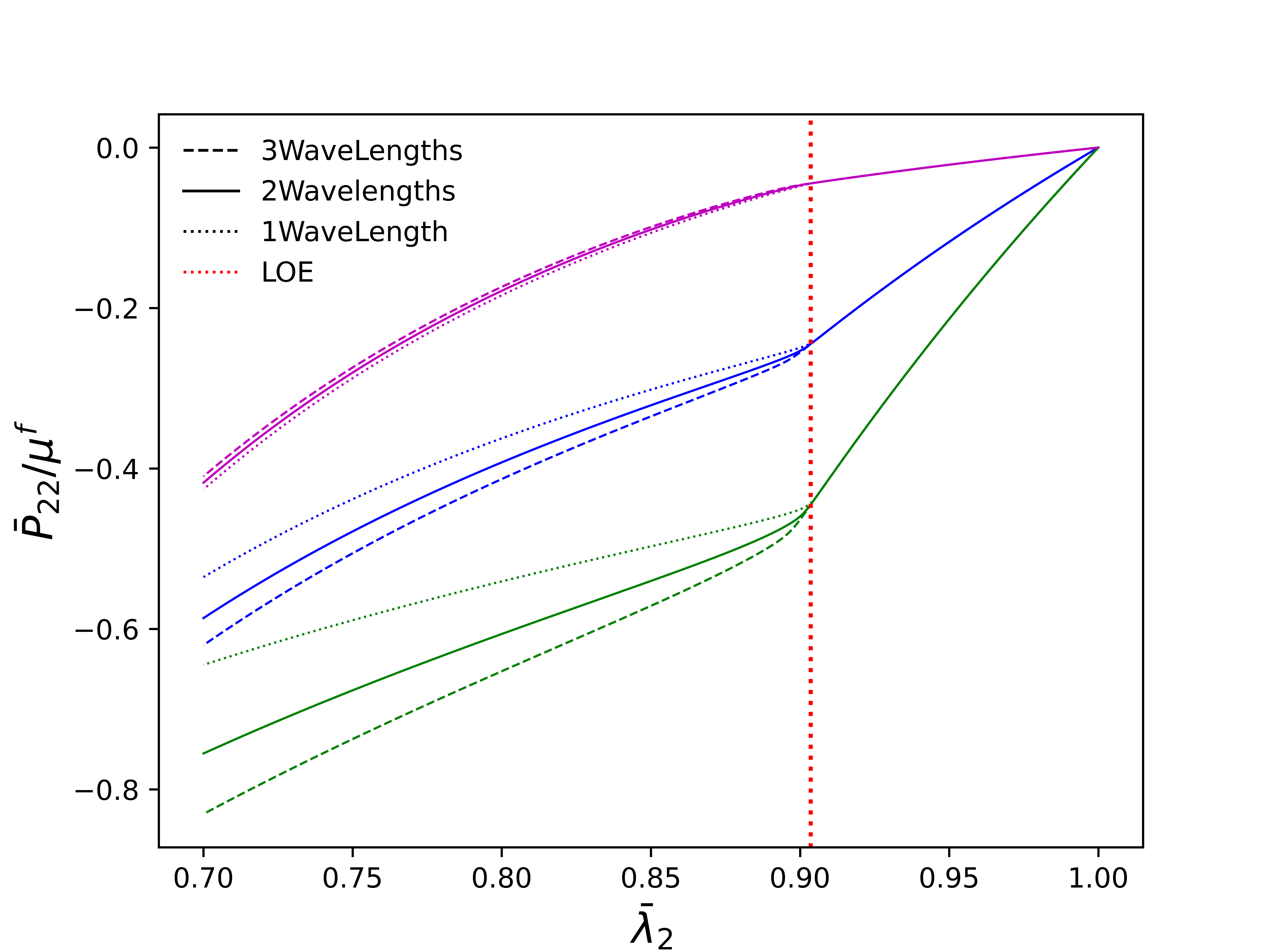}
\caption{Stress-Stretch response for geometrically imperfect models with different number of wavelengths. The material properties for the laminates, which are subjected to monotonic compression, are $\mu^{(f)}=10\mu^{(m)}$, $\kappa^{(f)}=100\mu^{(f)}$,$\kappa^{(m)}=100\mu^{(m)}$. The blue color is used for the composite, the green for the fiber phase and the pink for the matrix. The vertical red line indicates the loss of ellipticity in the analytical solution.\label{fig:wavelengthresponse}}
\end{figure}

Plotting the macroscopic response of the composites, Fig. \ref{fig:wavelengthresponse} reveals that unit cells with more wavelengths per unit of length are associated with an overall stiffer behavior.

\begin{figure}[h!tbp]
  \centering
  \subfloat[Averaged Strain energy densities]{\includegraphics[width=0.45\textwidth]{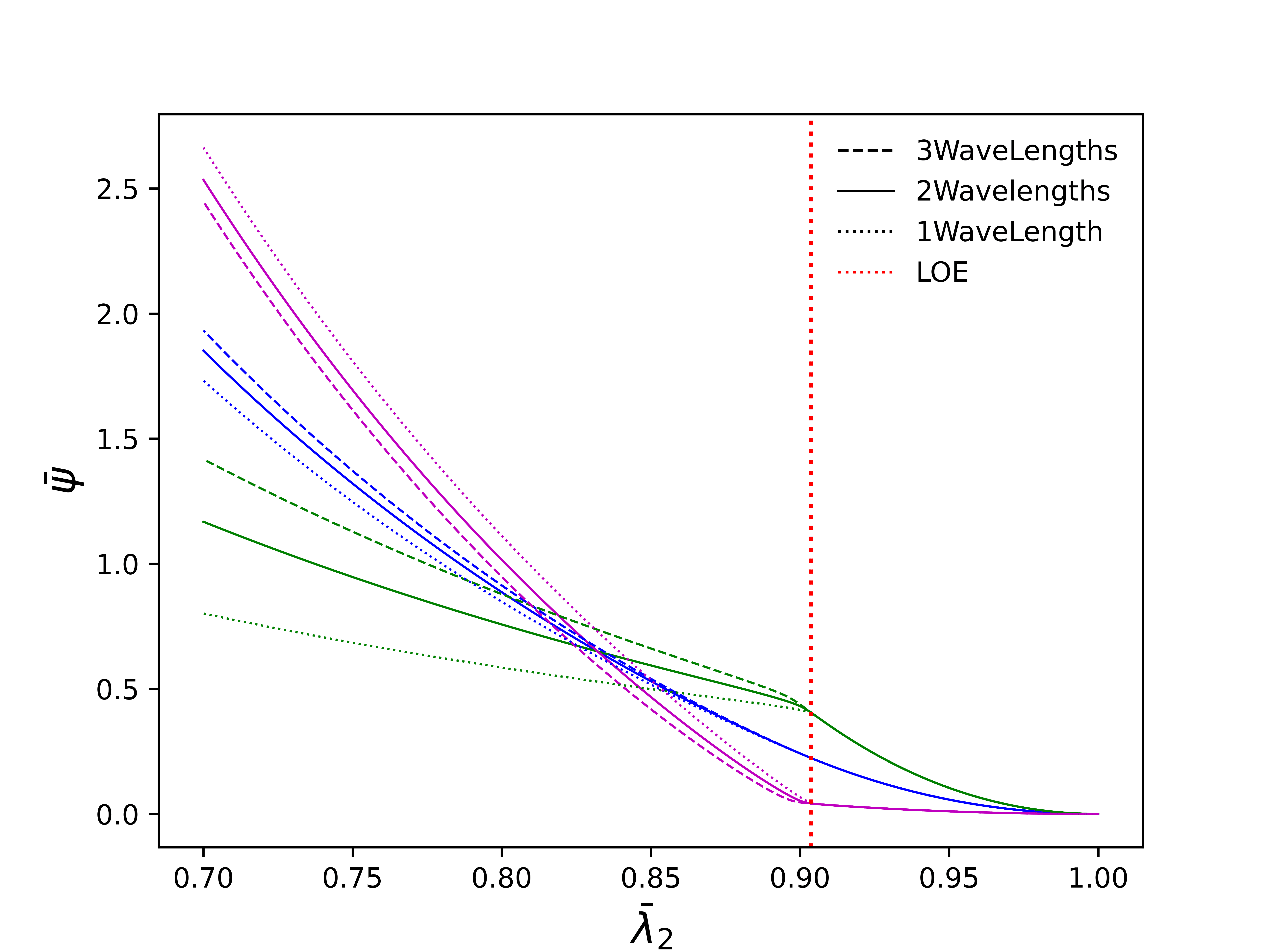}\label{fig:wavelengthEnergies}}
  \hspace{1mm}
  \subfloat[Non dimensionalized energies]{\includegraphics[width=0.45\textwidth]{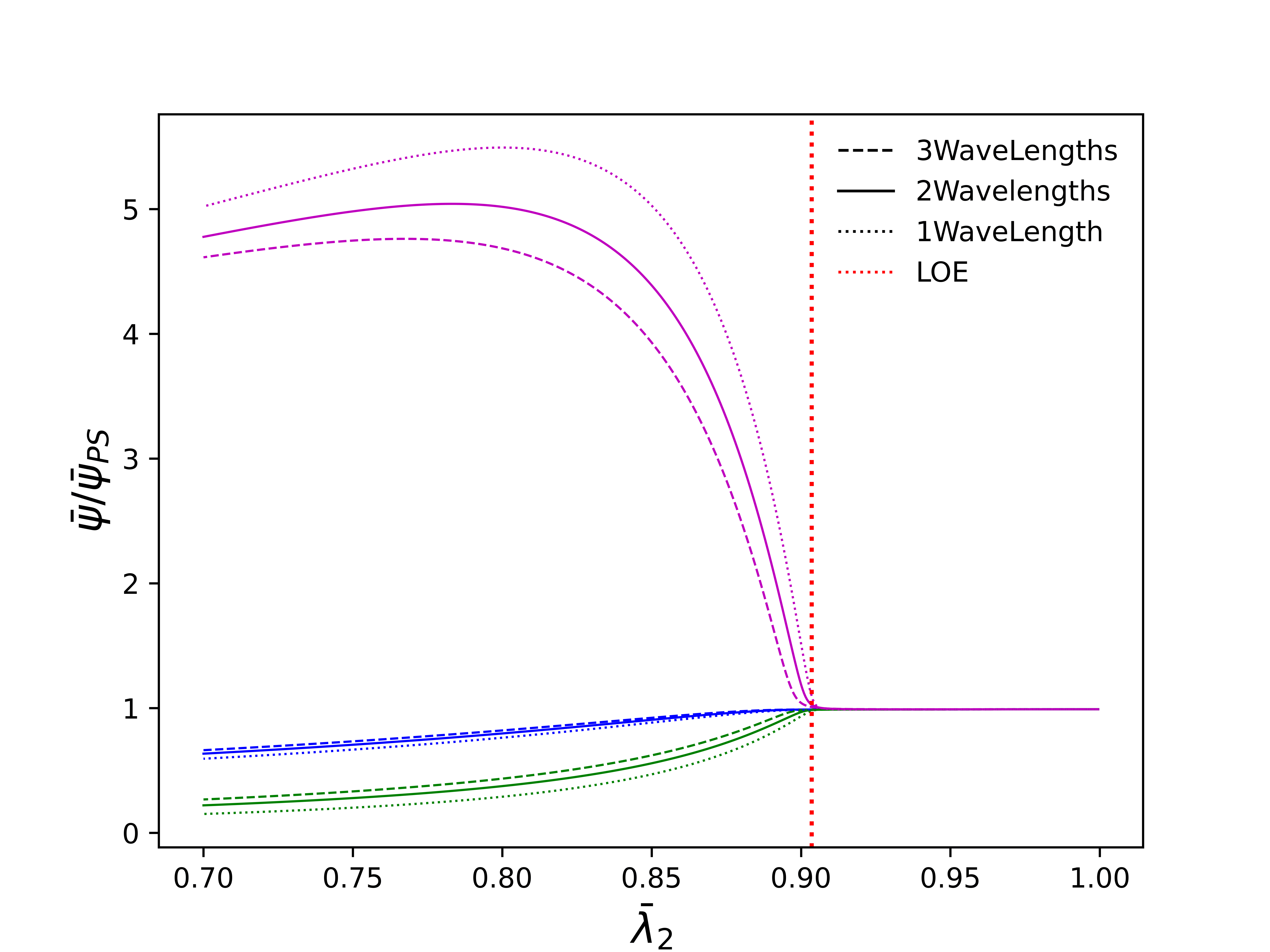}\label{fig:wavelengthNDEnergies}}
  \caption{Strain energies for geometrically imperfect models with different number of wavelengths. The material properties for the laminates, which are subjected to monotonic compression, are $\mu^{(f)}=10\mu^{(m)}$, $\kappa^{(f)}=100\mu^{(f)}$,$\kappa^{(m)}=100\mu^{(m)}$. The blue color is used for the composite, the green for the fiber phase and the pink for the matrix. The vertical red line indicates the loss of ellipticity in the analytical solution.\label{fig:wavelengthEnergiesBoth}}
\end{figure}

The plot of the strain energy, per unit of undeformed volume, shown in  Fig.\ref{fig:wavelengthEnergies}, reveals an additional aspect of the bifurcated solution. After domains start to emerge, indicated by the LOE,  a rapid redirection of the expended elastic work from the fiber to the matrix phase is observed. The bifurcated mode is such that favors the rotation of the stiffer phase to avoid compression in the direction of the loading, forcing the matrix to accommodate the deformation, making it, eventually, the energy richer phase which dominates the response. This tendency of a hyperelastic laminate to relieve stress through rotations was predicted in \cite{lopez2009microstructure} for perfect laminates. The introduction of the geometric imperfection permits the rotation to take place for aligned types of loading and as a consequence domains are formed. To get a better grip of this redistribution of energy, the strain energy density of the principal solution for the perfect laminates - of each phase - is used to non dimensionalize their geometrically imperfect counterparts. The results are in Fig.\ref{fig:wavelengthNDEnergies} and highlight the relationship $\Bar{\psi}(\boldsymbol{F}) \leq \Bar{\psi}(\boldsymbol{F}^{ps})$ between the bifurcated and the principal solution.

\newpage

\section{Numerical investigation of an elastoplastic laminate}\label{sec::plastic}
Since the implications of geometric imperfections on domain formation and the details of the macroscopic response were studied for purely Neo-Hookean phases, here we redirect our attention to the response of a laminate with a purely elastic (soft) matrix phase and an elastoplastic (stiff) fiber phase. This complements the study in Part I, and accordingly we will focus on both monotonic (aligned) loading in compression, and non-monotonic (aligned) loading cycles of tension --which induce the accumulation of plastic deformations--, elastic unloading and subsequent loading in compression. Unless noted, for the remainder of the section, the same geometrical parameters ($\alpha=10^{-4}L=10^{-2}H$, $w=1$) as in the baseline case are used. The following fundamental studies are presented: an initial yield limit analysis, a hardening analysis, and  finally a study of the effect of non-monotonic loading. The constitutive models presented in Section \ref{sec::methodology}.2 are utilized, and the analytical results for the unperturbed geometry are obtained using the expressions, for the principal path, LOE and post-bifurcation from Part I.

\subsection{Initial yield limit \label{subsec:InitYL}}
\begin{figure}[h!tbp]
  \centering
  \subfloat[Composite]{\includegraphics[width=0.47\textwidth]{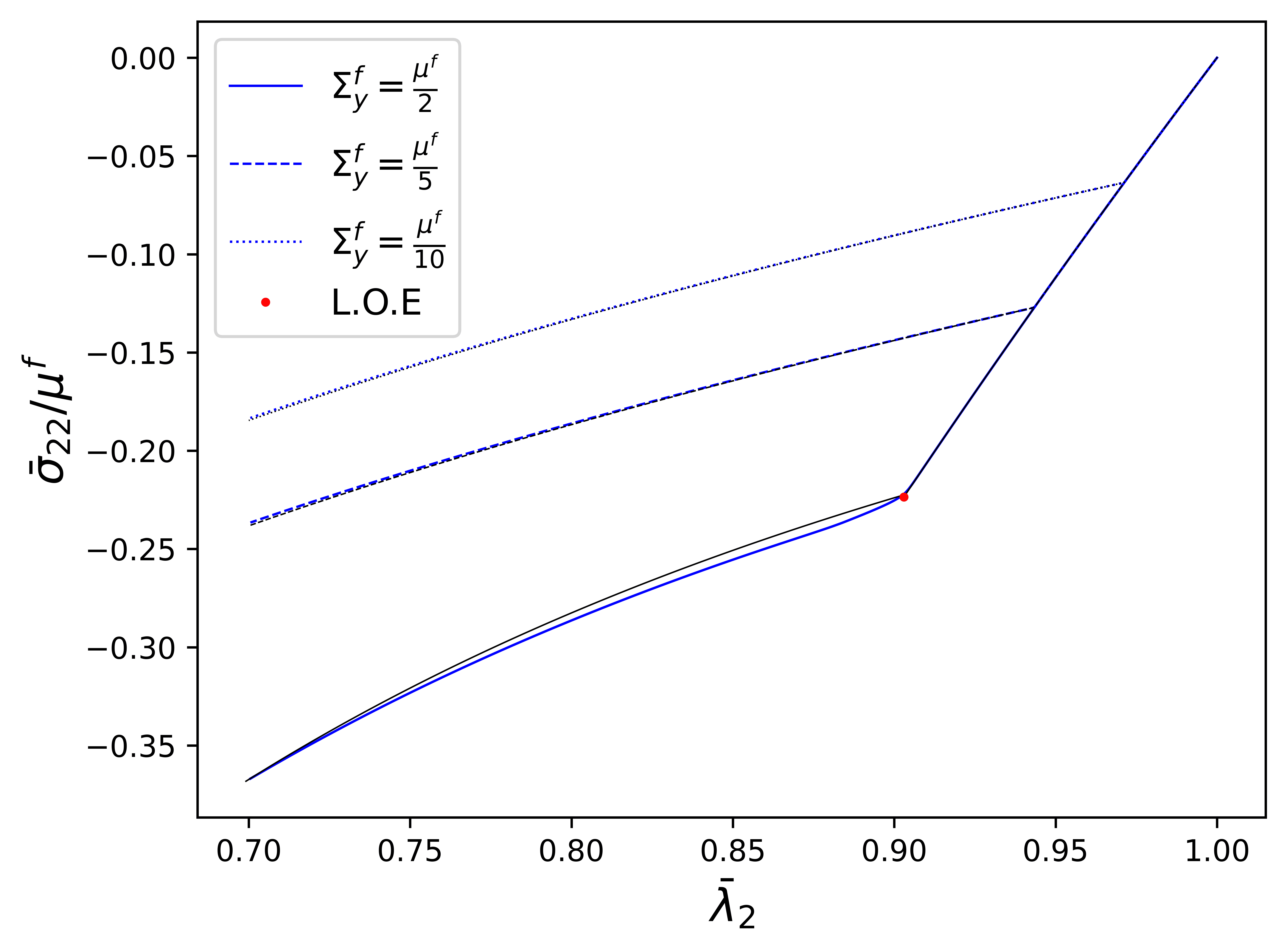}\label{fig:MYLC}}
  \hspace{0mm}
  \subfloat[Fiber phase]{\includegraphics[width=0.47\textwidth]{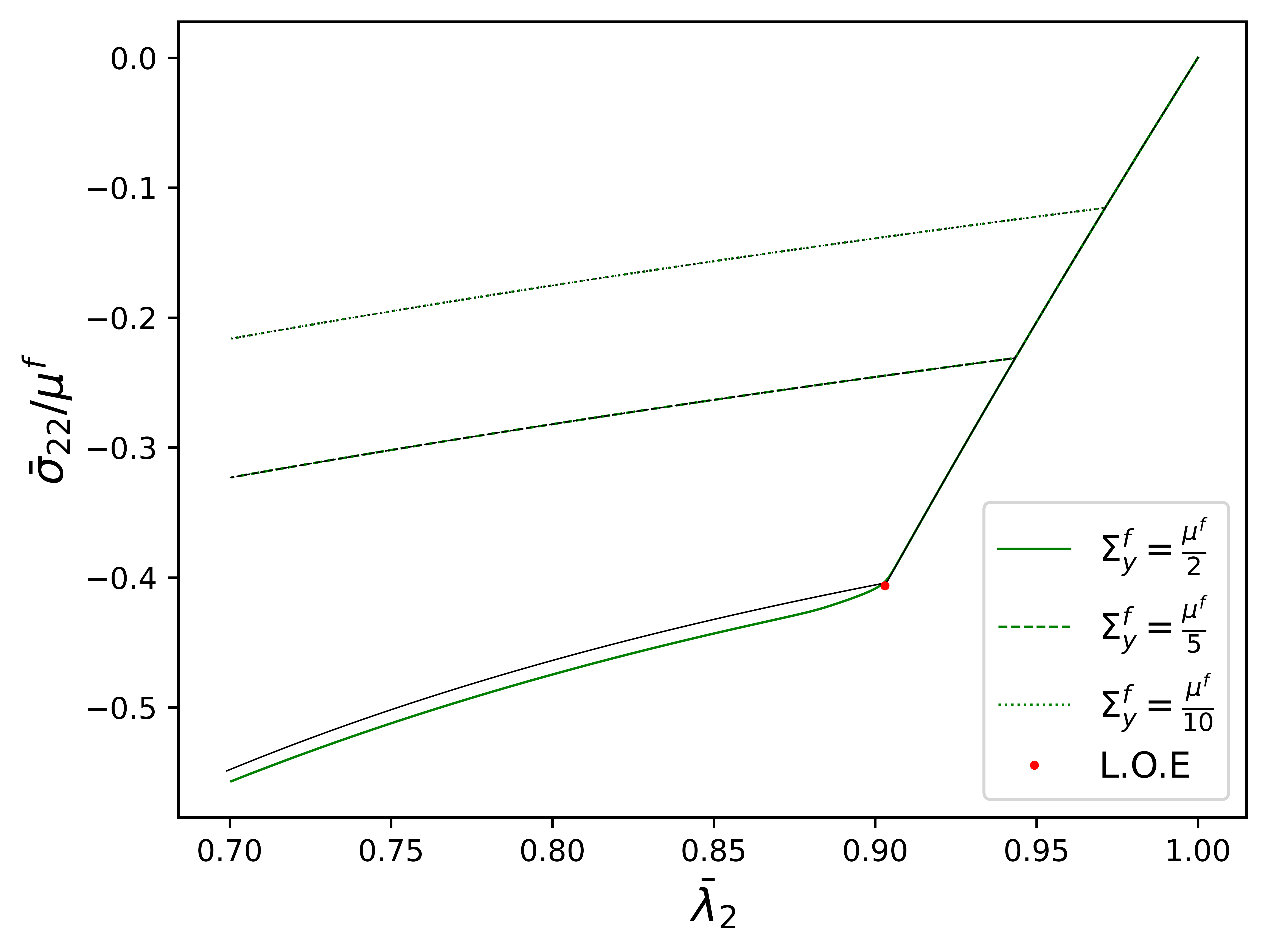}\label{fig:MYLF}}
  
  \vspace{2mm}
  
  \subfloat[Matrix phase]{\includegraphics[width=0.47\textwidth]{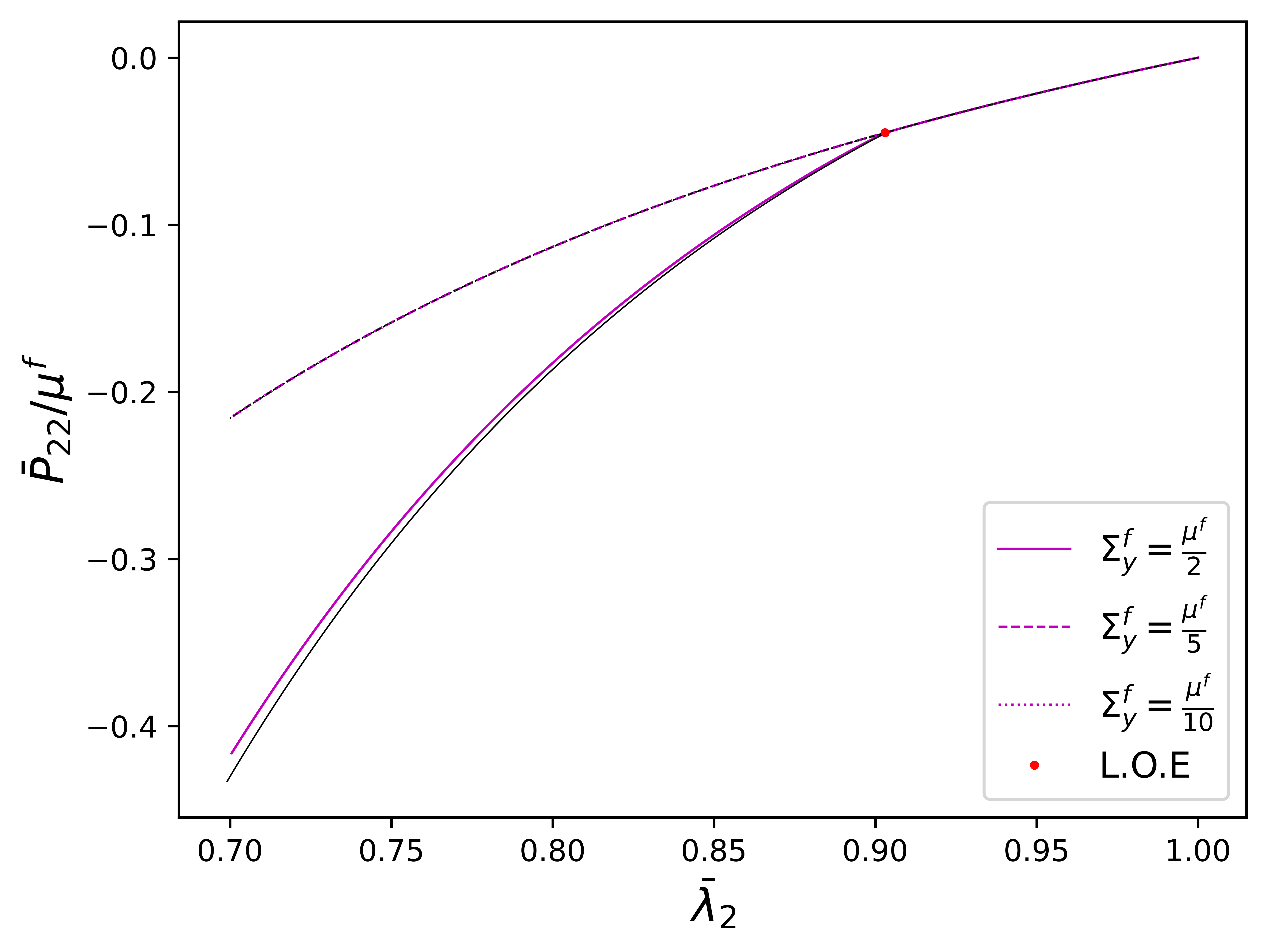}\label{fig:MYLM}}
  \hspace{0mm}
  \subfloat[Accumulated plastic strain in fiber phase]{\includegraphics[width=0.47\textwidth]{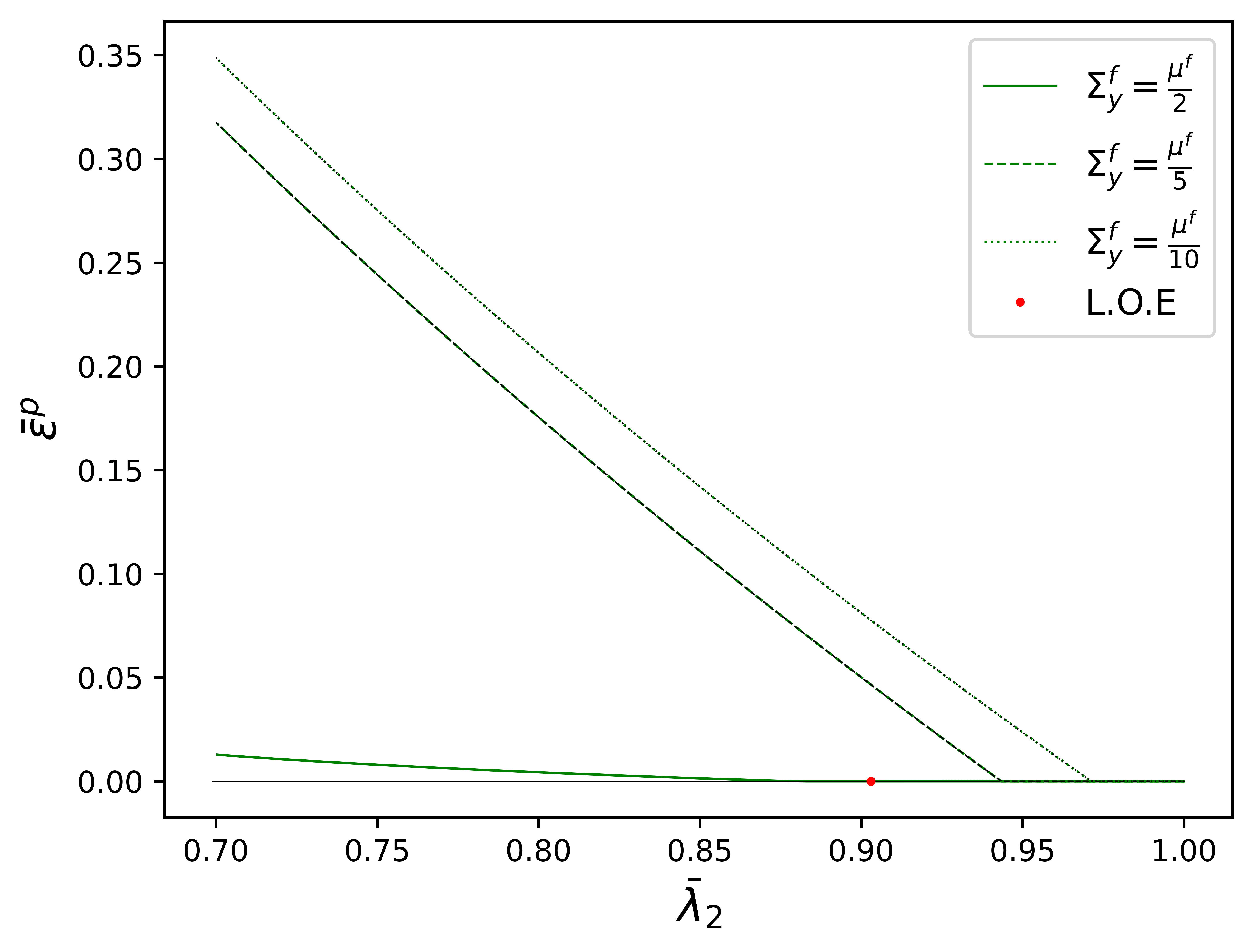}\label{fig:MYLep}}
  \caption{Results for laminates with different initial yield limit, the same hardening $h^{(f)}=\mu^{(f)}/4$ and the same elastic coefficients $\mu^{(f)}=10\mu^{(m)}$, $\kappa^{(f)}=100\mu^{(f)}$, $\kappa^{(m)}=100\mu^{(m)}$. With discontinuous lines are the cases where no domains were observed. The black lines represent the analytical solution of the perfect laminates. The red dot indicates the loss of ellipticity in the analytical solution.}\label{MYL}
\end{figure}

The impact of the initial yield limit is first examined, for monotonic compression. Fixing the hardening modulus to $h^{(f)}=\mu^{(f)}/4$ the results in Fig. \ref{MYL} are obtained, where the black lines correspond to the analytical solution for the perfect laminate. 

\begin{figure}[h!tbp]
  \centering
  \subfloat[$\Sigma_y^{(f)}=\mu^{(f)}/2, \Bar{\lambda}_2=0.903$ (LOE)]{\includegraphics[width=0.24\textwidth]{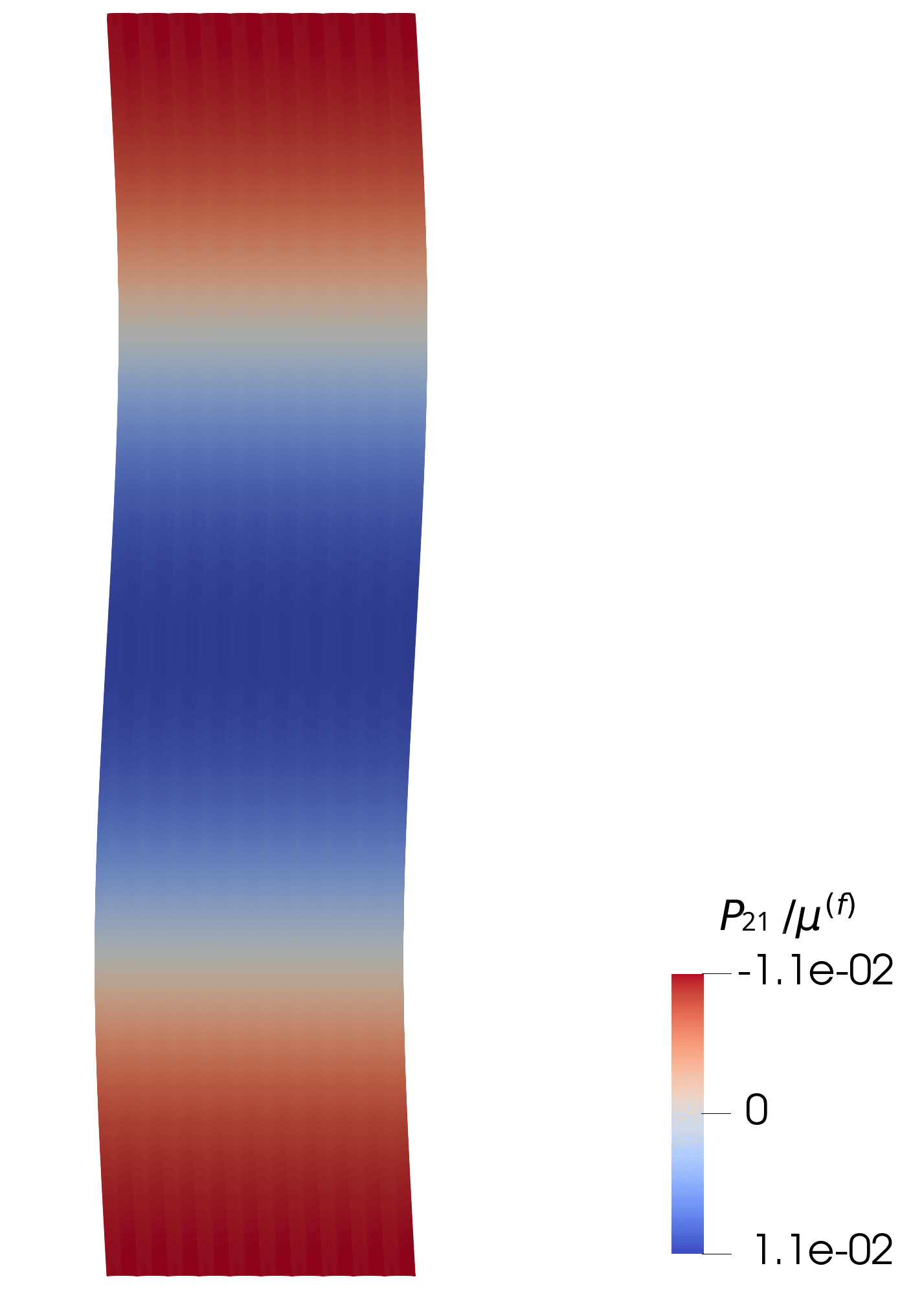}\label{fig:MYL2PKLOE}}
  \hspace{15mm}
  \subfloat[$\Sigma_y^{(f)}=\mu^{(f)}/2, \Bar{\lambda}_2=0.7$]{\includegraphics[width=0.24\textwidth]{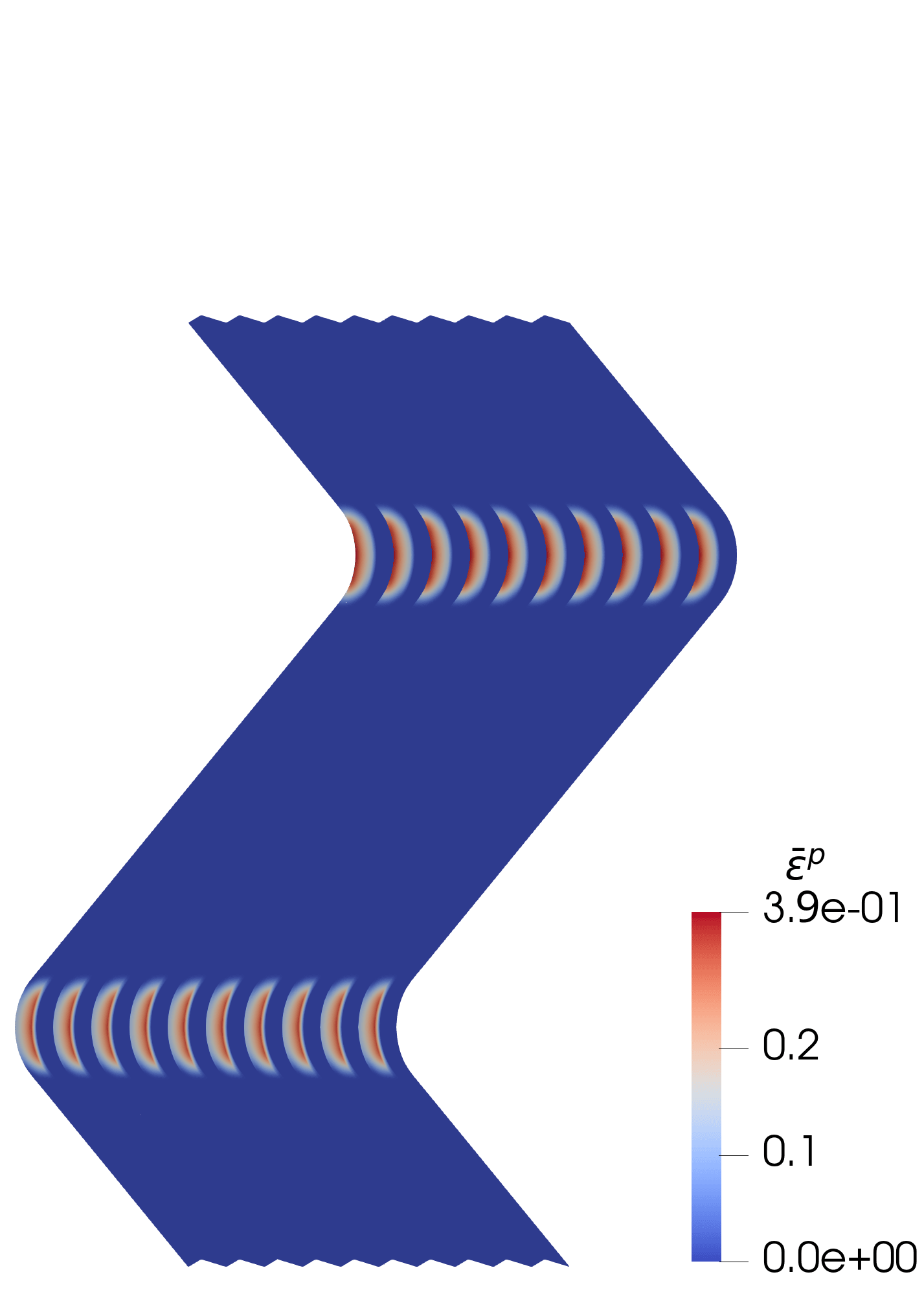}\label{fig:MYL2epL07}}
  
  \vspace{10mm}
  
  \subfloat[$\Sigma_y^{(f)}=\mu^{(f)}/5, \Bar{\lambda}_2=0.903$]{\includegraphics[width=0.24\textwidth]{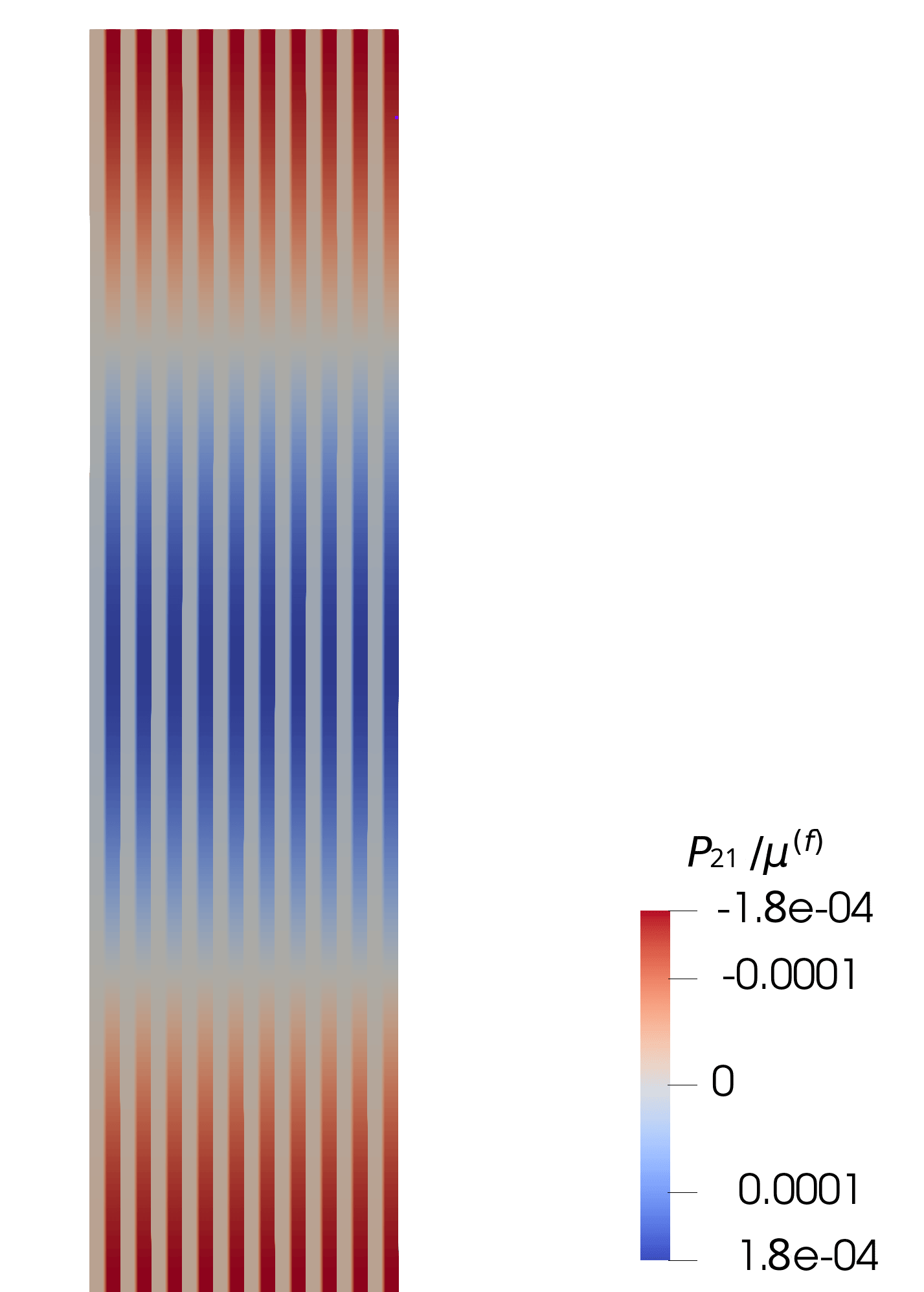}\label{fig:MYL5PK09}}
  \hspace{15mm}
  \subfloat[$\Sigma_y^{(f)}=\mu^{(f)}/5, \Bar{\lambda}_2=0.7$]{\includegraphics[width=0.24\textwidth]{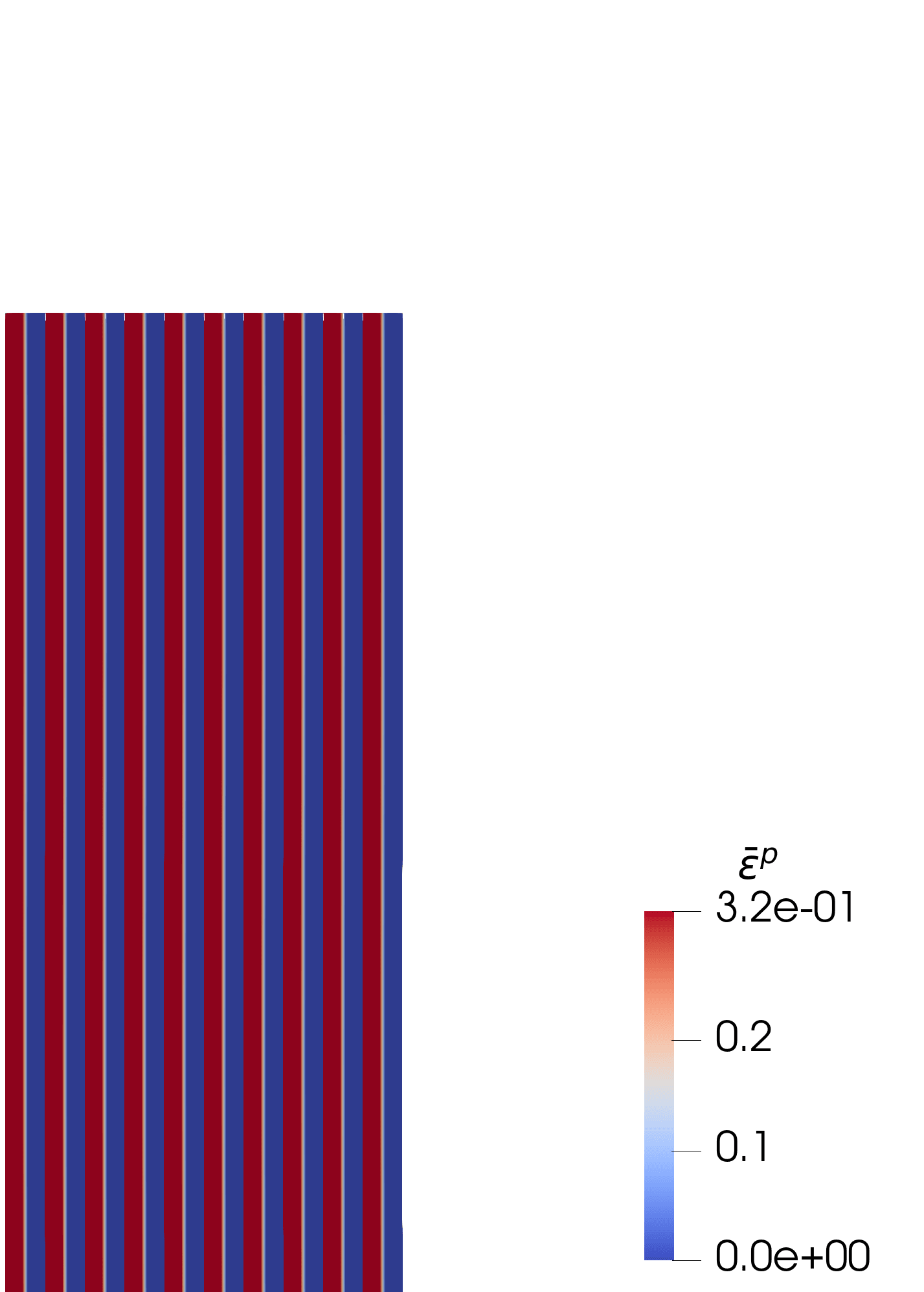}\label{fig:MYL5ep07}}
  \caption{Comparison of the deformed configurations between the composite with $\Sigma_y^{(f)}=\mu^{(f)}/2$ (top row) and $\Sigma_y^{(f)}=\mu^{(f)}/5$ (bottom row) that share the same hardening modulus and elastic coefficients. The contours of $P_{21}/\mu^{(f)}$ and accumulated plastic strain are plotted approximatly when ellipticity is lost for the analytical solution and at the final deformed state.}\label{YLmuf2VSYLmuf5}
\end{figure}

Attention in brought to Fig. \ref{fig:MYLM} where the response of the matrix phase is plotted. It has already been observed in the purely elastic case, that when domains are formed, the response of the matrix phase changes drastically. So it is expected that the model with the high initial yield limit in which LOE in the analytical solution is observed --under purely elastic loads-- has formed domains, contrary to the models with the low yield limits. Indeed, this is the case, as shown in Fig.\ref{YLmuf2VSYLmuf5} where the deformed state is depicted. It is also shown that in the imperfect body, when domains precede any plasticity, the intense localized bending in the transition regions induces plastic deformation, as is observed in Fig .\ref{fig:MYL2epL07} and the corresponding plot of accumulated plastic strain in Fig. \ref{fig:MYLep}. This result is not present in the analytical solution of Part I due to the assumption of constant per-phase fields which does not hold true here. On the other hand, when there is no domain formation, constant per-phase fields develop (see Figs \ref{fig:MYL5PK09}, \ref{fig:MYL5ep07}), in good agreement with the analytical solution --as is shown by the discontinuous lines in Fig. \ref{MYL}-- which maintains its elliptic character.

\subsection{Hardening}
\begin{figure}[h!tbp]
  \centering
  \subfloat[Composite]{\includegraphics[width=0.45\textwidth]{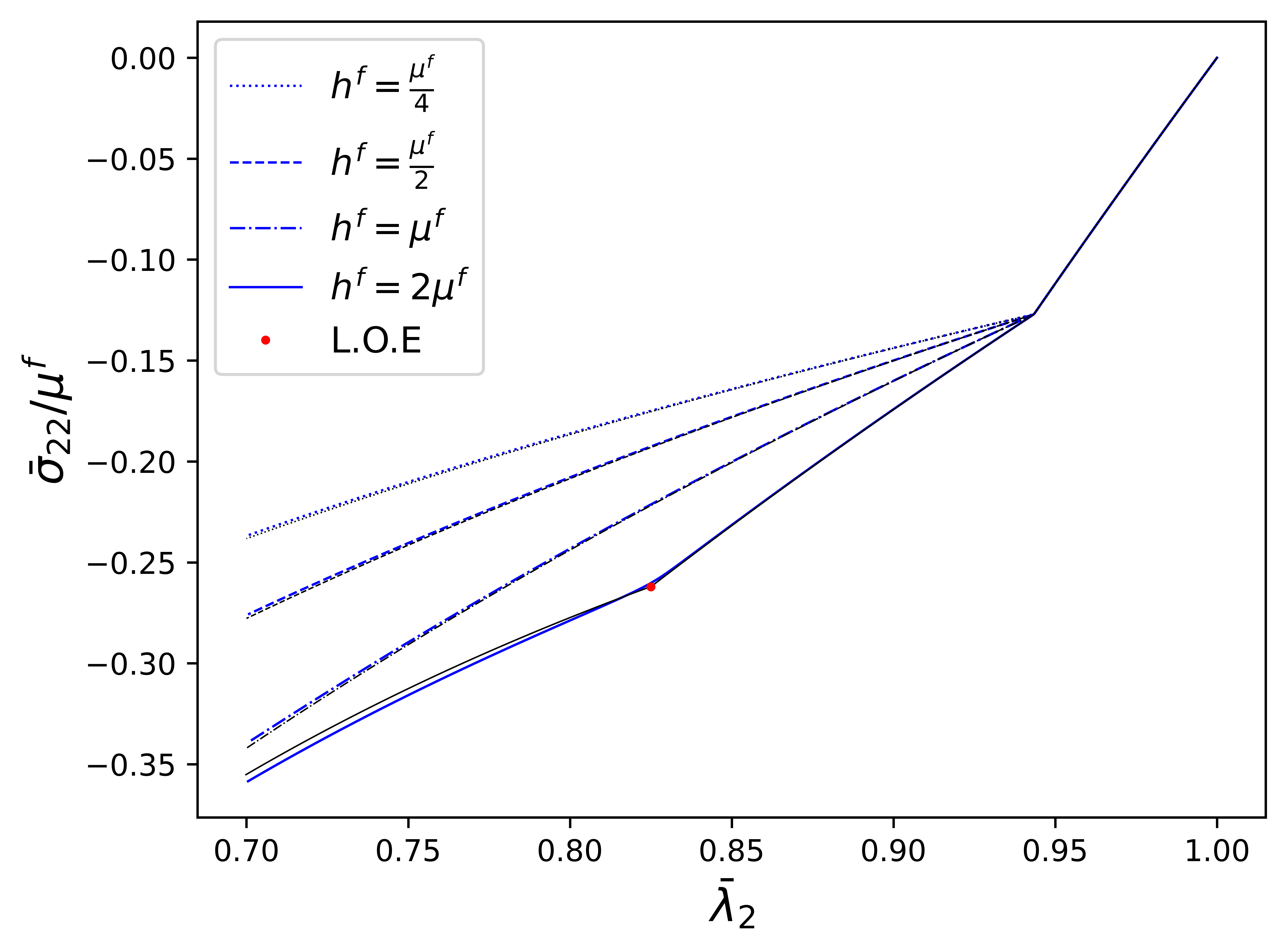}\label{fig:YL5HC}}
  \hspace{0mm}
  \subfloat[Fiber phase]{\includegraphics[width=0.45\textwidth]{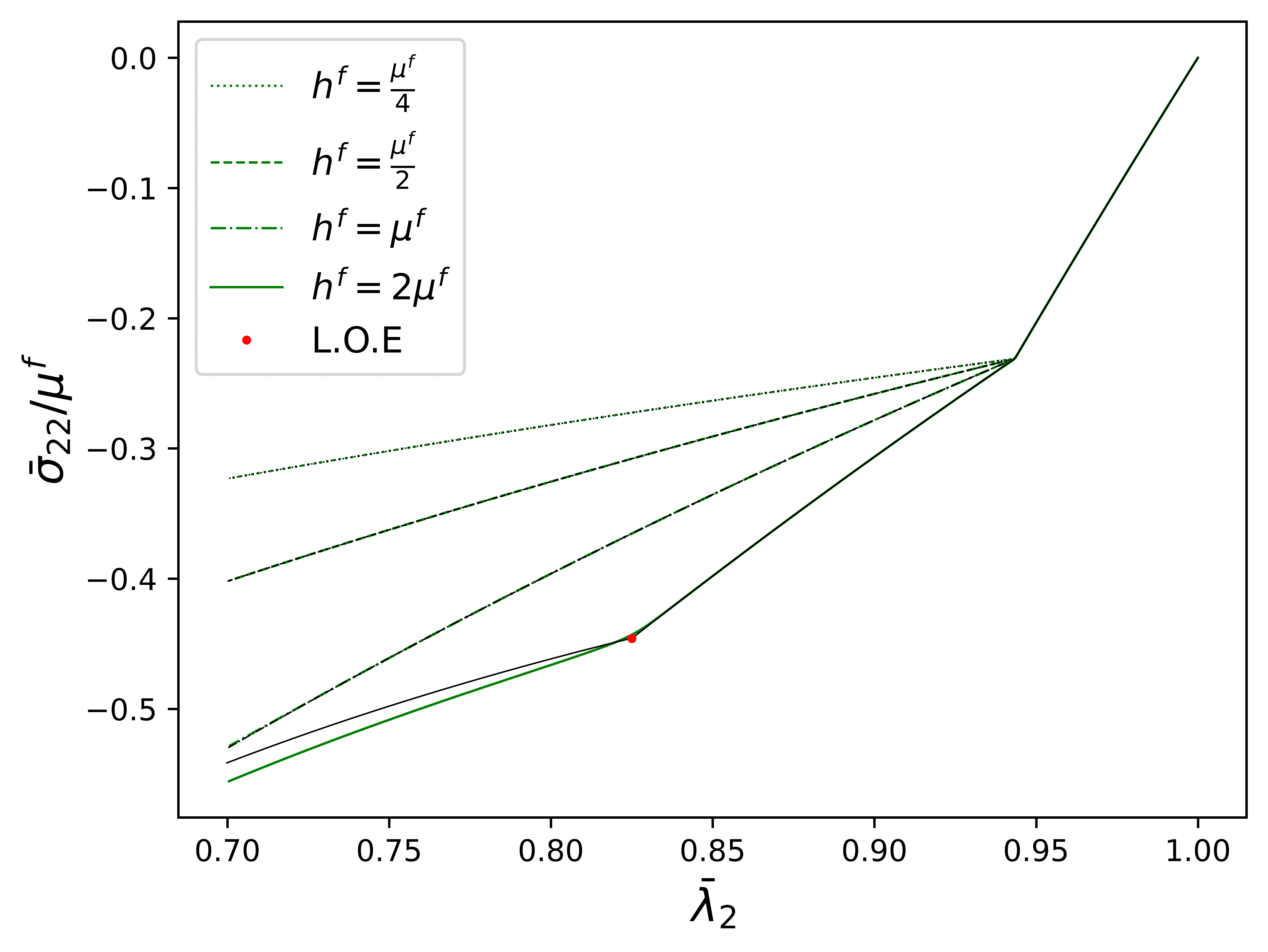}\label{fig:YL5HF}}
  
  \vspace{5mm}
  
  \subfloat[Matrix phase]{\includegraphics[width=0.45\textwidth]{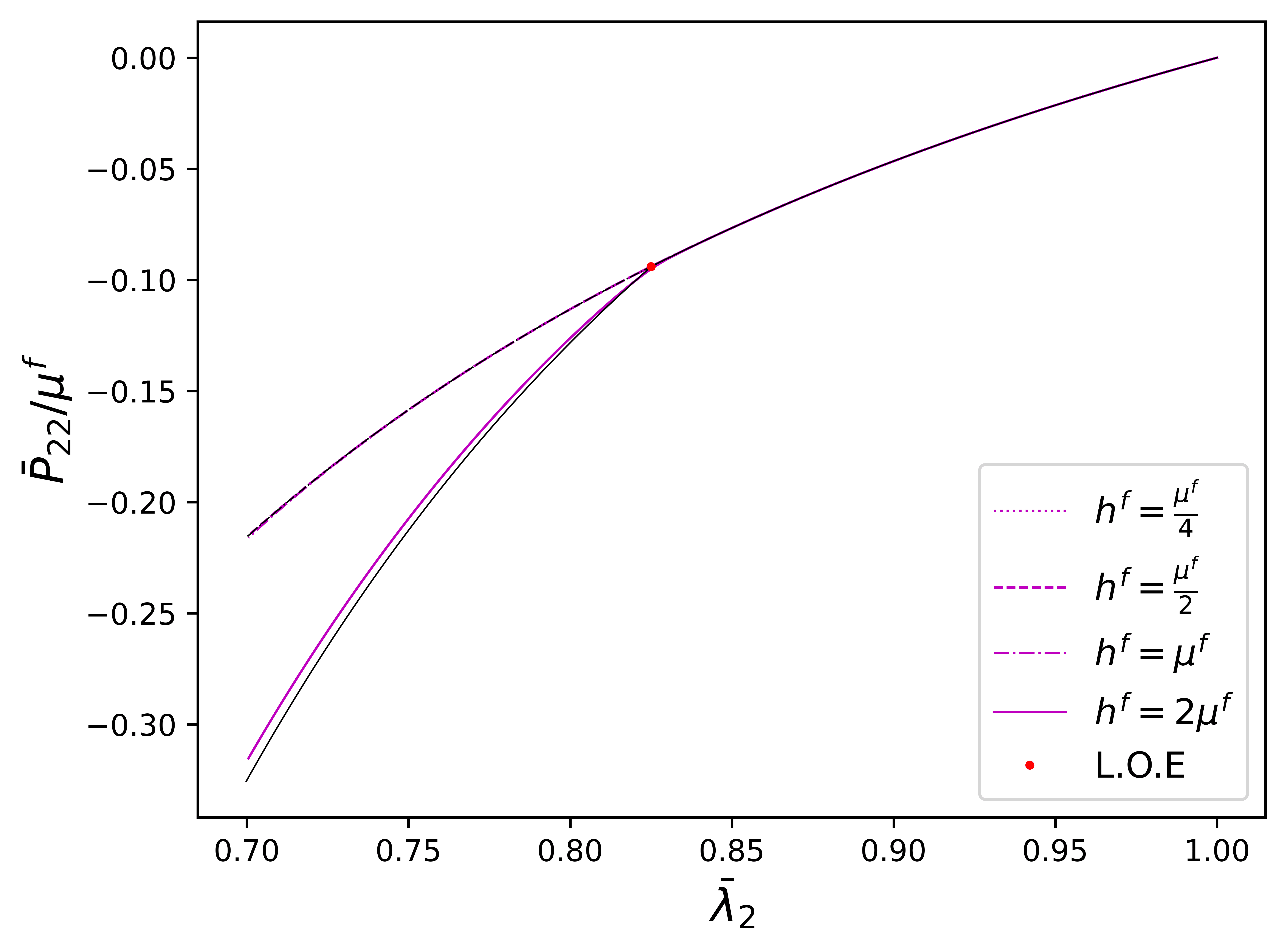}\label{fig:YL5HM}}
  \hspace{0mm}
  \subfloat[Accumulated plastic strain]{\includegraphics[width=0.45\textwidth]{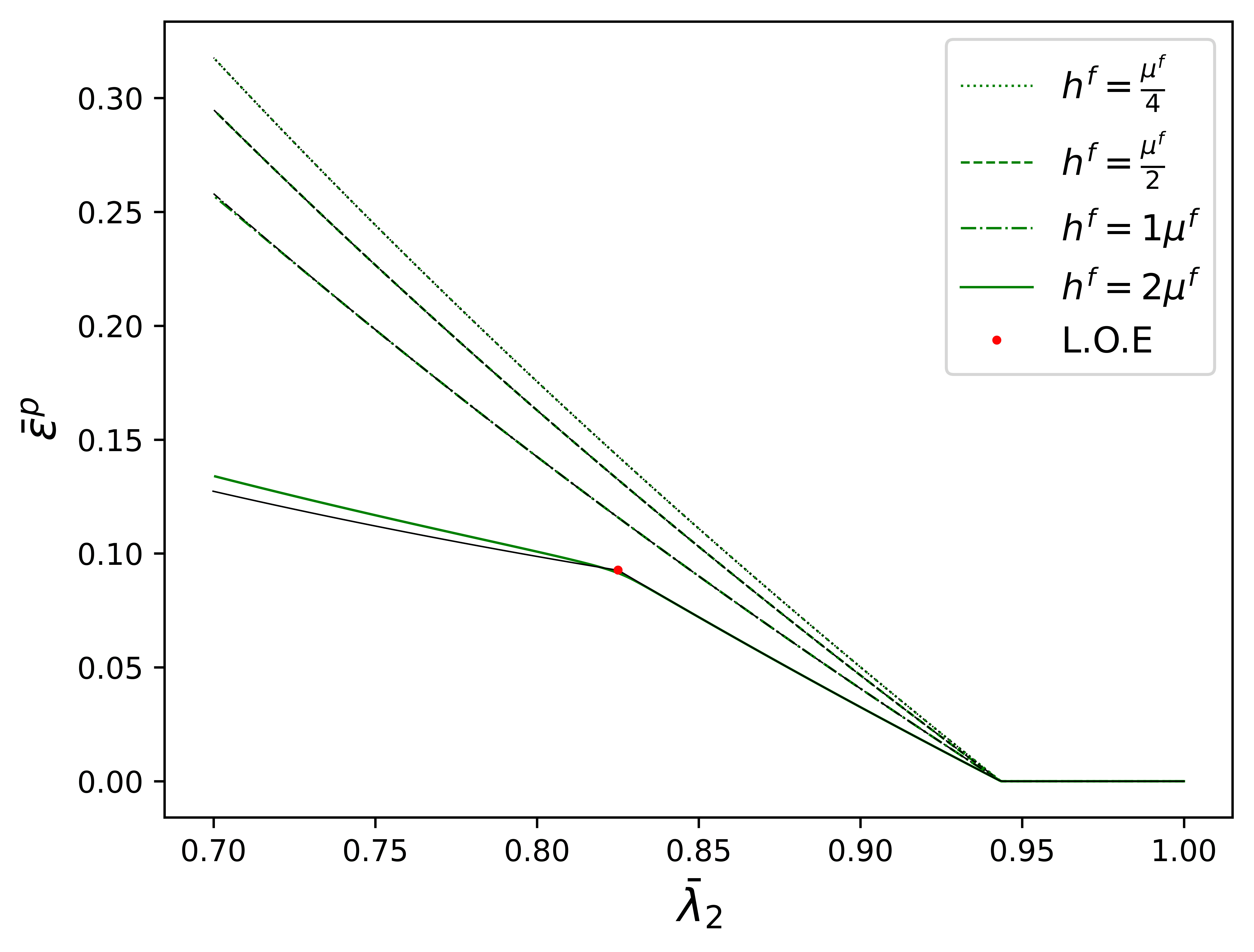}\label{fig:YL5Hep}}
  \caption{Results for laminates with different hardening, the same initial yield limit $\Sigma_y^{(f)}=\mu^{(f)}/5$ and the same elastic coefficients $\mu^{(f)}=10\mu^{(m)}$, $\kappa^{(f)}=100\mu^{(f)}$, $\kappa^{(m)}=100\mu^{(m)}$. With discontinuous lines are the cases where no domains were observed, while the black lines represent the analytical solution of the perfect laminates. The red dot indicates the loss of ellipticity in the analytical solution.}\label{fig:MCH}
\end{figure}

Up until this point, domains have been initiated and formed in bodies that were experiencing purely elastic deformations, whereas in the previous section it was shown that plasticity can in some cases suppress domain formation. In Part I a careful investigation of the impact that plastic phenomena have on the $\Tilde{\mathcal{L}}_{1212}$ component of the macroscopic incremental elastoplastic modulus --which when it reaches zero the analytical solution loses ellipticity-- showed their hardening effect --hence preventing the LOE-- during decreasing types of loadings (e.g. monotonic compression). This explains why lowering the yield limit, and thus allowing for plasticity to develop earlier, lead to no domain formation for the geometricaly imperfect models in section \ref{subsec:InitYL}, on the basis that the numerical solution is initially  in good agreement with the constant per-phase solution (on the principal path) in which case the LOE indicator accurately predicts domain formation. So the following question arises: can domains form under elastoplastic loads, during monotonic compression, for the imperfect body? Within the analytical framework of the unperturbed geometry in Part I, this question accepts an affirmative answer, given that the hardening of $\Tilde{\mathcal{L}}_{1212}$ is constrained. This can be achieved by increasing the hardening modulus and effectively allowing for the softening effect of elasticity on $\Tilde{\mathcal{L}}_{1212}$ --also shown in Part I-- to trigger LOE at an elastoplastically deformed state. Using the LOE criterion as a blueprint, we construct the following parametric study for varying hardening moduli, under constant yield limit $\mu^{(f)}/5$. The results are shown in Fig. \ref{fig:MCH}. Once again, the matrix response in Fig. \ref{fig:YL5HM}, highlights that only for $h^{(f)}=2\mu^{(f)}$ domains were formed, which is the only level of hardening that permits the loss of ellipticity. 

\begin{figure}[h!tbp]
  \centering
  \subfloat[$\Bar{\lambda}_2=0.825$ (LOE)]{\includegraphics[width=0.25\textwidth]{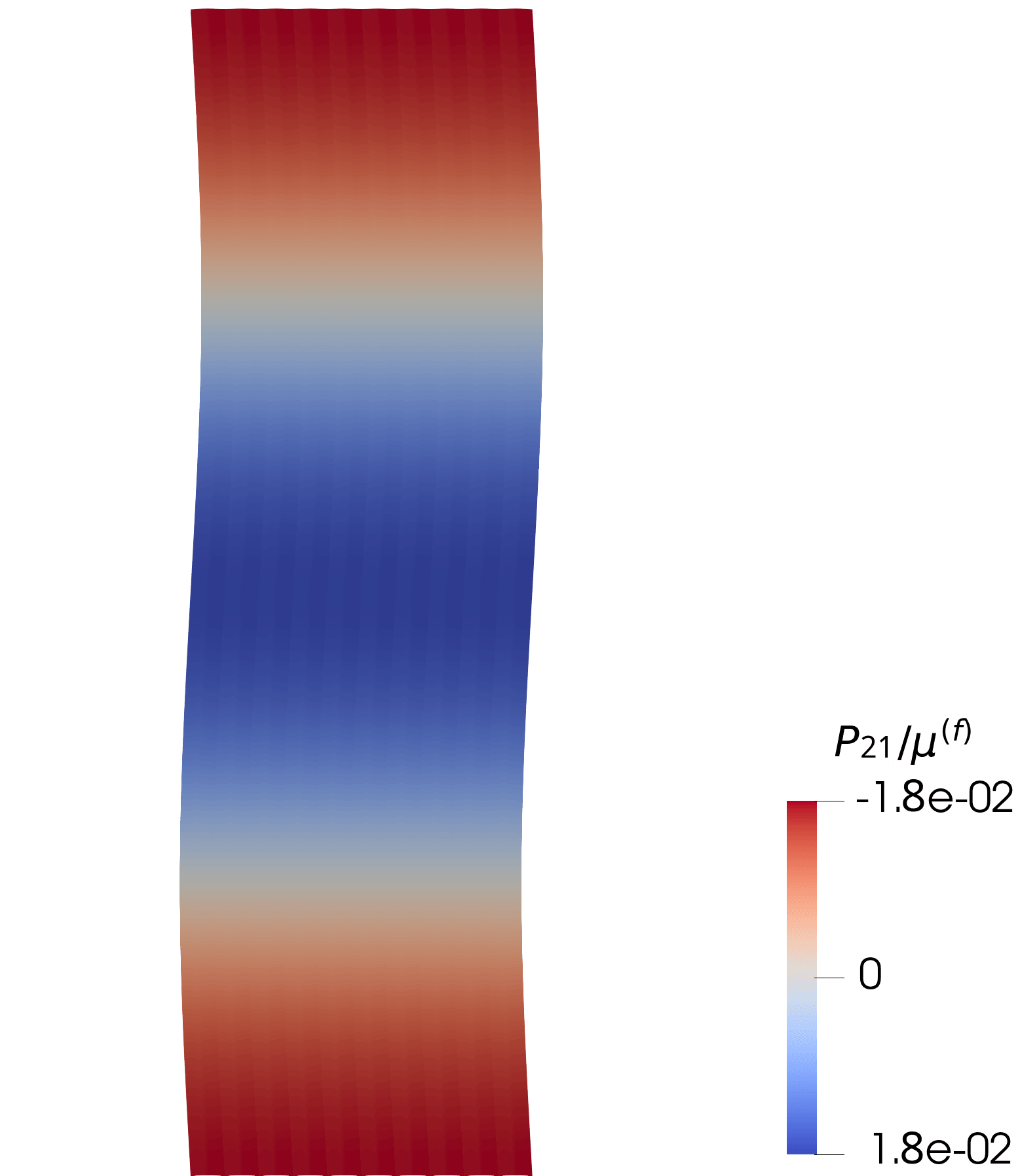}\label{fig:HYL5epLOE}}
  \hspace{15mm}
  \subfloat[$\Bar{\lambda}_2=0.825$ (LOE)]{\includegraphics[width=0.25\textwidth]{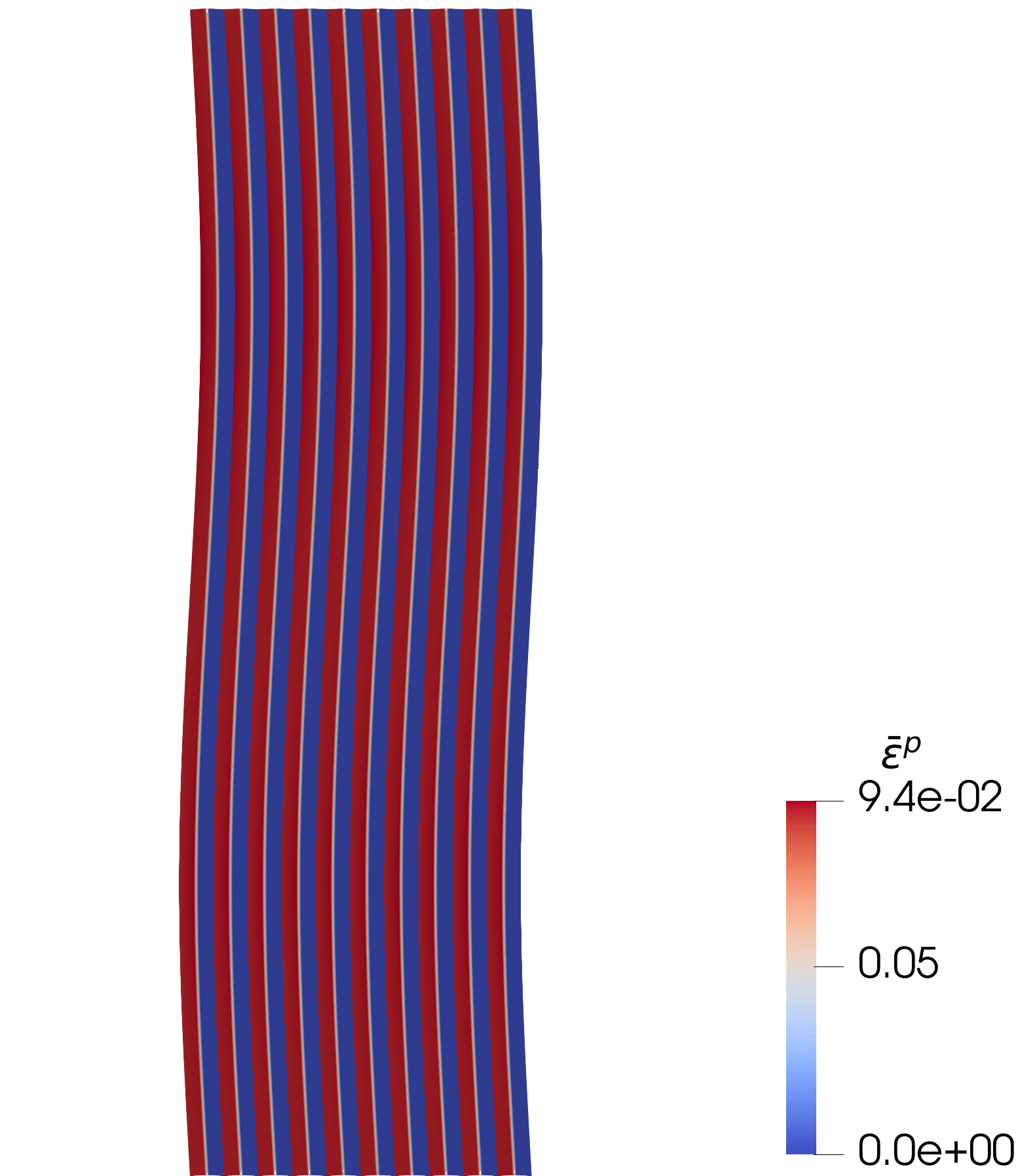}\label{fig:HYL5ACLOE}}
  
  \vspace{2mm}
  
  \subfloat[$\Bar{\lambda}_2=0.7$]{\includegraphics[width=0.25\textwidth]{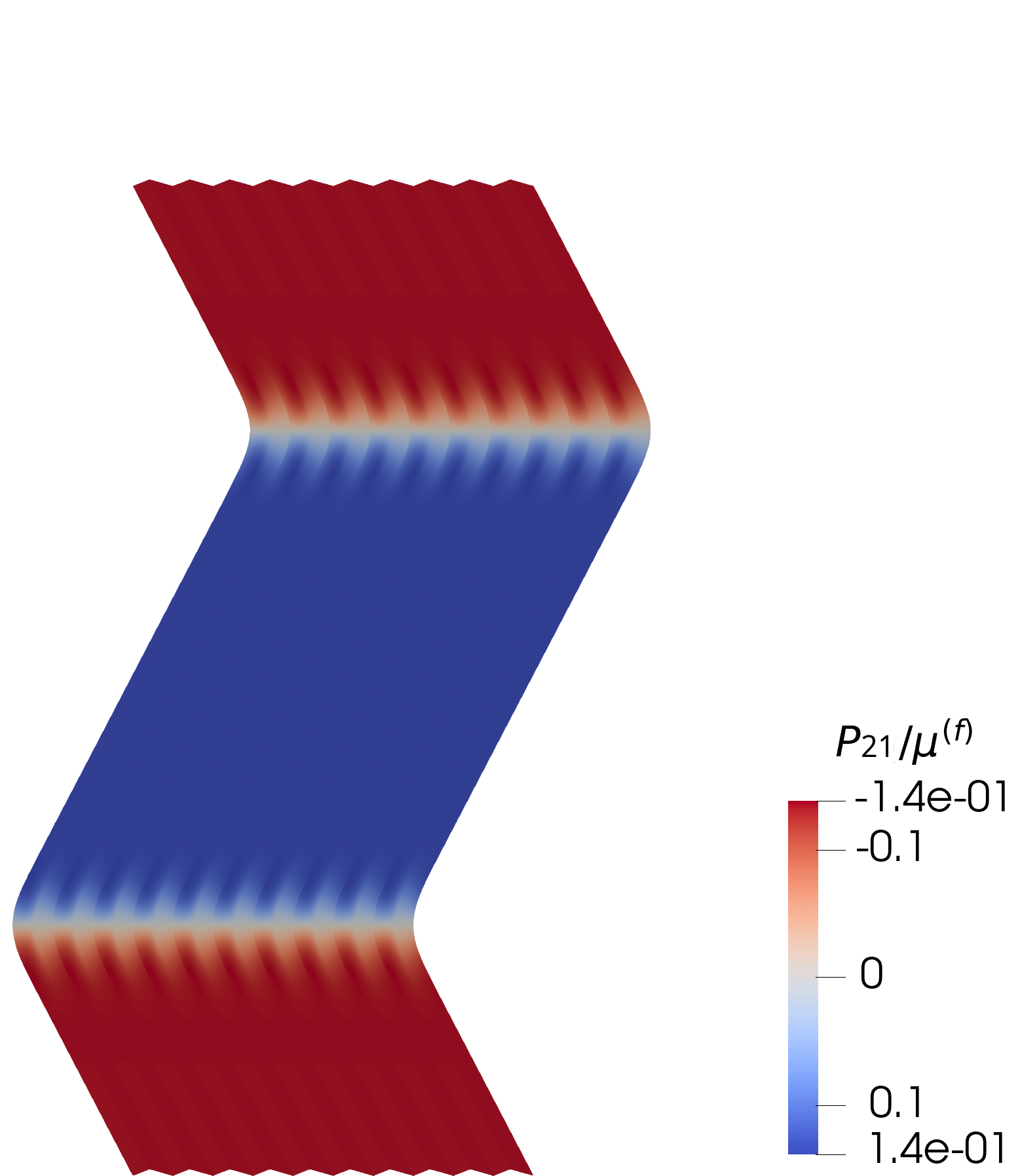} }
  \hspace{15mm}
  \subfloat[$\Bar{\lambda}_2=0.7$]{\includegraphics[width=0.25\textwidth]{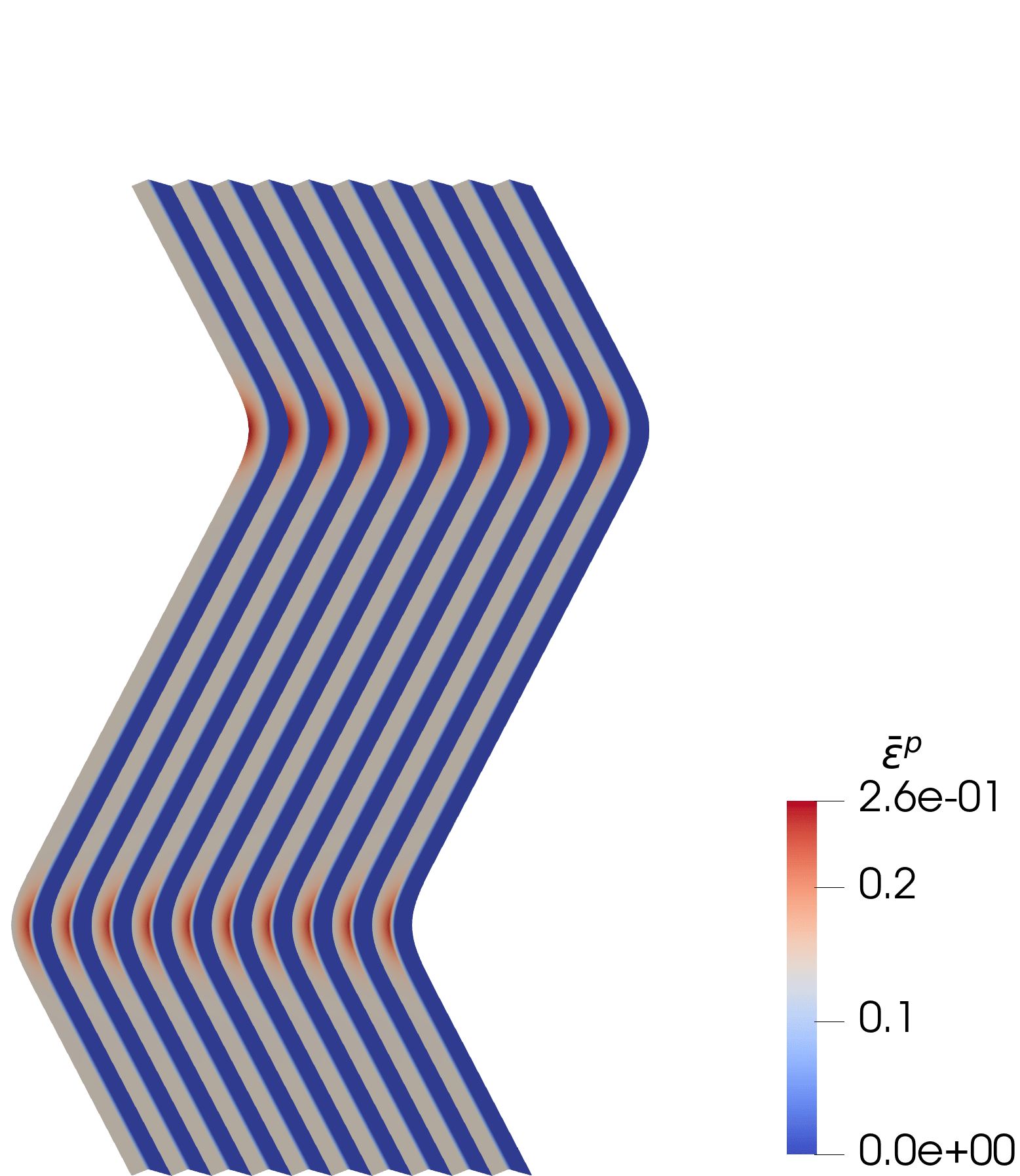}\label{fig:HYL5ep07}}
  \caption{Deformed configurations for the composite with $\Sigma_y^{(f)}=\mu^{(f)}/5$, $h^{(f)}=2\mu^{(f)}$, $\mu^{(f)}=10\mu^{(m)}$, $\kappa^{(f)}=100\mu^{(f)}$, $\kappa^{(m)}=100\mu^{(m)}$. The contours of $P_{21}/\mu^{(f)}$ and accumulated plastic strain are plotted at $\bar{\lambda}_2=0.903$ and at the final deformed state.}\label{fig:HYL5AC07}
\end{figure}

Due to the low amplitude geometric imperfection used, the numerical solution, prior to bifurcation, exhibits approximately constant per-phase fields of stress and strain and so plastic deformation is almost uniformly accumulated in the fiber phase, as is shown in Fig. \ref{fig:HYL5ACLOE}.  Despite the fact that domain formation, with its accompanied regions of intense bending lead to locally higher stresses and hence more accumulated plastic strain, highlighted by Fig. \ref{fig:HYL5ep07} and the plot at Fig. \ref{fig:YL5Hep}, their emergence could be rather beneficial for the overall state of the body. In the more conventional composites of \cite{kyriakides1995compressive}, studied both numerically and experimentally, the onset of an alternative solution was met with a sudden and catastrophic failure of the material. When domains can be formed, the overall response, for the given perturbation and loading conditions, remains stable. In fact there is macroscopic relaxation, thus shielding the material from abrupt failure. This observation also suggests that potential material damage/failure, which is not considered in this work, will be constrained inside the transition regions, in agreement with the experimental observations of \cite{veres2012designed}. From a different perspective, increased accumulated plastic strain could be an indication of fiber damage in transition regions, in the case where damage is a function of the accumulated plastic strain. 

To conclude this section, domains can emerge under elastoplastic compressive loads in imperfect lamiantes. The results are  consistent with the bifurcation analysis of Part I which dictates that sufficiently high elastic strains need to be allowed, so that their softening effect on $\Tilde{\mathcal{L}}_{1212}$ can trigger LOE, which is found to predict domain formation. This can be achieved by either intensifying elastic phenomena (e.g. by increasing the shear modulus contrast) or by inhibiting plasticity (e.g. increasing the hardening modulus). The consideration of geometric imperfections gives rise to inhomogeneous fields, due to the formation of transition regions acting as the phase boundaries of the rank-2 laminate. It is inside these regions --not present in the analytical solution-- where high bending stresses develop, intensifying the accumulation of plastic strain.

\subsection{Tension to compression\label{TenToComp}}
In all of the studies considered so far, the loading program was chosen to be monotonic compression. For the purely elastic composites, nothing of interest lays in the tensile stretch regime. The consideration of an elastoplastic fiber phase alters this viewpoint by incorporating dependency on the history of the deformation and accumulation of plastic strains during tensile loading. With the current loading scenario, an attempt to emulate loading cycles reminiscent of physiological tendon operation is performed. 

\begin{figure}[h!tbp]
  \centering
  \subfloat[Composite]{\includegraphics[width=0.45\textwidth]{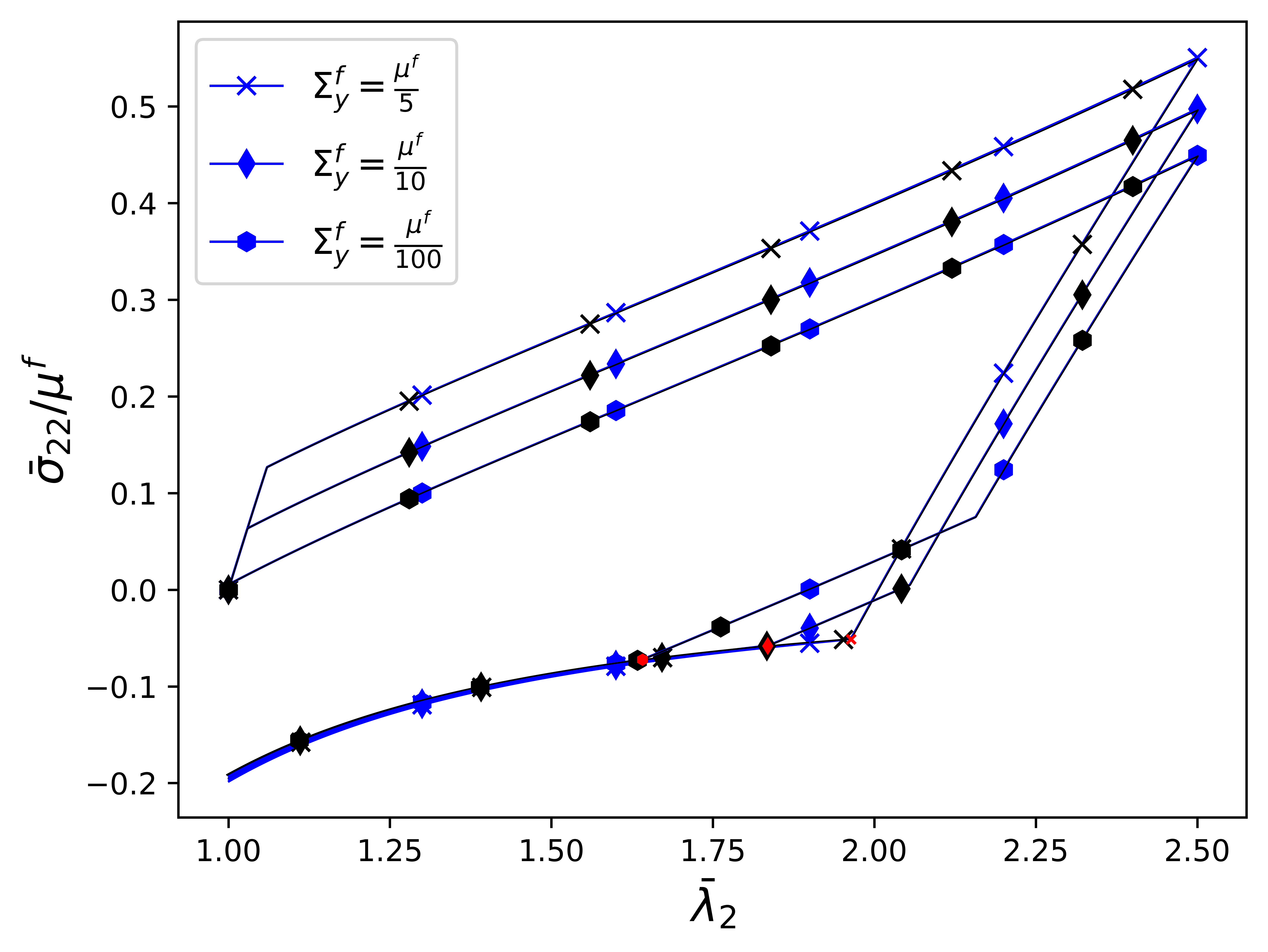}\label{fig:TTCYLcomp}}
  \hspace{0mm}
  \subfloat[Fiber phase]{\includegraphics[width=0.45\textwidth]{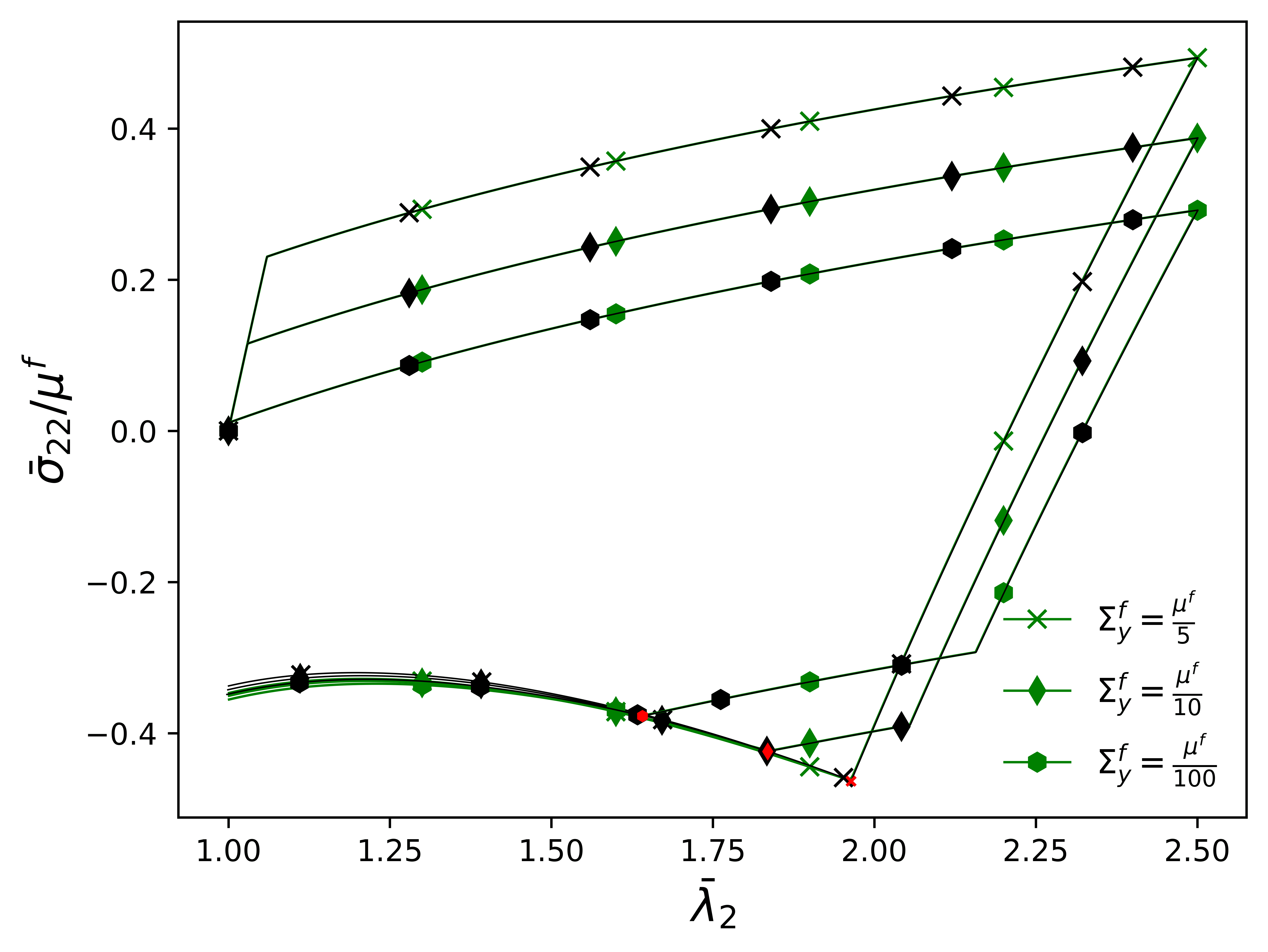}\label{fig:TTCYLfiber}}
  
  \vspace{1mm}
  
  \subfloat[Matrix Phase]{\includegraphics[width=0.45\textwidth]{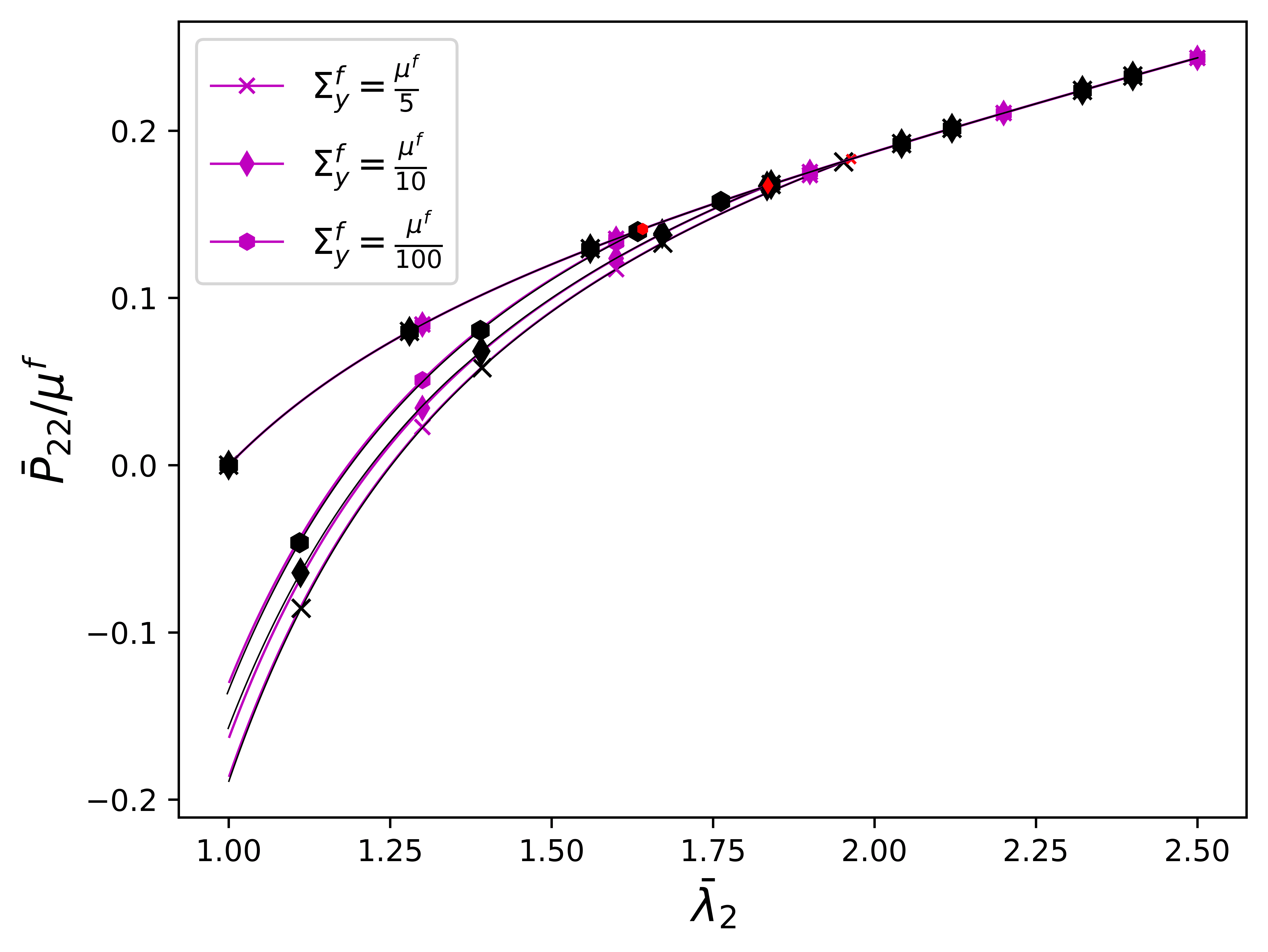}\label{fig:TTCYLmatrix}}
  \hspace{0mm}
  \subfloat[Accumulated plastic strain]{\includegraphics[width=0.45\textwidth]{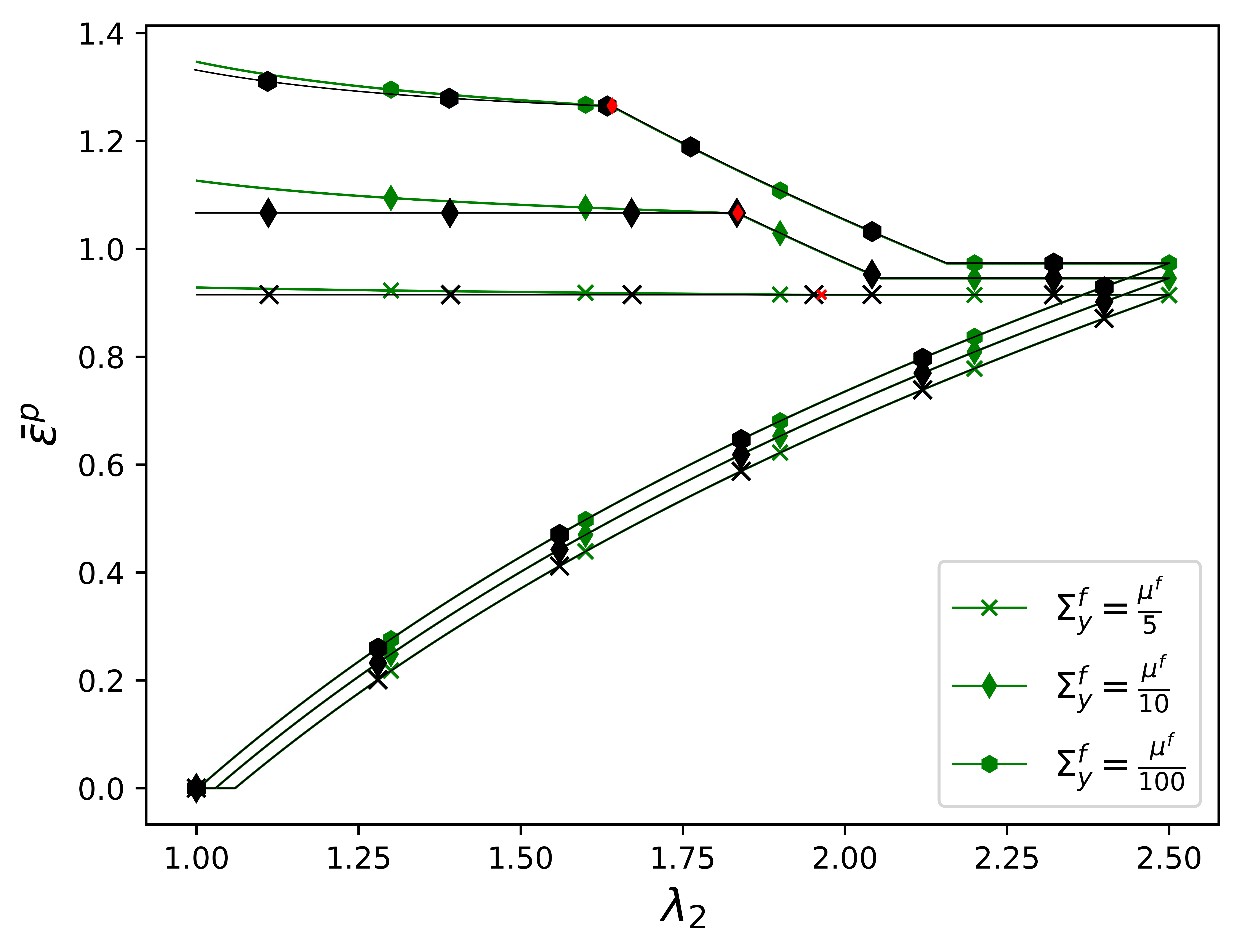}\label{fig:TTCYLep}}
  \caption{Results for laminates with different initial yield limit, the same hardening $h^{(f)}=\mu^{(f)}/4$ and the same elastic coefficients $\mu^{(f)}=10\mu^{(m)}$, $\kappa^{(f)}=100\mu^{(f)}$, $\kappa^{(m)}=100\mu^{(m)}$. With discontinuous lines are the cases where no domains were observed. The black lines represent the analytical solution. The red dot indicates the loss of ellipticity in the analytical solution }\label{fig:NonMonotLinePlts}
\end{figure}

In this section we consider loading paths that consist of a varying maximum tensile stretch ($\bar{\lambda}_2^{\text{max}}$), followed by  unloading and subsequent compression. The effect of the yield limit is explored, for a fixed hardening modulus $h^{(f)}=\mu^{(f)}/4$ and $\bar{\lambda}_2^{\text{max}}=2.5$ (see the appendix for the analogous study on the hardening modulus). The numerical results are shown in Fig. \ref{fig:NonMonotLinePlts} and as already shown in Part I, within the tensile stretch regime, prior plastic deformation can lead to macroscopic LOE which has been an accurate predictor of domain formation in our imperfect body. Following the discussion on the previous section, materials that are dominated by plastic deformations, i.e have lower yield limit, form domains at a later stage of the reverse loading, when sufficient elastic strains have developed. It is interesting to note that when domains fully develop, the responses coincide (see Fig. \ref{fig:TTCYLcomp}). This is attributed to the dominant role of the matrix phase in the post bifurcation regime and the response of the fiber phase to subsequent yielding, both of which are common features among the different materials considered. 

\begin{figure}[h!tbp]
  \centering
  \subfloat[$\bar{\lambda}_2<1$]{\includegraphics[width=0.35\textwidth]{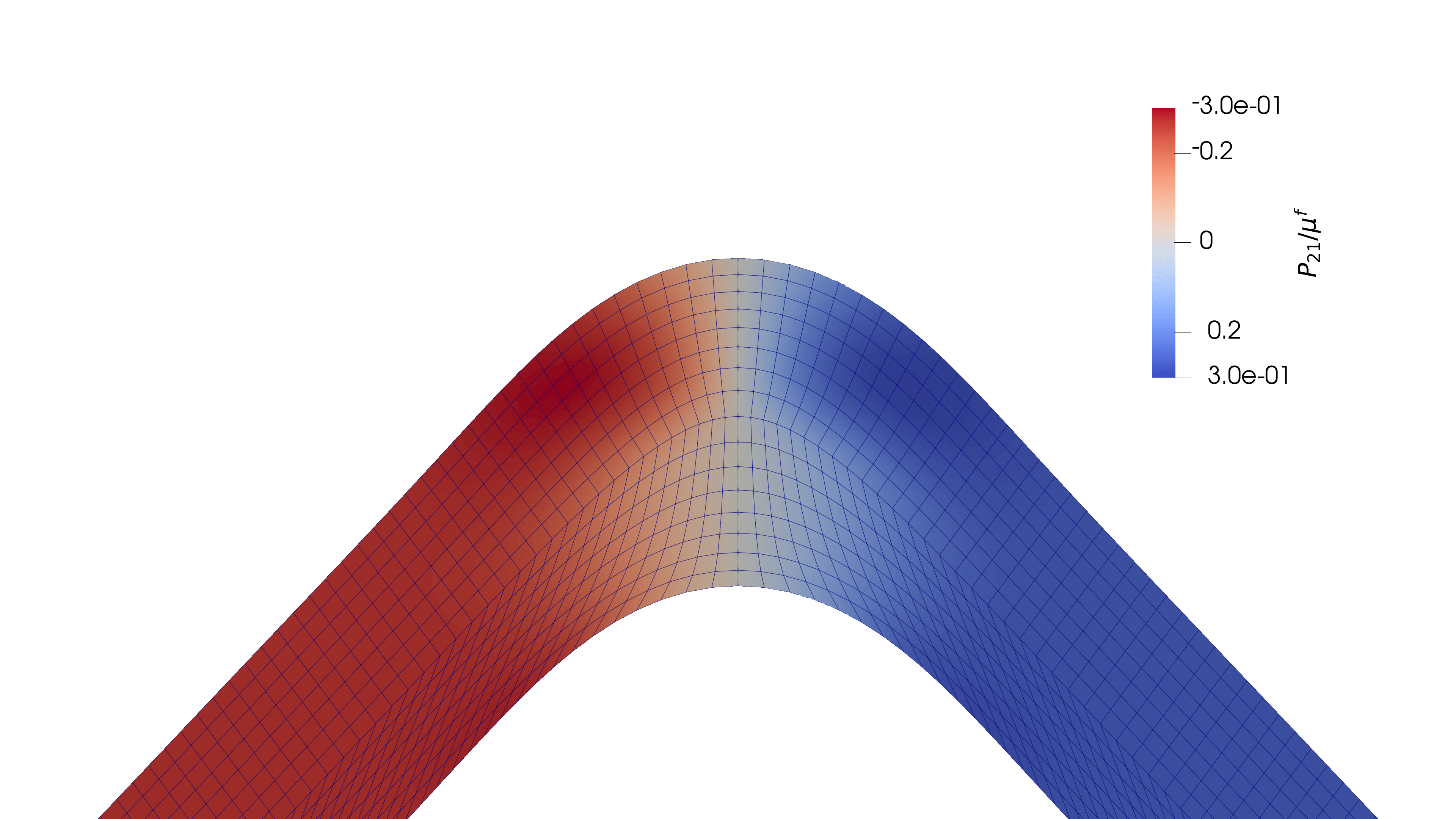}\label{fig:CompLEO}}
  \hspace{0mm}
   \subfloat[$\bar{\lambda}_2>1$]{\includegraphics[width=0.35\textwidth]{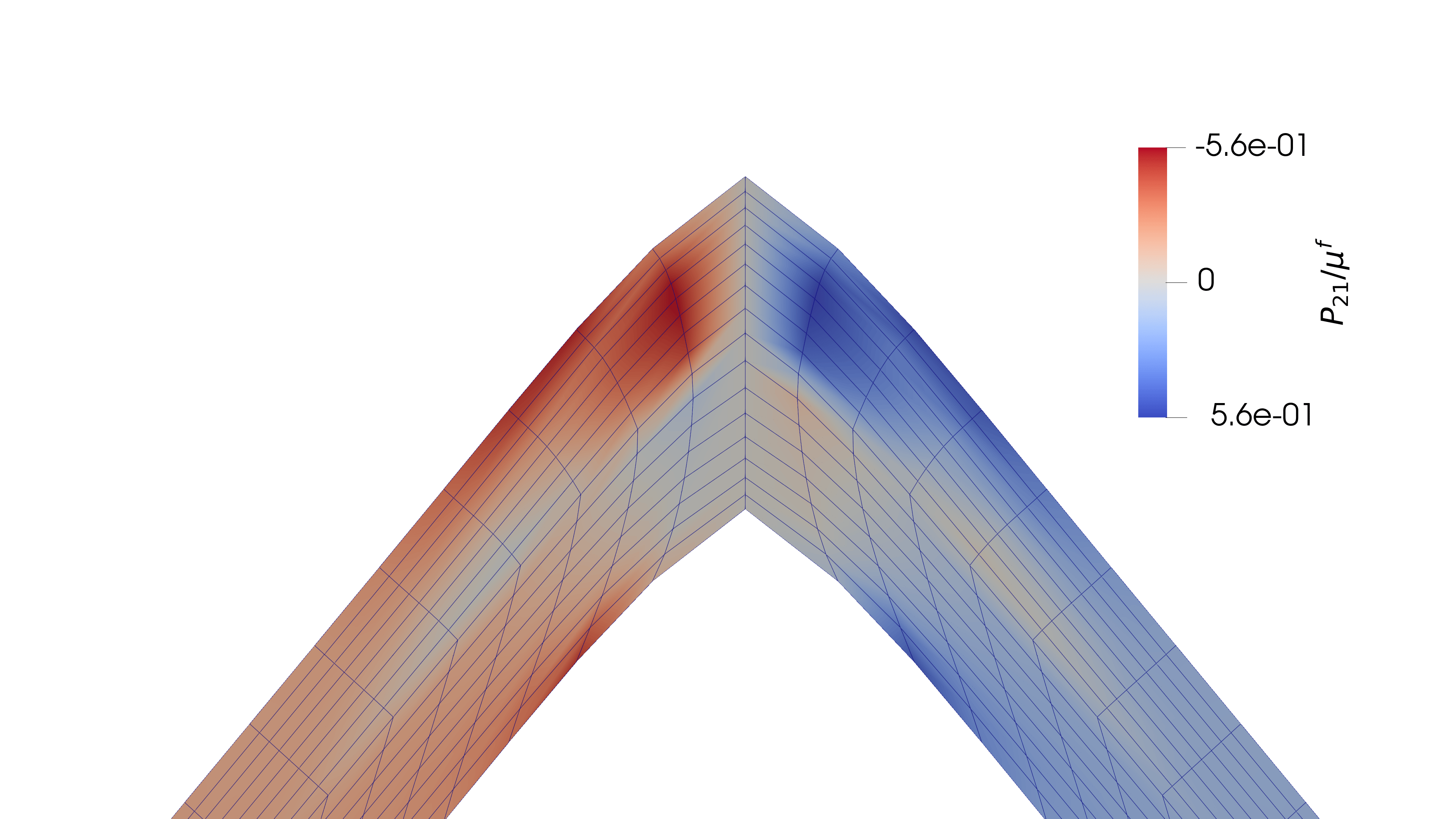}\label{fig:TensLEO}}
 \caption{Transition regions for 2 laminates that were subjected to different loading programs. The left corresponding to monotonic compression while the right to non-monotonic loading. The contours of $P_{21}/\mu^{(f)}$ are depicted.\label{fig:TranReg}}
\end{figure}

An important consequence of considering geometric imperfection and plasticity in non-monotonic loading scenarios is highlighted in Fig. \ref{fig:TranReg}, where the transition regions of 2 elastoplastic composites are shown. On the left hand side of Fig. \ref{fig:TranReg} the model with $\Sigma_y^{(f)}=\mu^{(f)}/5$, $h^{(f)}=2\mu^{(f)}$ which was subjected to monotonic compression  in the previous section is depicted. Despite the fact that  the fiber phase experienced approximately uniform yielding prior to domain formation, the trend of finite curvature regions, consistent with the purely elastic case in compression, is retained. On the other hand, domains that were formed in the cyclic program, where plasticity was accumulated under tensile stretches, shown in Fig.\ref{fig:TensLEO}, appear to be sharper. The distinguishing factor between the two cases is the mismatch, between the plastic and the elastic tensile strains, in the the unloading part of the non monotonic sequence.

The final study to be performed is centered around the effects of the maximum applied stretch during tension. The same investigation is performed, for the unperturbed geometry, in Part I where the fundamental result is that the continuous softening of $\Tilde{\mathcal{L}}_{1212}$ during the tensile elastoplastic step is prolonged when $\bar{\lambda}_2^{\text{max}}$ is increased and hence potential LOE is encountered at increased magnitudes of stress and strain. In the reasonable range of $\bar{\lambda}_2^{\text{max}}$ that was considered in Part I, LOE was observed in tensile normal strain in the direction of the loading -- in which case domain formation has been verified by the numerical results-- but never under tensile normal stresses (in the direction of the loading). Since so far, the LOE of the analytical solution has been a trustworthy indicator of domain formation, we increase the maximum applied tensile stretch to $\bar{\lambda}_2^{\text{max}}=4$ in order to investigate if the perturbed system is able to form domains under tensile normal stresses. 
The response of the composite, under increasing amplitude of the sinusoidal perturbation $\Delta X_1^S$, is shown in Fig. \ref{fig:VaringLamda}. (see the Appendix for a study under decreasing amplitude of the perturbation)

\begin{figure}[ht!bp]
  \centering
    \includegraphics[width=0.45\linewidth]{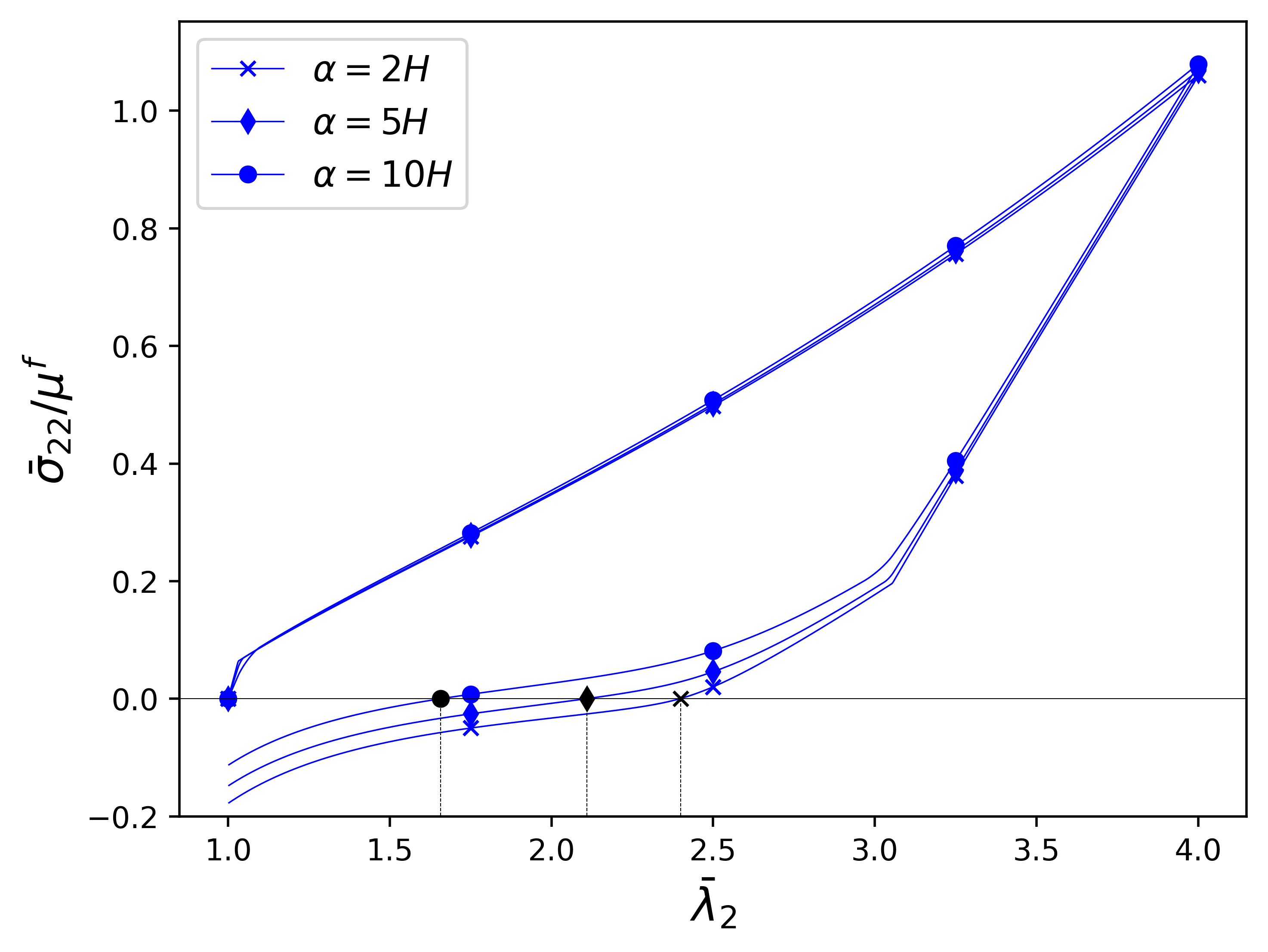}
    \caption{\label{fig:VaringLamda} Results for laminates with the same plastic parameters $h^{(f)}=\mu^{(f)}/4$, $\Sigma_y^{(f)}=\mu^{(f)}/10$, the same elastic coefficients $\mu^{(f)}=10\mu^{(m)}$, $\kappa^{(f)}=100\mu^{(f)}$, $\kappa^{(m)}=100\mu^{(m)}$ and different amplitudes of the sinusoidal geometric perturbation.} 
\end{figure}

\begin{figure}[h!tbp]
  \centering
  \subfloat[$\alpha=2H$, $\bar{\lambda}_2=2.4, \bar{\sigma}_{22}>0$]{\includegraphics[width=0.28\textwidth]{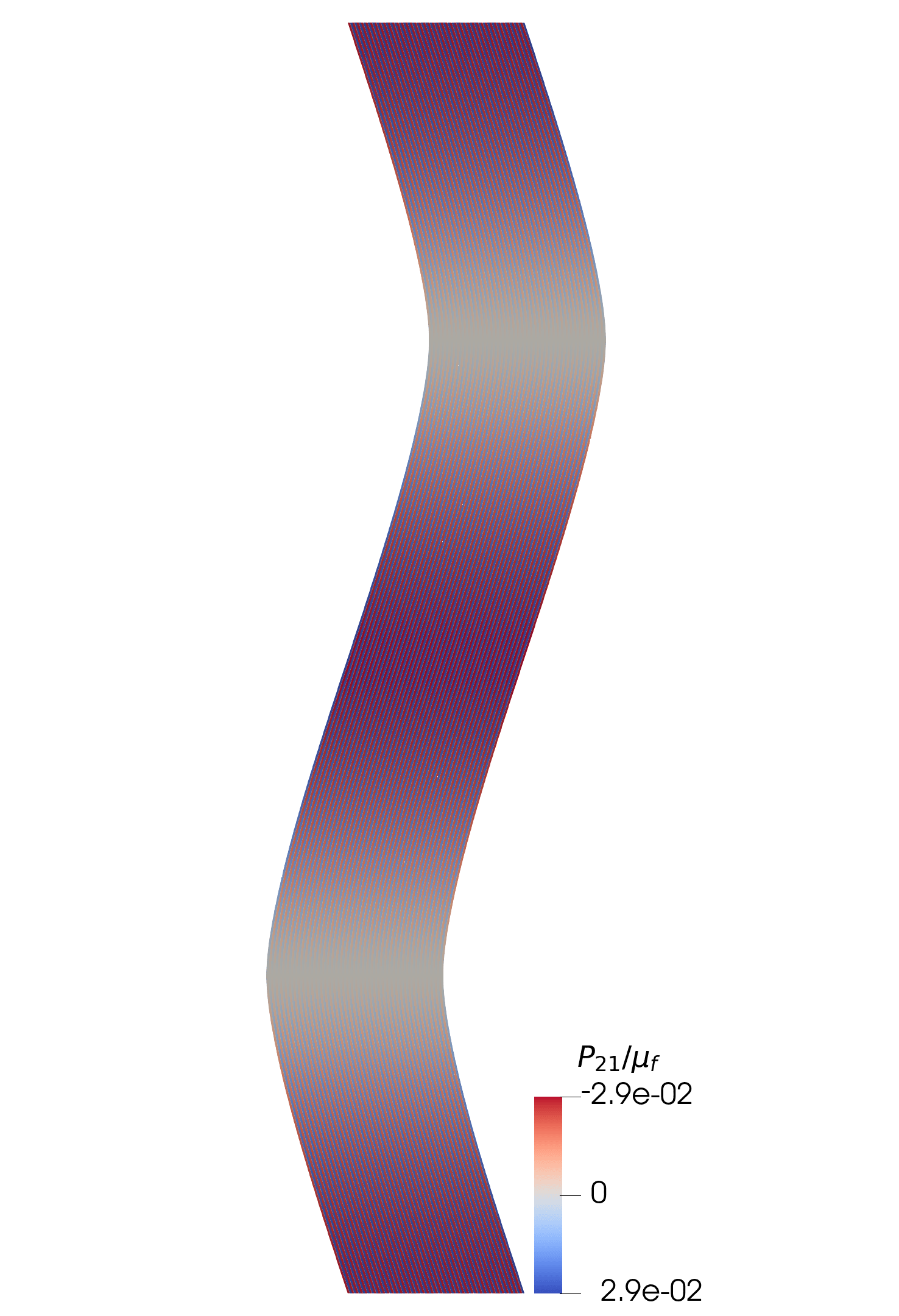}\label{}}
  \hspace{0mm}
   \subfloat[$\alpha=5H$, $\bar{\lambda}_2=2.1, \bar{\sigma}_{22}>0$]{\includegraphics[width=0.28\textwidth]{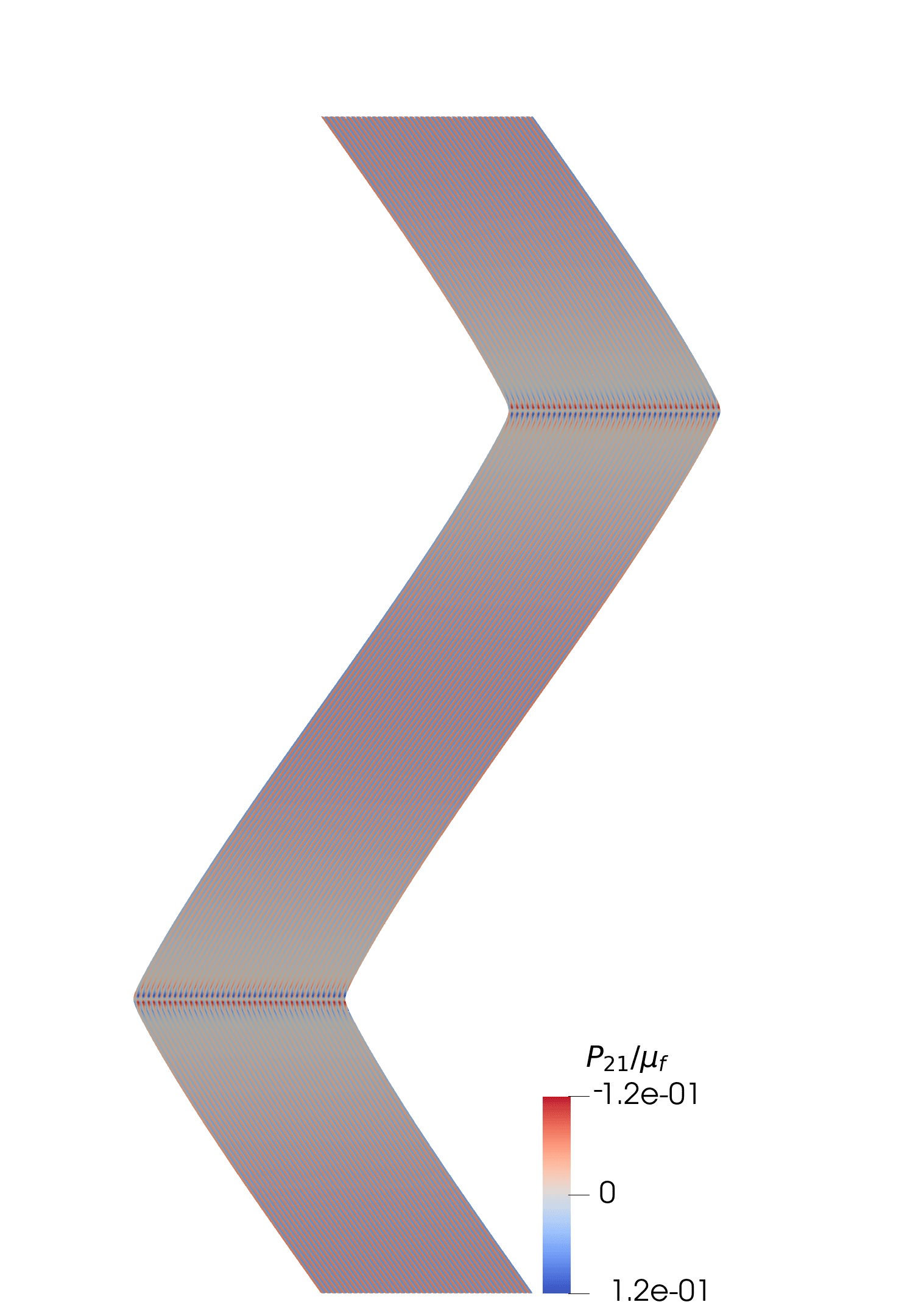}\label{}}
   \hspace{0mm}
   \subfloat[$\alpha=10H$, $\bar{\lambda}_2=1.6, \bar{\sigma}_{22}>0$]{\includegraphics[width=0.28\textwidth]{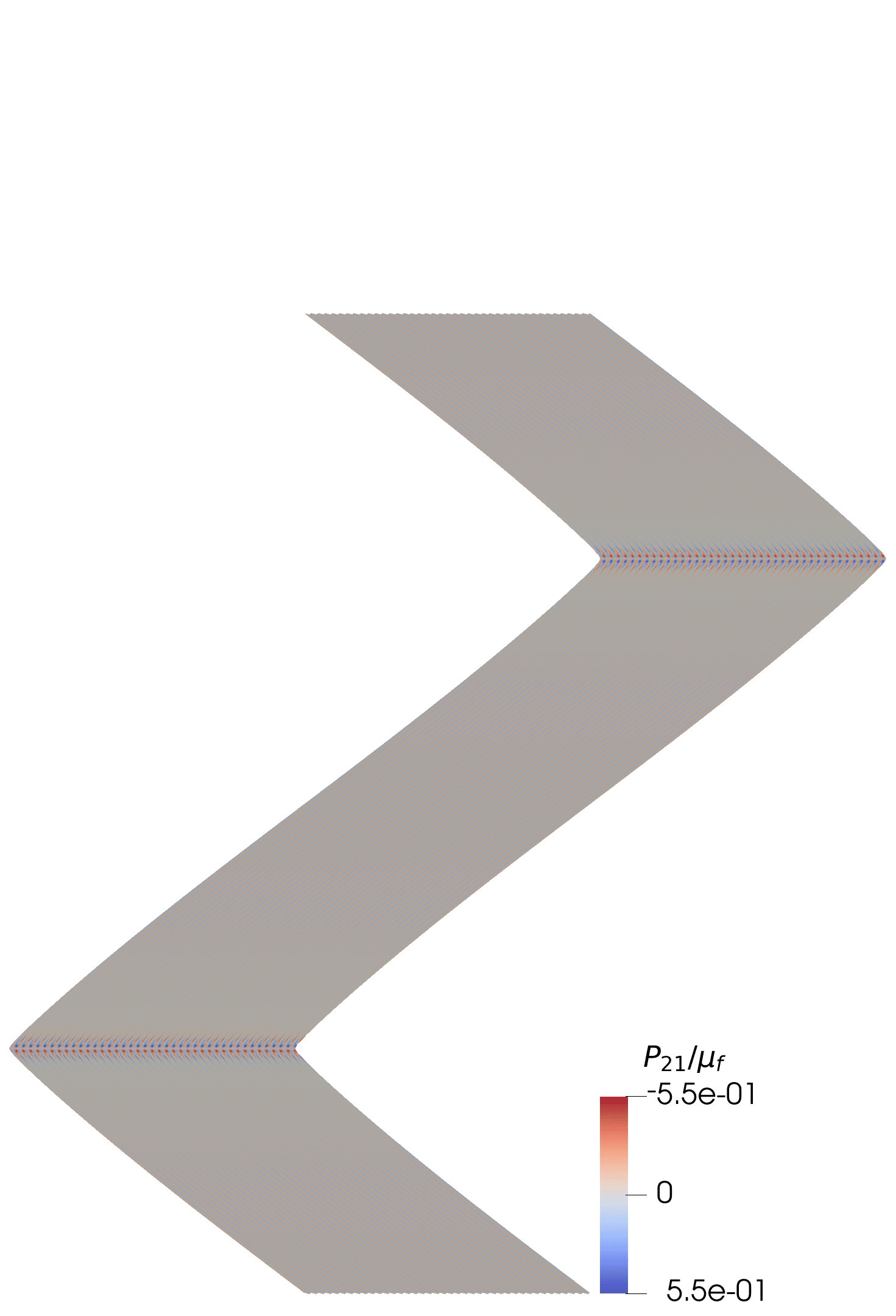}\label{}}
   
   \vspace{3mm}
  
  \subfloat[$\alpha=2H$, $\bar{\lambda}_2=1, \bar{\sigma}_{22}<0$]{\includegraphics[width=0.28\textwidth]{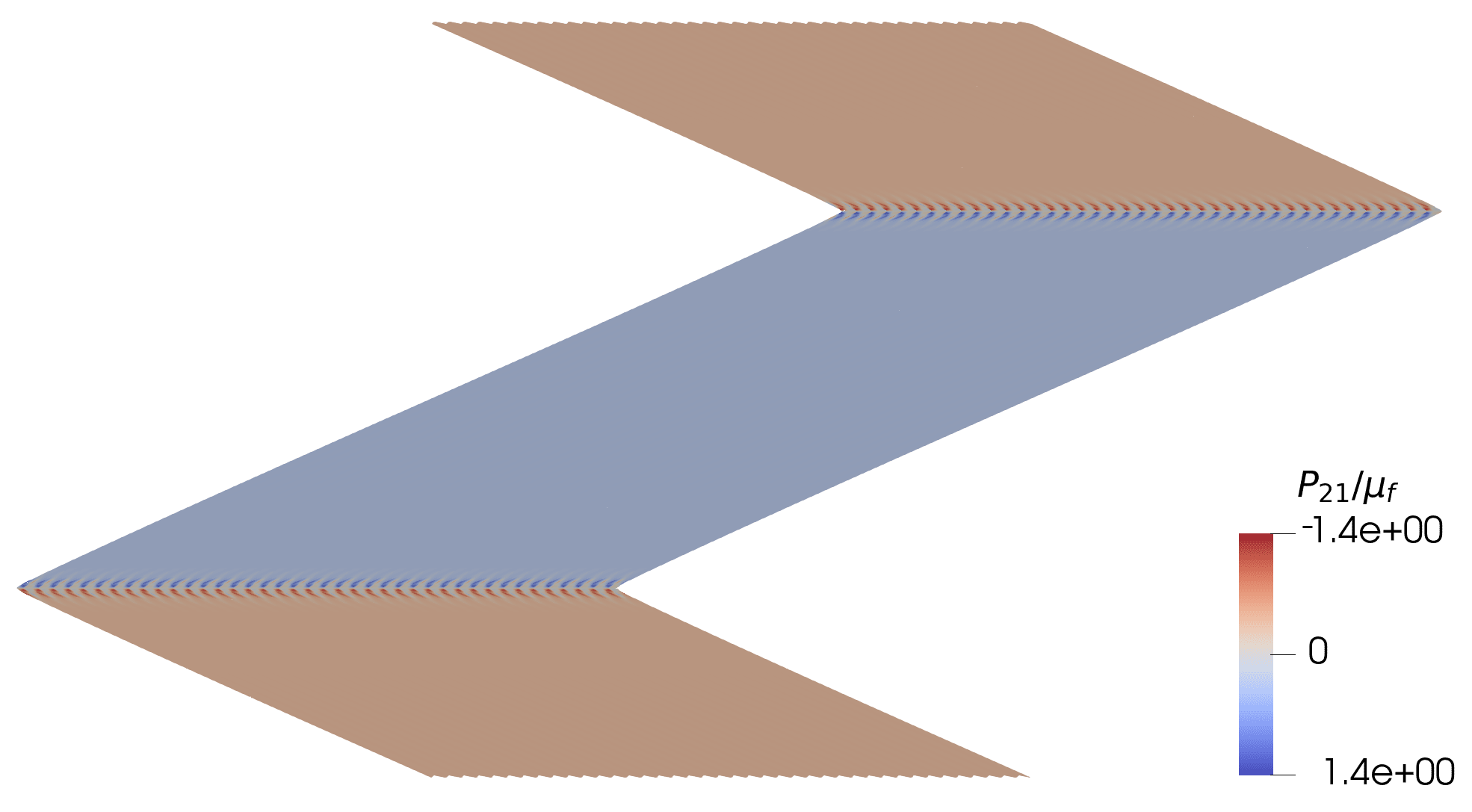}\label{}}
  \hspace{0mm}
  \subfloat[$\alpha=5H$, $\bar{\lambda}_2=1, \bar{\sigma}_{22}<0$]{\includegraphics[width=0.28\textwidth]{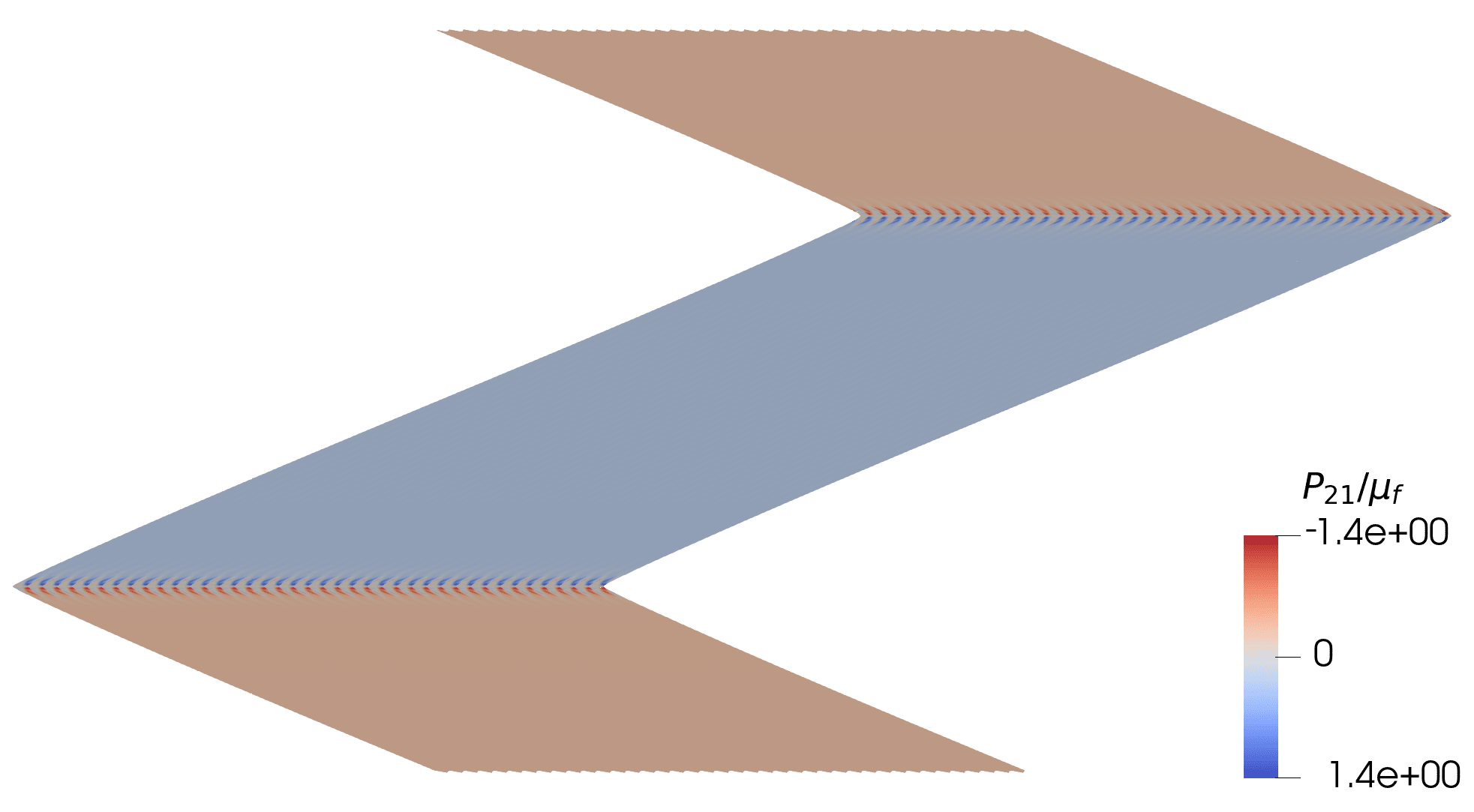}}
  \hspace{0mm}
   \subfloat[$\alpha=10H$, $\bar{\lambda}_2=1, \bar{\sigma}_{22}<0$]{\includegraphics[width=0.28\textwidth]{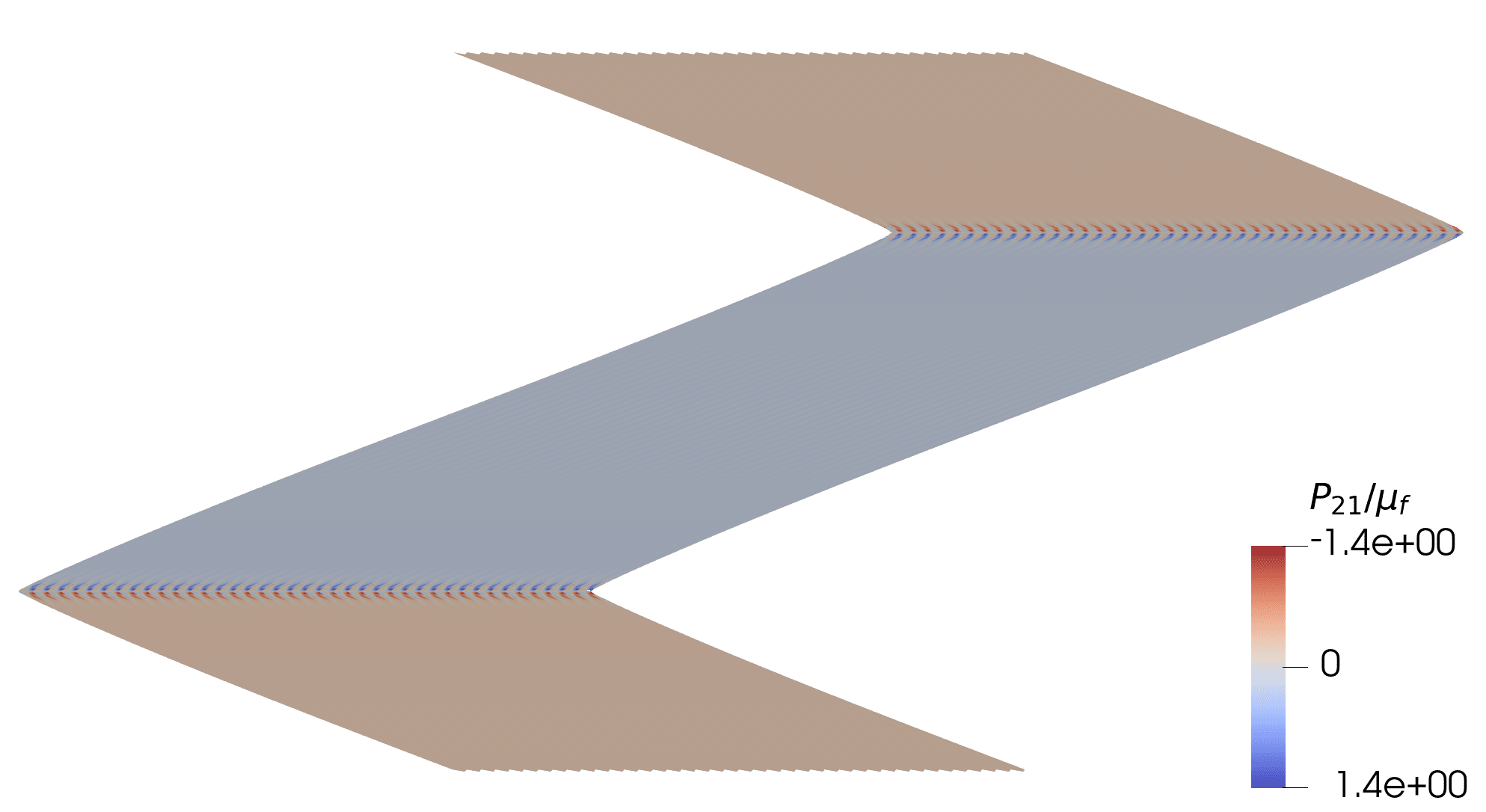}\label{}}
 \caption{Deformed configurations for the elastoplastic laminates of figure \ref{fig:VaringLamda} for the 3 different levels of geometric imperfections.\label{fig:DomainsInTensionALl}}
\end{figure}

The high amplitude geometric perturbations induce inhomogeneous fields from the beginning of the loading sequence, differentiating the responses observed in Fig. \ref{fig:VaringLamda}, despite the identical elastoplastic characteristics. During the reverse elastoplastic loading, the imperfect laminates considered allow for an earlier transition into a state where domains are observed -- note that this is contrary to the LOE predictions from Part I where the bifurcation of the perfect laminate is expected always in macroscopic compression --. More specifically, the last loading increments of the numerical simulation in which the bodies were in macroscopic tensile stresses, corresponding to the black markers of Fig. \ref{fig:VaringLamda}, are presented at the top row of Fig. \ref{fig:DomainsInTensionALl}.  40 unit cells have been used for the visualization of the final RVE.  As loading becomes compressive, domain formation is more clear even for low amplitude cases, as is shown at the bottom row of Fig. \ref{fig:DomainsInTensionALl}. The finding  specimens with large wavelength amplitude can exhibit domain formation in nonmonotonic loading programs, and specifically under tensile unloading, aligns with the emergence of domains in tendon where loading is primarily tensile in the direction of the collagen fibers and crimp is prominent.%Despite their geometric characteristics, the composites should not be treated as rank-2 laminates at this stage, as no distinct mesolayers have formed yet. Further application of the loading program gives, eventually, a better approximation of the produced rank-2 laminate, as is shown at the bottom row of Fig. \ref{fig:DomainsInTensionALl}.

\section{Concluding remarks}
\label{sec::conclusion}
In this work, we have been primarily concerned with the effect of geometric imperfections on the macroscopic response and domain formation in soft biological composites that exhibit fiber plasticity, with tendons being the archetype material in mind.  Physiologically,  these tissues are predominantly subjected to aligned  loading in the direction of the fibers, and considering plane strain loading conditions we simplify the problem to consider a laminate composite. Interestingly, the microstructure of these tissues exhibits imperfections at a hierarchy of scales, and critically, at the fiber scale their undeformed state exhibit a wavy pattern often referred to as crimp. We motivate our study with the results of the companion article, Part I, where the unperturbed geometry of a perfect two phase laminate was considered to explore the effect of fiber plasticity. We utilize the same constitutive laws presented in Part I, and follow similar loading programs to examine the response.   

The results in section \ref{sec::lastic} were centered around exploring domain formation presented in \cite{furer2018macroscopic} and extended to the elastoplastic case in Part I. For that, purely hyper-elastic phases in monotonic compression were studied. Two different perturbations were considered, a simple sinusoidal imperfection, and a more complex decaying case. Domains were formed in the former case and for low imperfection amplitudes their emerge was predicted to coincide with the loss of ellipticity of the governing equations for the unperturbed geometry. The principal path and post-bifurcation path were also consistent between the perfect and imperfect cases, confirming the softening mechanisms for the post-bifurcation response accommodated by rotation of the stiff fiber phase and increased shear in the matrix phase. Having in mind that the system under consideration is that of a tendon, there are intrinsic length scales --such as the finite diameter of a fibril-- that need to be taken into account by the modeling process. In that direction, a qualitative analysis of the influence that geometrical variables (such as the amplitude of the imperfection, the thickness to length ratio of phase and the number of wavelengths included in a unit cell) have on the overall behavior was conducted. For sinusoidal imperfections with characteristics motivated from tendon (in the vicinity of thickness to length ratios of $H/L=100$), the response of the imperfect specimen deviates from the analytical predictions as the effect of the transition zones that arise during domain formation is prominent. The intense bending within that region was found to cause a significant increase on the local stresses. On the other hand, the response in the bulk of each phase of the rank-2 laminate was consistent with the analytical results for the principal path and post-bifurcation. 

Following, an elastoplastic fiber phase was considered, and  monotonic as well as  nonmonotonic loading programs were examined for only sinusoidal imperfections. The numerical results were in line with the theoretical work of Part I and domain formation was observed. Their emergence for low amplitude imperfections closely matched the loss of ellipticity in the governing equations for the perfect laminate geometry and the competition between the elastic and plastic response was highlighted. Under set of assumptions, domain formation can be interpreted to mitigate excessive plasticity accumulation and allow for continuous operation. The intense bending in the transition regions causes the local increase of stress which results into a higher local plastic strain accumulation. These regions is where the material is expected to exhibit significant damage, something that has already been observed experimentally in tendons.     

However, as the imperfection amplitude is increased and becomes of the order to the layer width, or greater, domains begin to develop at macroscopically tensile stresses, which is in agreement with the fact that the loading of soft biological materials such as tendons and ligaments is tensile in nature. 
Thus, the findings of this work suggest strongly that plasticity and geometric imperfections of collagen fibers may play a key role on the onset and evolution of domains in actual soft biological composites.

\section{Acknowledgments}

D.I., F.F.F. and N.B. acknowledge the support by the National Science Foundation under grant
no. CMMI-2038057.

\section*{Appendix A}
\label{FDef}

\setcounter{figure}{0}
\renewcommand\thefigure{A\arabic{figure}}
\renewcommand{\theequation}{A.\arabic{equation}}
\setcounter{equation}{0}
\subsection*{Geometrical characteristics: Amplitude study}
\begin{figure}[ht!bp]
  \centering
    \includegraphics[width=0.60\linewidth]{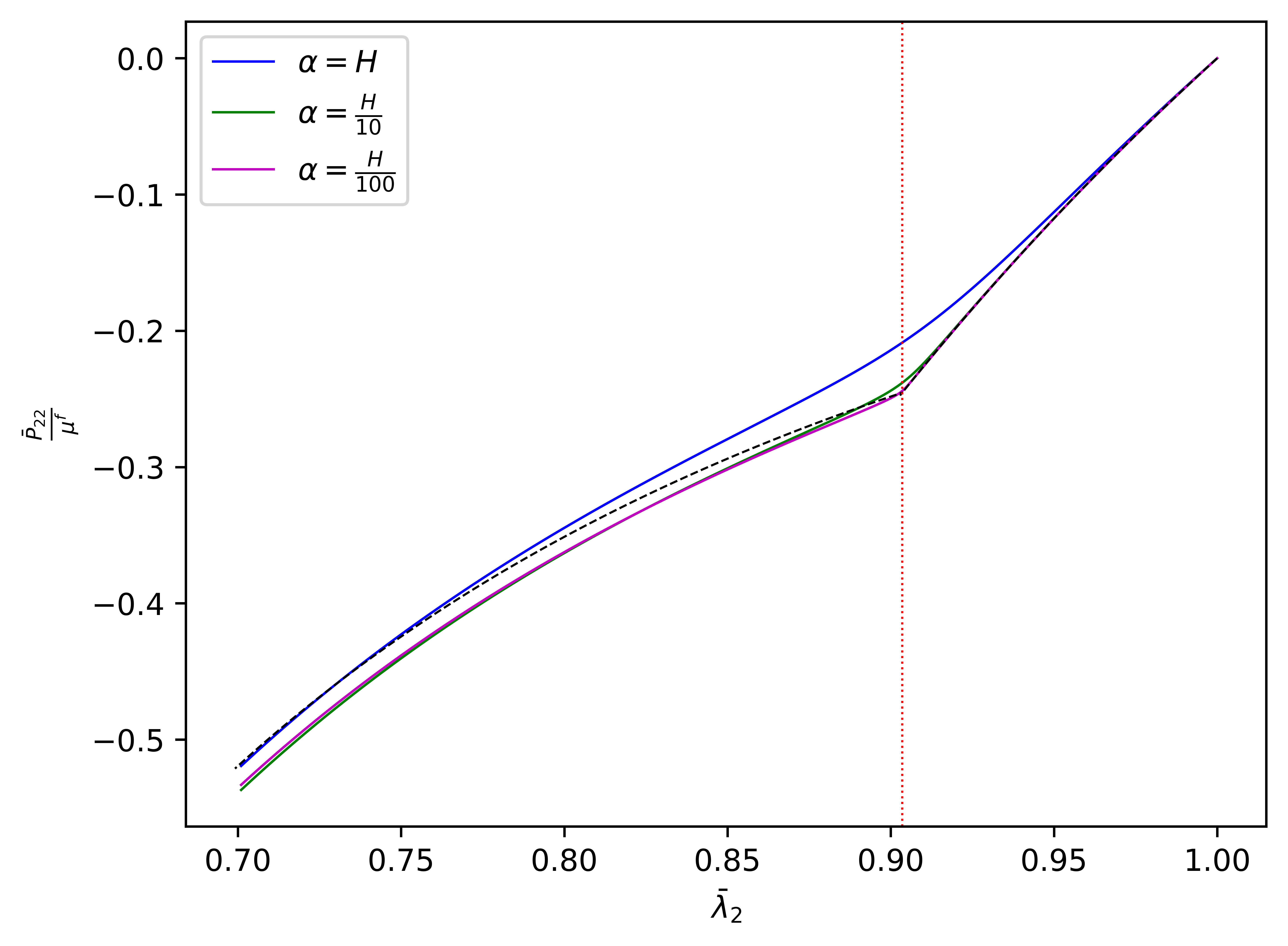}
    \caption{\label{fig:AmpAnalysis} response of the purely elastic composite with $\mu^{(f)}=10\mu^{(m)}$, $\kappa^{(f)}=100\mu^{(f)}$, $\kappa^{(m)}=100\mu^{(m)}$, under monotonic compression and a decreasing amplitude on the sinusoidal imperfection. The black dashed line is the analytical solution and the red, dotted, vertical line signals the L.O.E} 
\end{figure}

As is mentioned in the main part of this  work, the imperfect model in hand exhibits a continuous response , rather than a sudden bifurcation, when subjected in both monotonic and nonmonotonic loading programs. Decreasing the amplitude of the sinusoidal perturbation (see Fig. \ref{fig:AmpAnalysis}) results in an asymptotic convergence to the analytical solution prior to bifurcation while at the same time a better description of the abrupt change exhibited by the analytical solution is obtained, when macroscopic ellipticity is lost for the unperturbed specimen. Interestingly, for the largest imperfection considered, the response softens significantly earlier compared to the LOE point. This is critical in understanding how imperfections might impact the response of the composite. In the case presented only one wavelength is included in the unit cell. After LOE, the reader is reminded that the emergence of phase boundaries stiffens the response. 

\subsection*{Geometrical characteristics: Layer thickness study}
\begin{figure}[h!tbp]
  \centering
\includegraphics[width=0.5\textwidth]{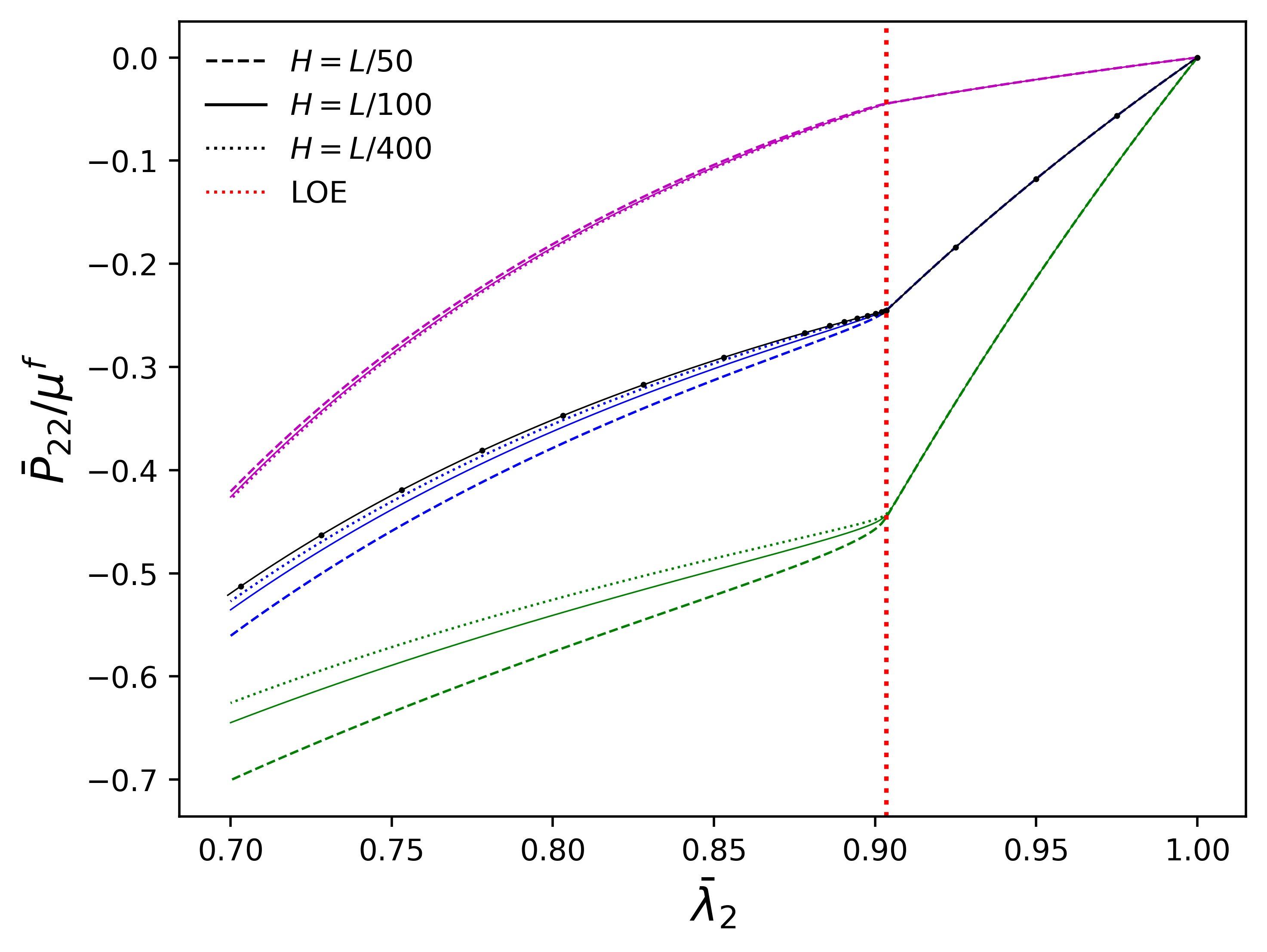}
\caption{Stress-Stretch response for geometrically imperfect models with different thickness to length ratios. The material properties for the laminates, which are subjected to monotonic compression, are $\mu^{(f)}=10\mu^{(m)}$, $\kappa^{(f)}=100\mu^{(f)}$,$\kappa^{(m)}=100\mu^{(m)}$. The blue color is used for the composite, the green for the fiber phase and the pink for the matrix. The analytical solution for the composite has been included with black markers. The vertical red line indicates the loss of ellipticity in the analytical solution.\label{fig:ResponseThickness}}
\end{figure}

\begin{figure}[h!tbp]
  \centering
  \subfloat[Averaged Strain energy densities]{\includegraphics[width=0.45\textwidth]{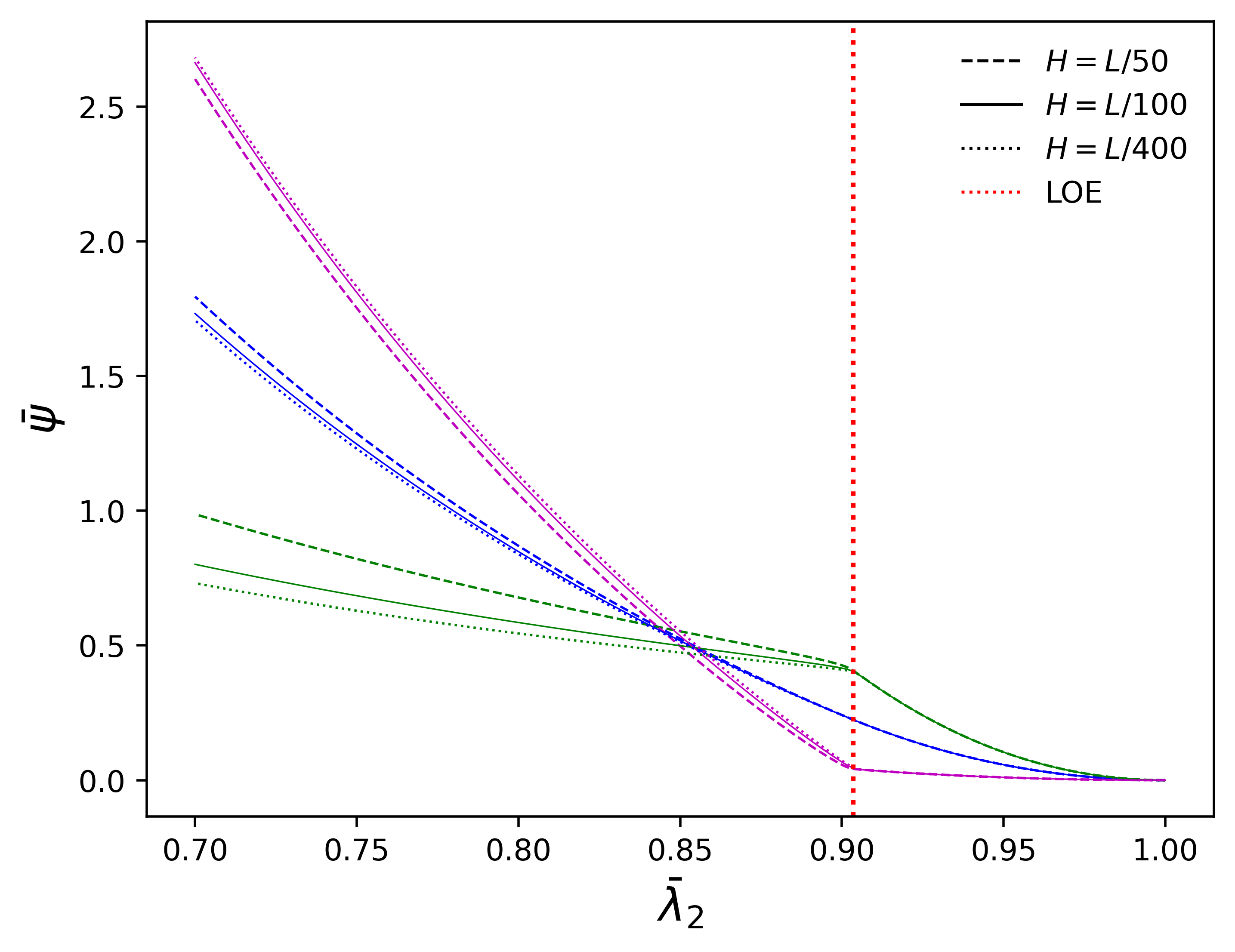}\label{fig:ThicnessEnergies}}
  \hspace{2mm}
  \subfloat[Non dimensionalized energies]{\includegraphics[width=0.45\textwidth]{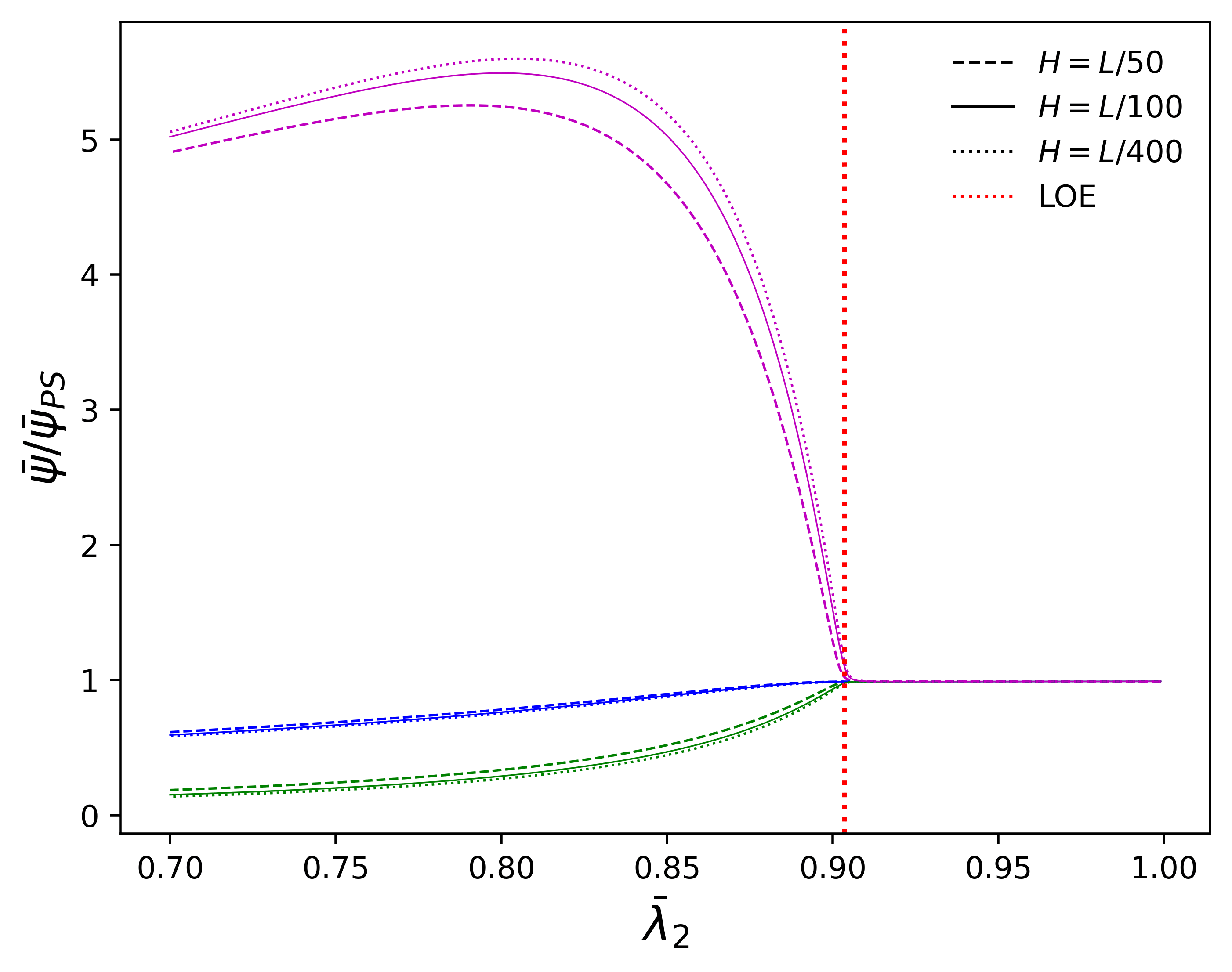}\label{fig:ThicnessNoNDimEnergies}}
  \caption{Strain energies for geometrically imperfect models with different different thickness to length ratios. The material properties for the laminates, which are subjected to monotonic compression, are $\mu^{(f)}=10\mu^{(m)}$, $\kappa^{(f)}=100\mu^{(f)}$,$\kappa^{(m)}=100\mu^{(m)}$. The blue color is used for the composite, the green for the fiber phase and the pink for the matrix. The vertical red line indicates the loss of ellipticity in the analytical solution.\label{fig:ThicknessEnergiesBoth}}
\end{figure}

This is a complementary study to the wavelength analysis, presented in the main text, in the sense that now the size of the transition regions is studied instead of their number. This is done by varying the ratio $H_r/L$ of the thickness of layer to its length, for a unit cell with $H_m = H_f = H$. The sinusoidal geometry is used with $\alpha=L\cdot 10^{-4}$, $w=1$ and the phases are modeled as hyper elastic with the same material parameters as is section \ref{subsec:WaveLength}. The response is shown in Fig. \ref{fig:ResponseThickness}. As has already been interpreted in the main part, these transition regions account for the stiffer (compared to the analytical solution) response. The same conclusion can be drawn from Fig. \ref{fig:ResponseThickness} where as the ratio $H/L$ gets progressively smaller, the size of the transition regions, relative to the size of the unit cell, gets smaller and the numerical solution of the perturbed system tends to the analytical solution of the unperturbed system, which constitutes a lower bound for the response. Similarly to the energy plots shown in Fig. \ref{fig:wavelengthEnergies}, the results in Fig. \ref{fig:ThicknessEnergiesBoth} are presented. It is again concluded that as the size of the transition regions gets smaller, their energetic contribution decreases. 

\subsection*{Geometrical characteristics: Spatially varying imperfection}
For the present work, two more families of geometric imperfections --that are inspired by the work of Kyriakides, Arseculeratne, Perry, Liechti-- were examined to access the sensitivity of domain formation. More specifically, the two families are:

\begin{align}
    &\Delta X_1^{(1)}(X_1,X_2)= -\left[\alpha+\hat{\alpha}\cos(\frac{2\pi X_2}{Ln})\right] \cdot \sin(2\pi X_2/L)\cdot \exp{(-\beta\abs{X_1}/H_{\text{tot}})} \\[10pt]
     &\Delta X_1^{(2)}(X_1,X_2) = -\left[\alpha+\hat{\alpha}\cos(\hat{y} \cdot 2\pi X_2/L)\right]\cdot \sin(2\pi X_2/L)\cdot \sin(2\pi X_1/H_{\text{tot}})
\end{align}

\begin{figure}[h!tbp]
  \centering
  \subfloat[$\bar{\lambda}_2=0.95$]{\includegraphics[width=0.49\textwidth]{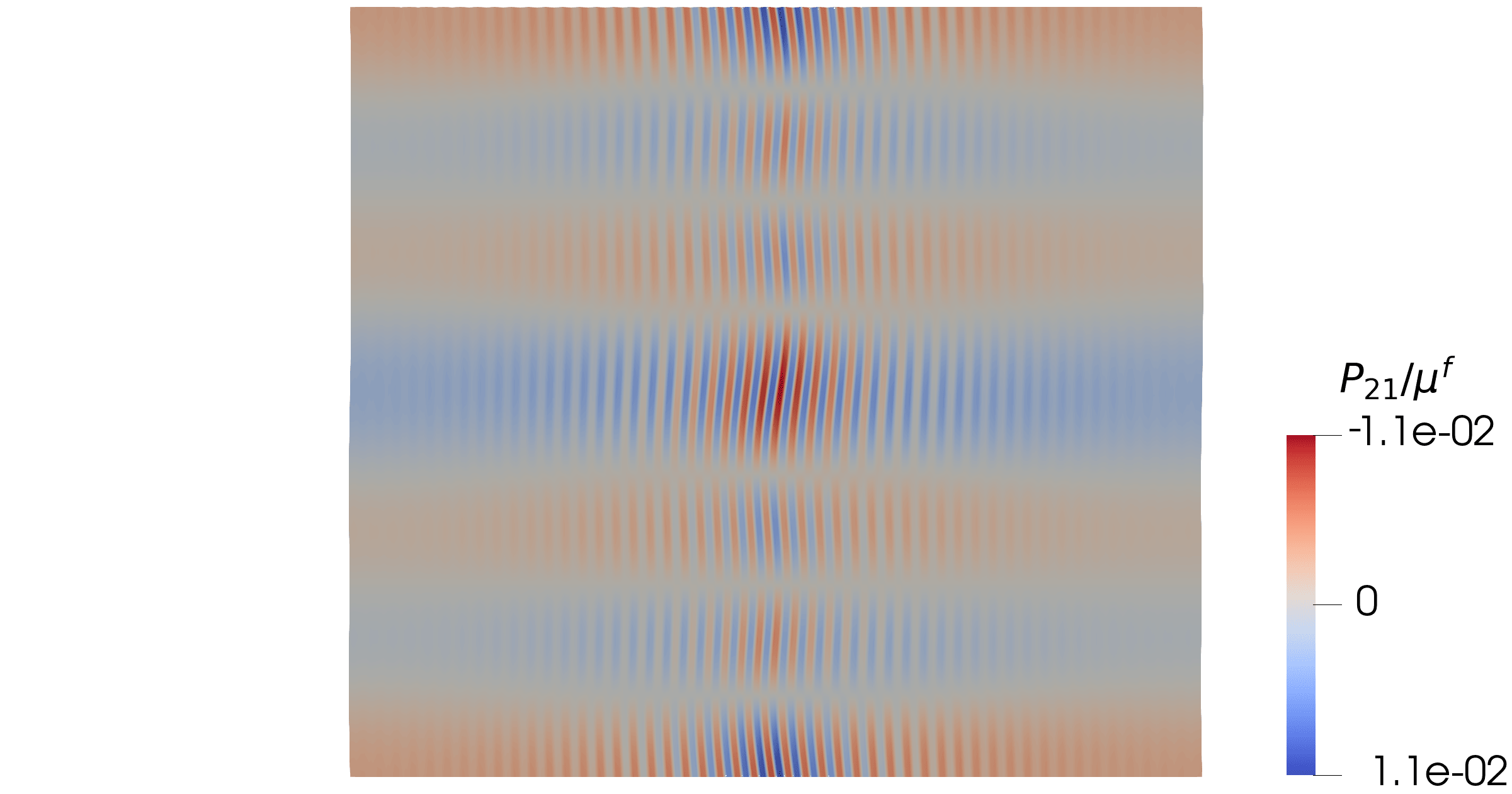}}
  \hspace{0mm}
  \subfloat[$\bar{\lambda}_2=0.903$ (LOE)]{\includegraphics[width=0.49\textwidth]{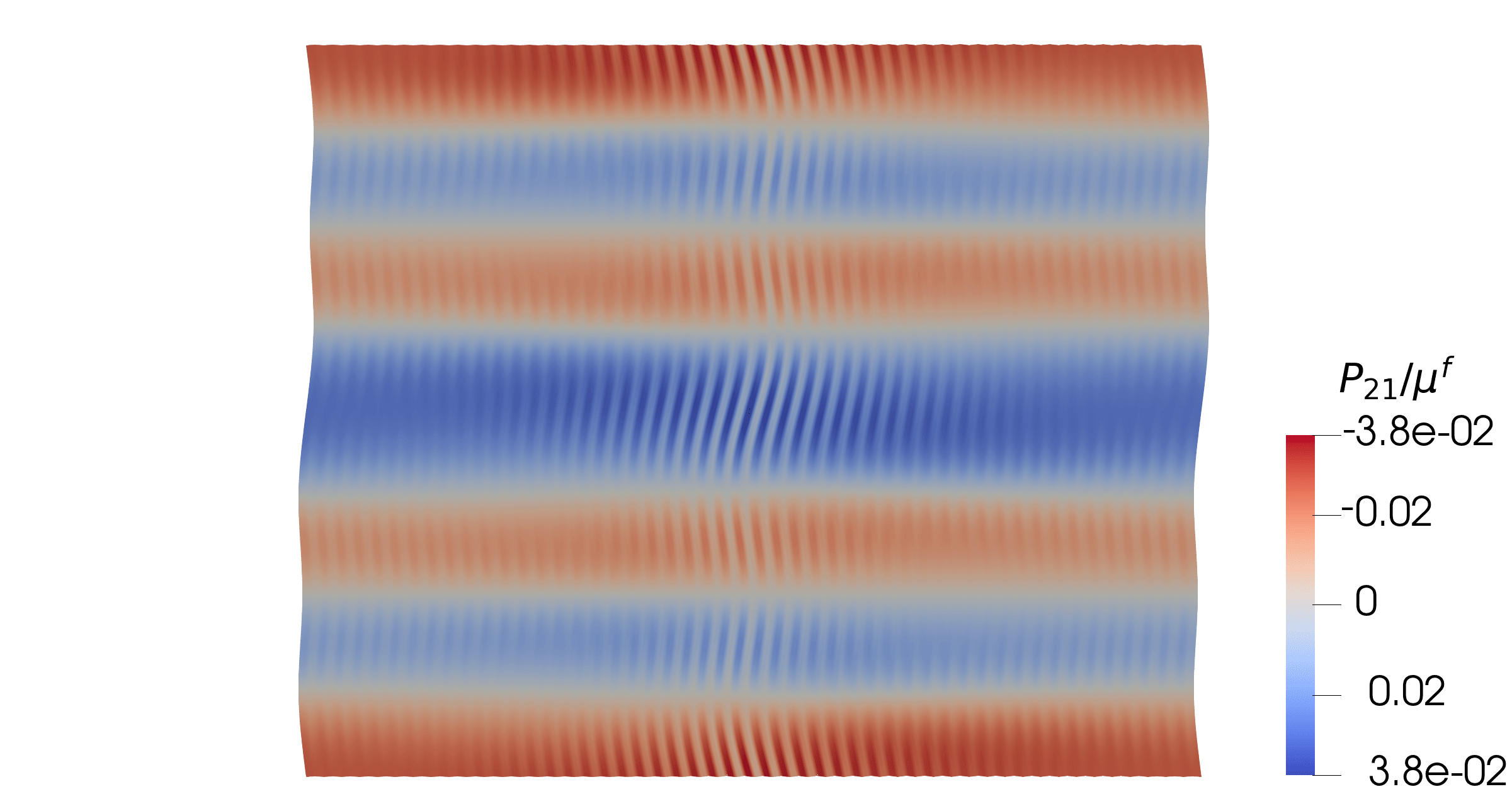}}
  
  \vspace{5mm}
  
  \subfloat[$\bar{\lambda}_2=0.7$]{\includegraphics[width=0.49\textwidth]{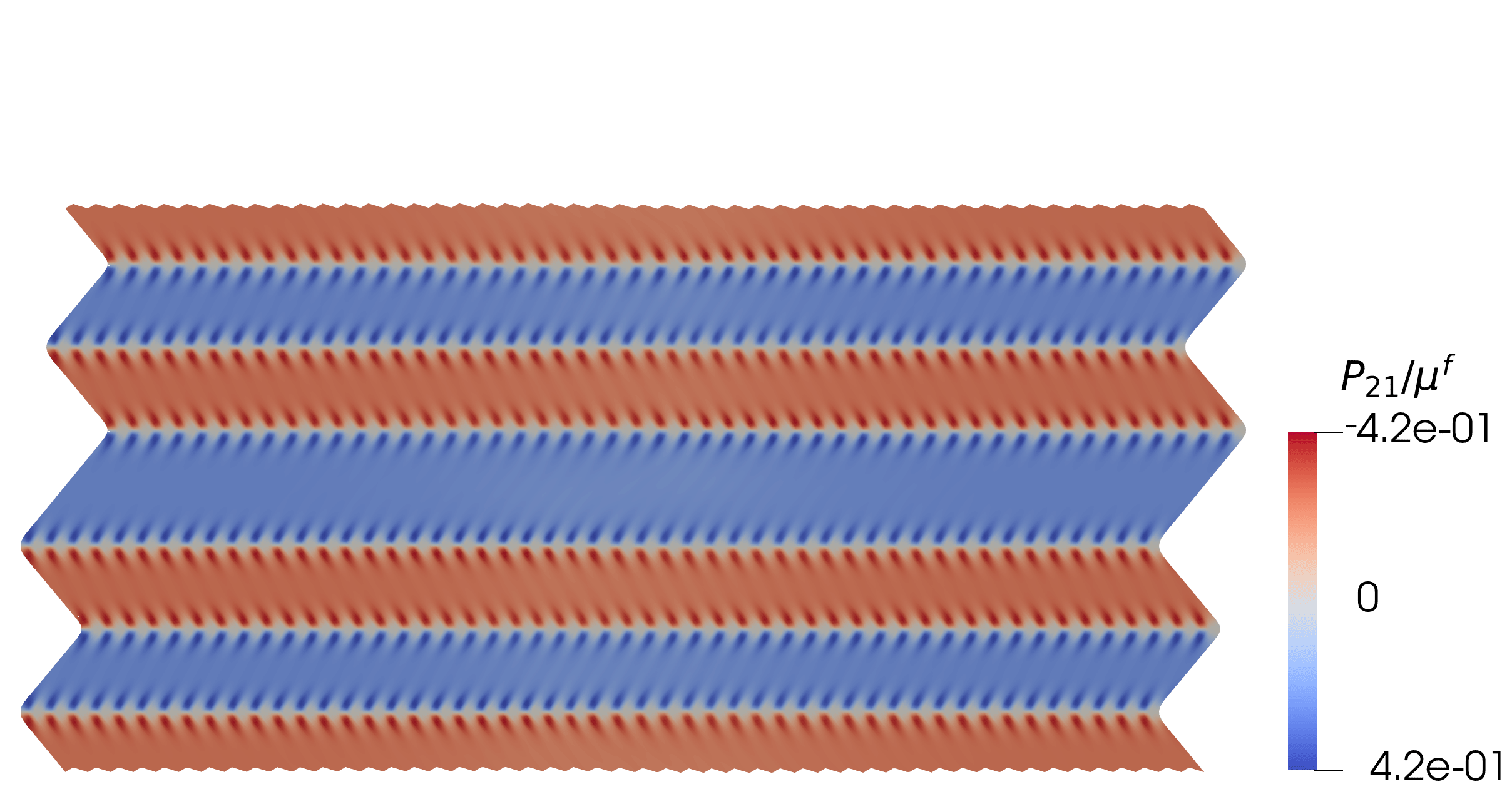}}
  \hspace{0mm}
  \caption{Evolution of the deformation for the perturbed geometry $\Delta X_1^{(1)}$, under monotonic compression, for hyper elastic phase with $\mu^{(f)}=10\mu^{(m)}$, $\kappa^{(f)}=100\mu^{(f)}$, $\kappa^{(m)}=100\mu^{(m)}$.\label{fig:VaryingStructured}}
\end{figure}

With $\Delta X_1^{(1)}$ the goal is to have the imperfection concentrate at the mid-plane of the domain and then diminish when the perpendicular distance from it increases. The term inside the square brackets adds a (symmetric) variation on the amplitude of the imperfection. For this simulation a body with $L=2$, $H_m=H_f=H=L/100=\alpha$ and 100 layers of the laminate were used, hence the total height $H_{\text{tot}}=L$. The parameter $\beta=8$ controls the decay in the $X_1$-direction, while the parameters $\hat{\alpha}=0.8*\alpha$, $n=0.5$ determine the magnitude and the frequency of the secondary variation in the $X_2$ direction, where the coordinate system of Fig. \ref{FEMunitCell} is being used, with the origin placed on the mid point of the left face. The evolution of the deformed state is shown in Fig.\ref{fig:VaryingStructured}.

\begin{figure}[h]
  \centering
    \includegraphics[width=0.48\linewidth]{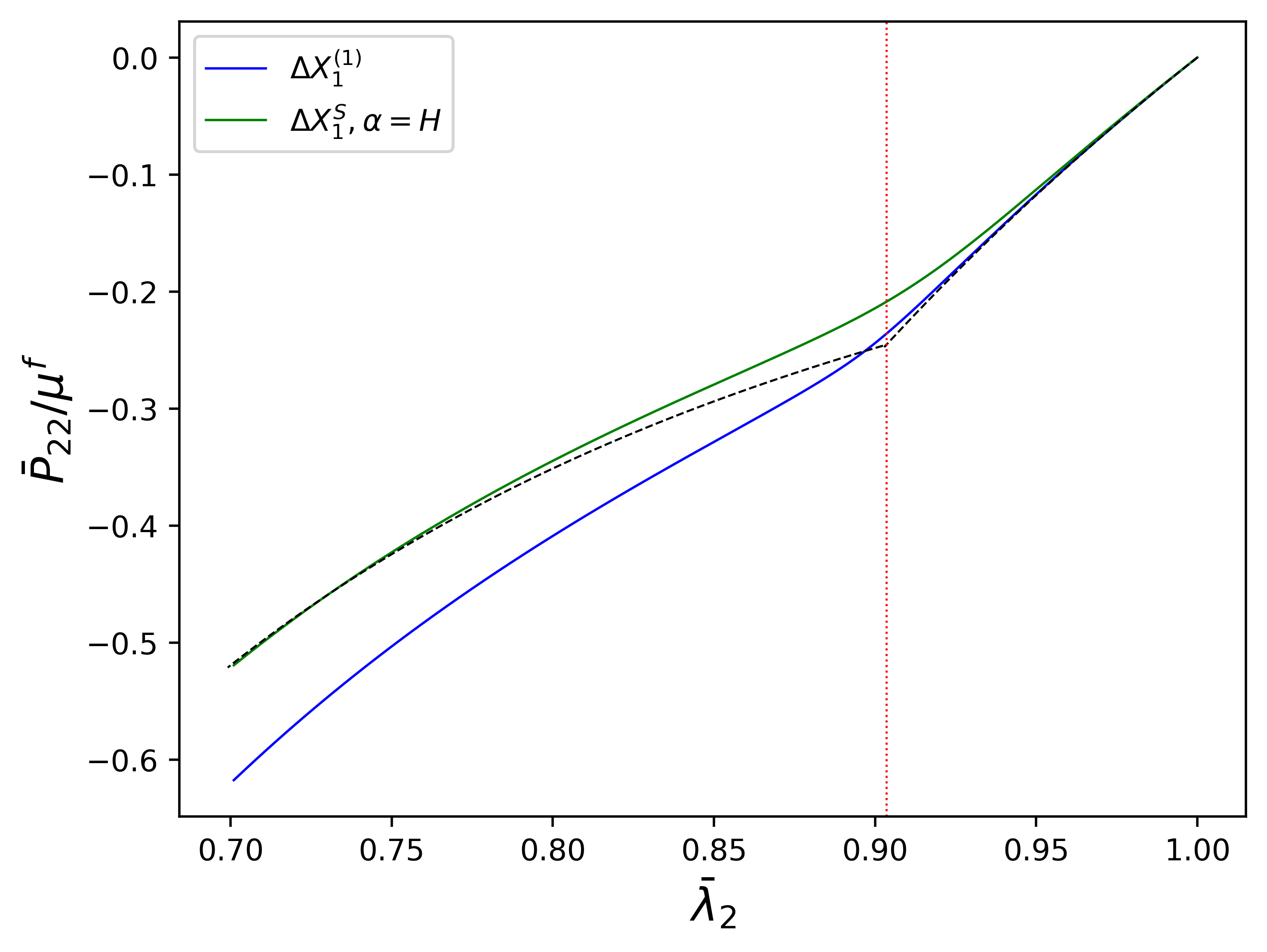}
    \caption{\label{fig:ResponceVaryingStructured} 
    Comparison between $\Delta X_1^S$ and $\Delta X_1^{(1)}$ for purely elastic laminates, subjected to monotonic compression, with elastic coefficients $\mu^{(f)}=10\mu^{(m)}$, $\kappa^{(f)}=100\mu^{(f)}$,$\kappa^{(m)}=100\mu^{(m)}$. The black lines represent the analytical solution of the unperturbed laminates. The vertical red line indicates the loss of ellipticity in the analytical solution.} 
\end{figure}

Despite the added complexity, the perturbation of the geometry is found to have a predictable influence to domain formation, as it is shown in Fig.\ref{fig:VaryingStructured}, where transition regions between the phases of the produced rank-2 laminate have formed around the  regions of high curvature in the undeformed perturbed RVE. The response of the composite is plotted in Fig. \ref{fig:ResponceVaryingStructured} and compared with the response obtained from the sinusoidal geometry with amplitude $\alpha=H$. The decay of the concentrated severity in the mid plane of $\Delta X_1^{(1)}$ gives rise to an overall response which is closer to the constant per-phase analytical solution --plotted with dashed black lines in Fig. \ref{fig:ResponceVaryingStructured}-- prior to bifurcation, compared with the sinusoidal geometry. The secondary amplitude variation introduces more wavelengths to the RVE and as established in the main part, this induces an overall stiffening to the response, confirmed once again here.

\begin{figure}[h!tbp]
  \centering
  \subfloat[$\bar{\lambda}_2=0.95$]{\includegraphics[width=0.49\textwidth]{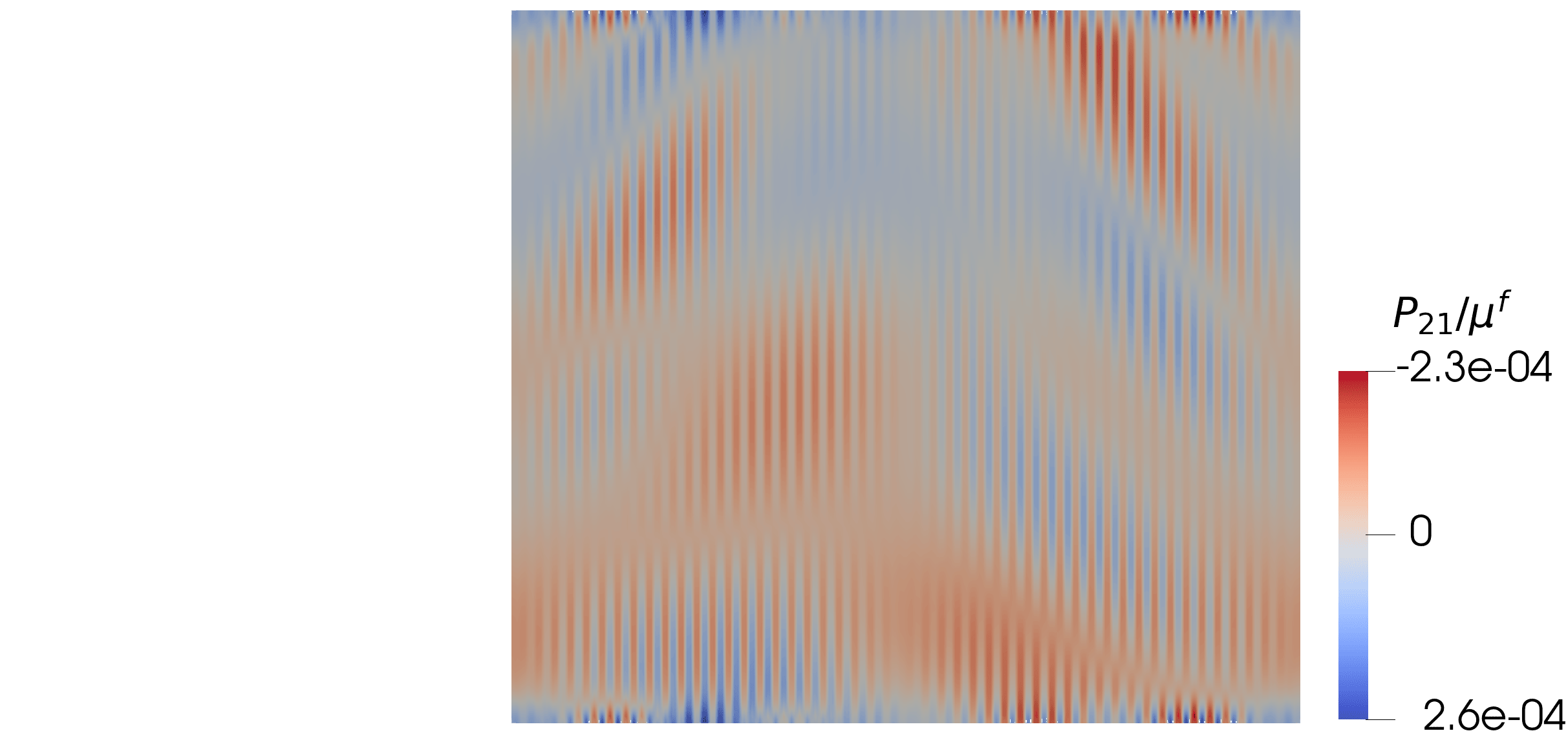}}
  \hspace{0mm}
  \subfloat[$\bar{\lambda}_2=0.903$ (LOE)]{\includegraphics[width=0.49\textwidth]{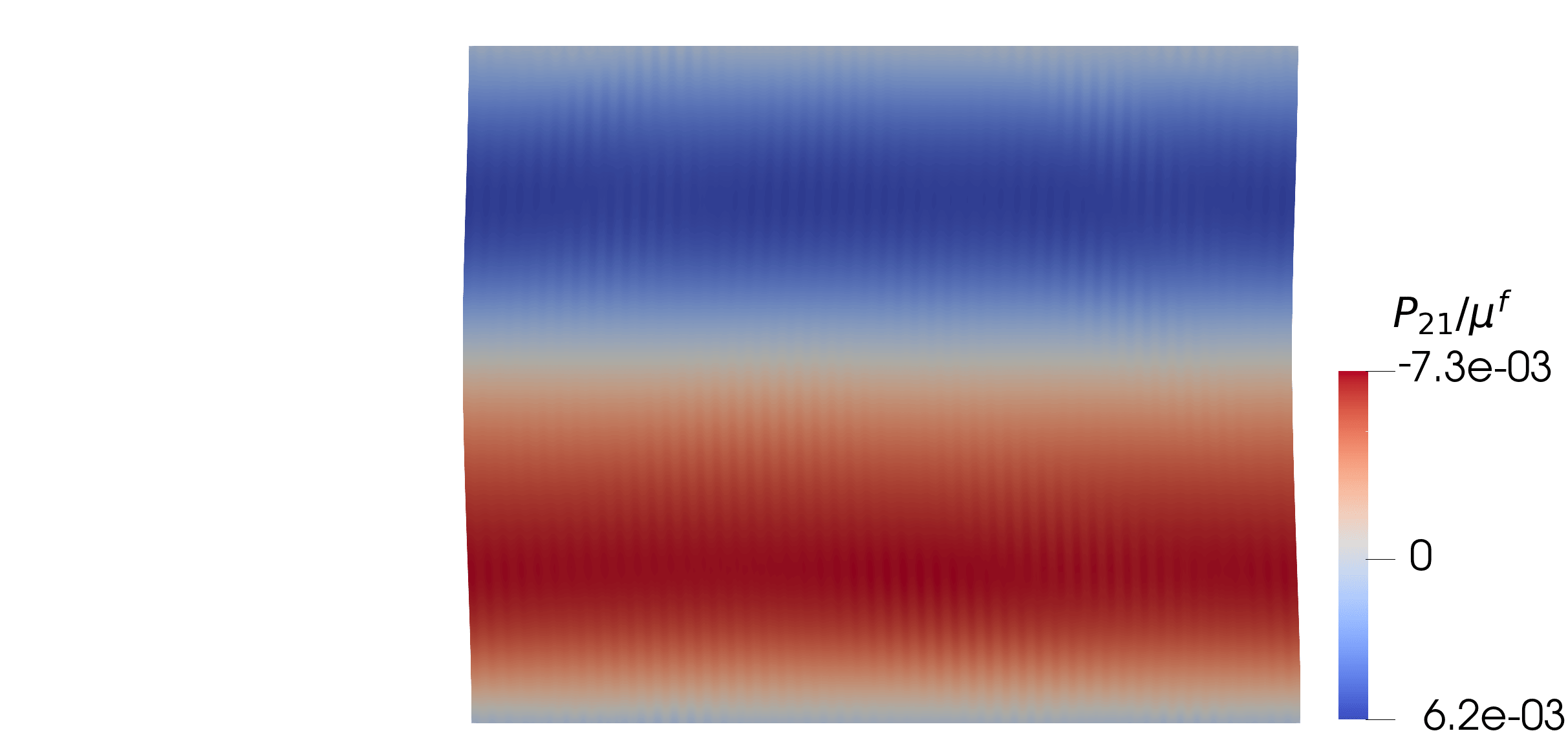}}
  
  \vspace{0mm}
  
  \subfloat[$\bar{\lambda}_2=0.7$]{\includegraphics[width=0.49\textwidth]{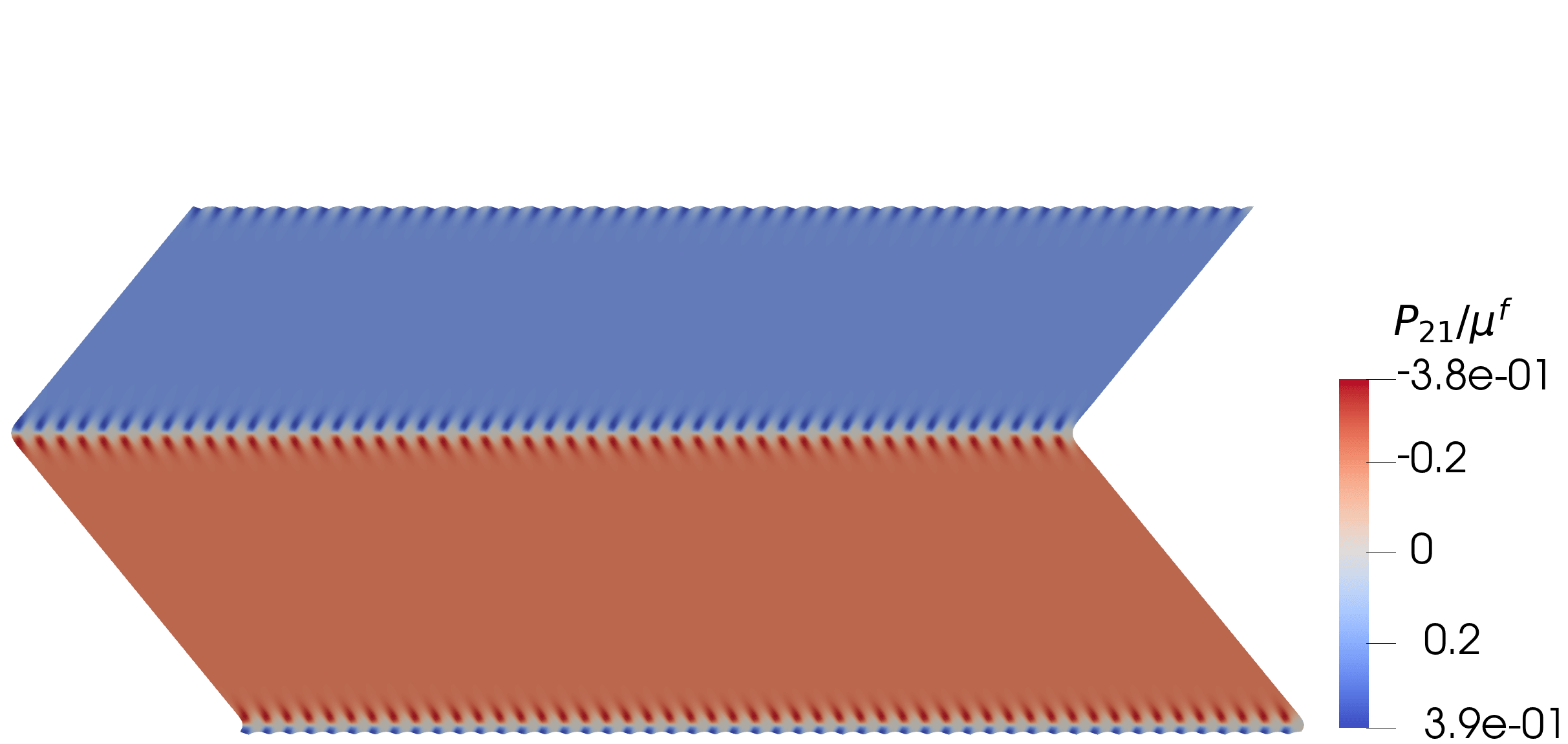}}
  \caption{Evolution of the deformation for the perturbed geometry $\Delta X_1^{(2)}$, under monotonic compression, for hyper elastic phase with $\mu^{(f)}=10\mu^{(m)}$, $\kappa^{(f)}=100\mu^{(f)}$, $\kappa^{(m)}=100\mu^{(m)}$.\label{fig:VaryingUnstructured}}
\end{figure}

 In order to perturb the geometry in a less ordered fashion, $\Delta X_1^{(2)}$ is used, which varies the amplitude of the perturbation in both directions in an unstructured way. Using the same parameters as in $\Delta X_1^{(1)}$, except now $\alpha=H/100$ and the origin of the coordinate system is place on node (4) of Fig. \ref{FEMunitCell}, the results are showcased in Fig.\ref{fig:VaryingUnstructured} for $\hat{y}=2X_2$. The conclusion is that given a (pseudo-)arbitrary wavy perturbation, as the deformation progress, one (or multiple) of the higher curvature regions will allow for a domain like structure to be developed and this will dominate the behavior of the RVE.

\begin{figure}[h]
  \centering
    \includegraphics[width=0.45\linewidth]{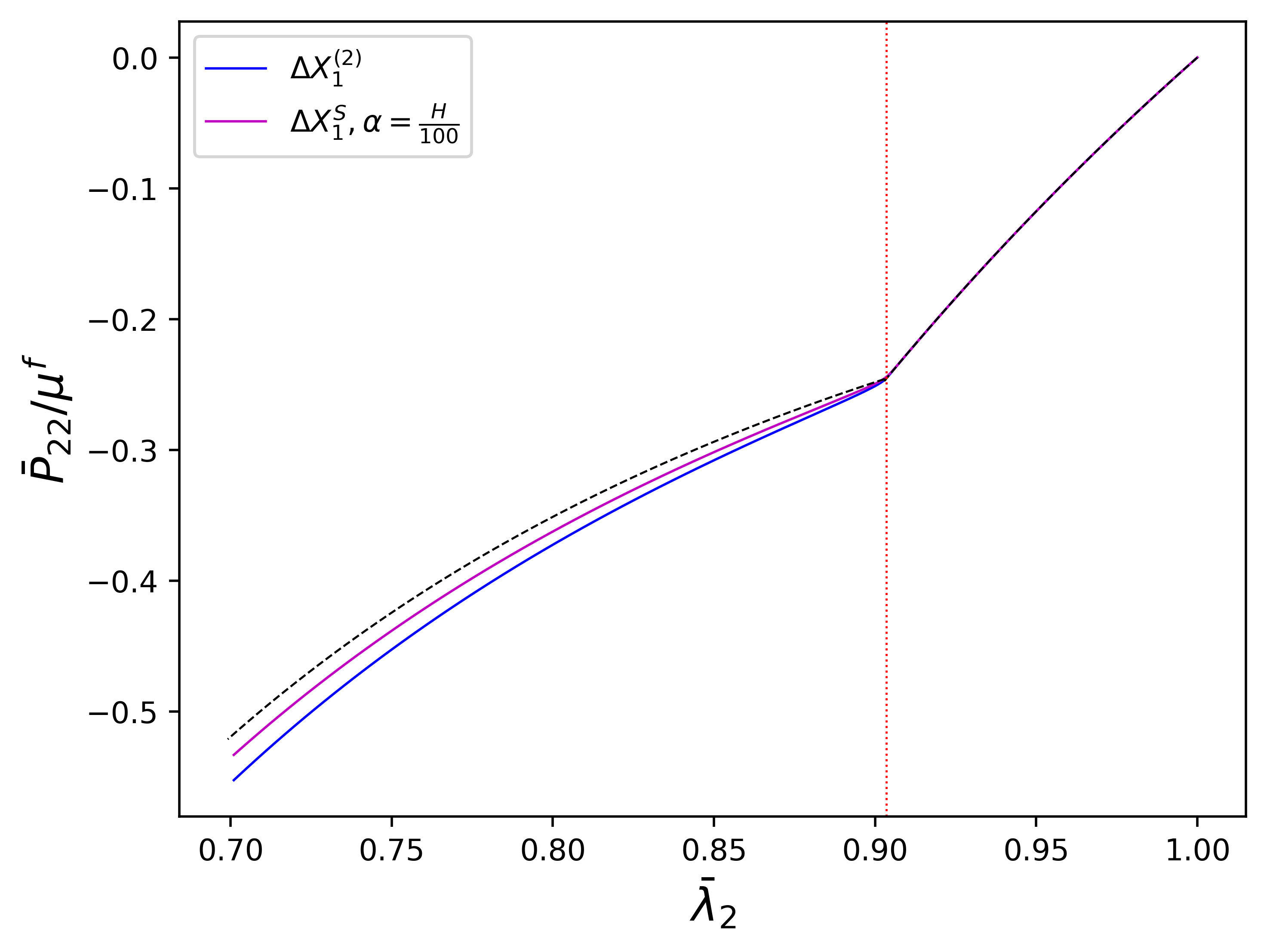}
    \caption{ 
    Comparison between $\Delta X_1^S$ and $\Delta X_1^{(2)}$ for purely elastic laminates, subjected to monotonic compression, with elastic coefficients $\mu^{(f)}=10\mu^{(m)}$, $\kappa^{(f)}=100\mu^{(f)}$,$\kappa^{(m)}=100\mu^{(m)}$. The black lines represent the analytical solution of the unperturbed laminates. The vertical red line indicates the loss of ellipticity in the analytical solution.\label{Response_VaryingUnstructured}} 
\end{figure}

The response of the composite is compared with the sinusoidal imperfection ($\alpha=H/100$) and the unperturbed geometry in Fig. \ref{Response_VaryingUnstructured}. 

\subsection*{Elastic Parameters}
\begin{figure}[h!tbp]
  \centering
  \subfloat[Varying $\mu^{(f)}/\mu^{(m)}$ under constant ratio $\kappa^{(r)}/\mu^{(r)}$]{\includegraphics[width=0.40\textwidth]{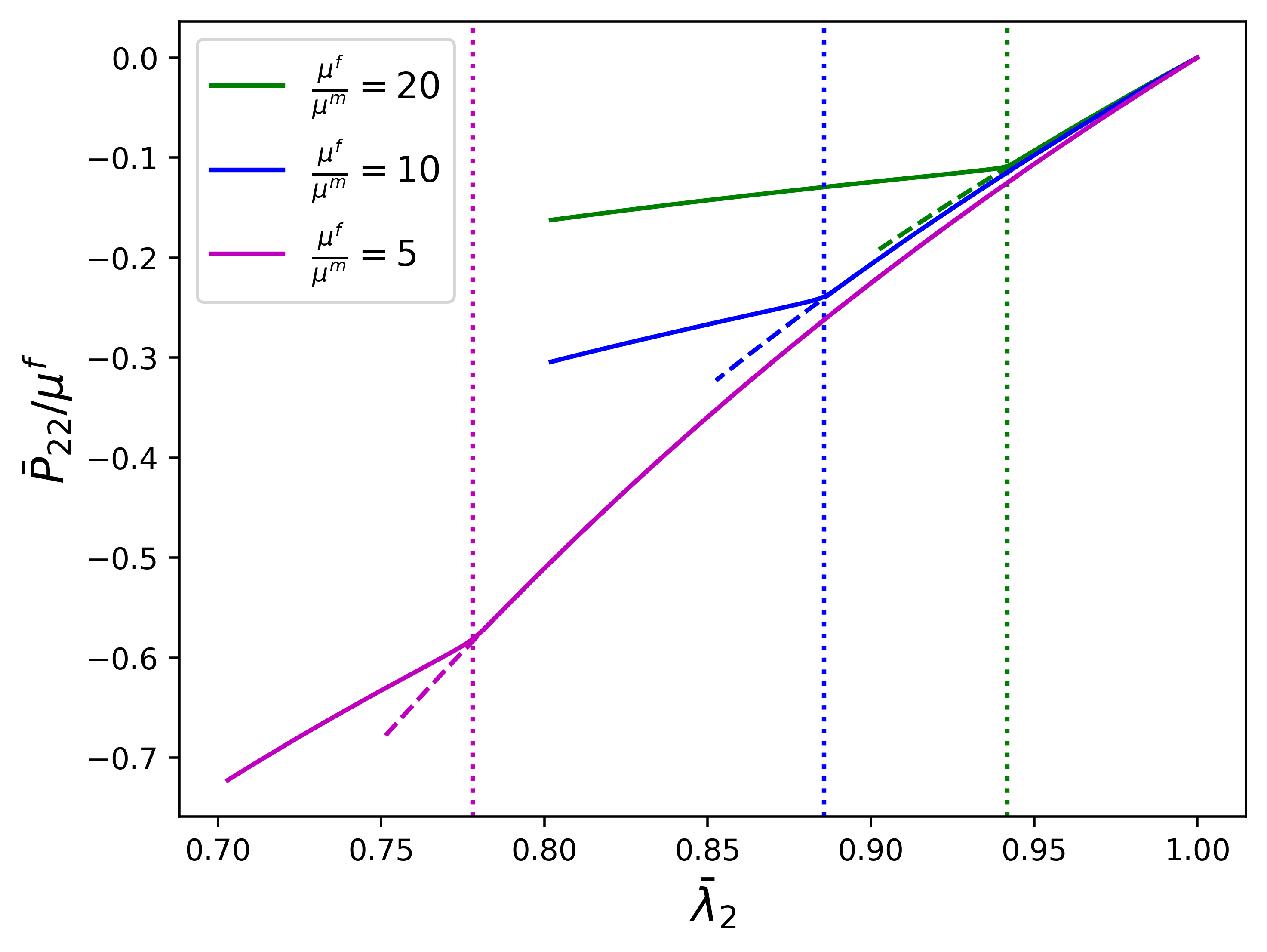}\label{fig:ShearMod}}
  \hspace{3mm}
  \subfloat[Varying ratio $\kappa^{(r)}/\mu^{(r)}$ and constant $\mu^{(f)}/\mu^{(m)}$]{\includegraphics[width=0.40\textwidth]{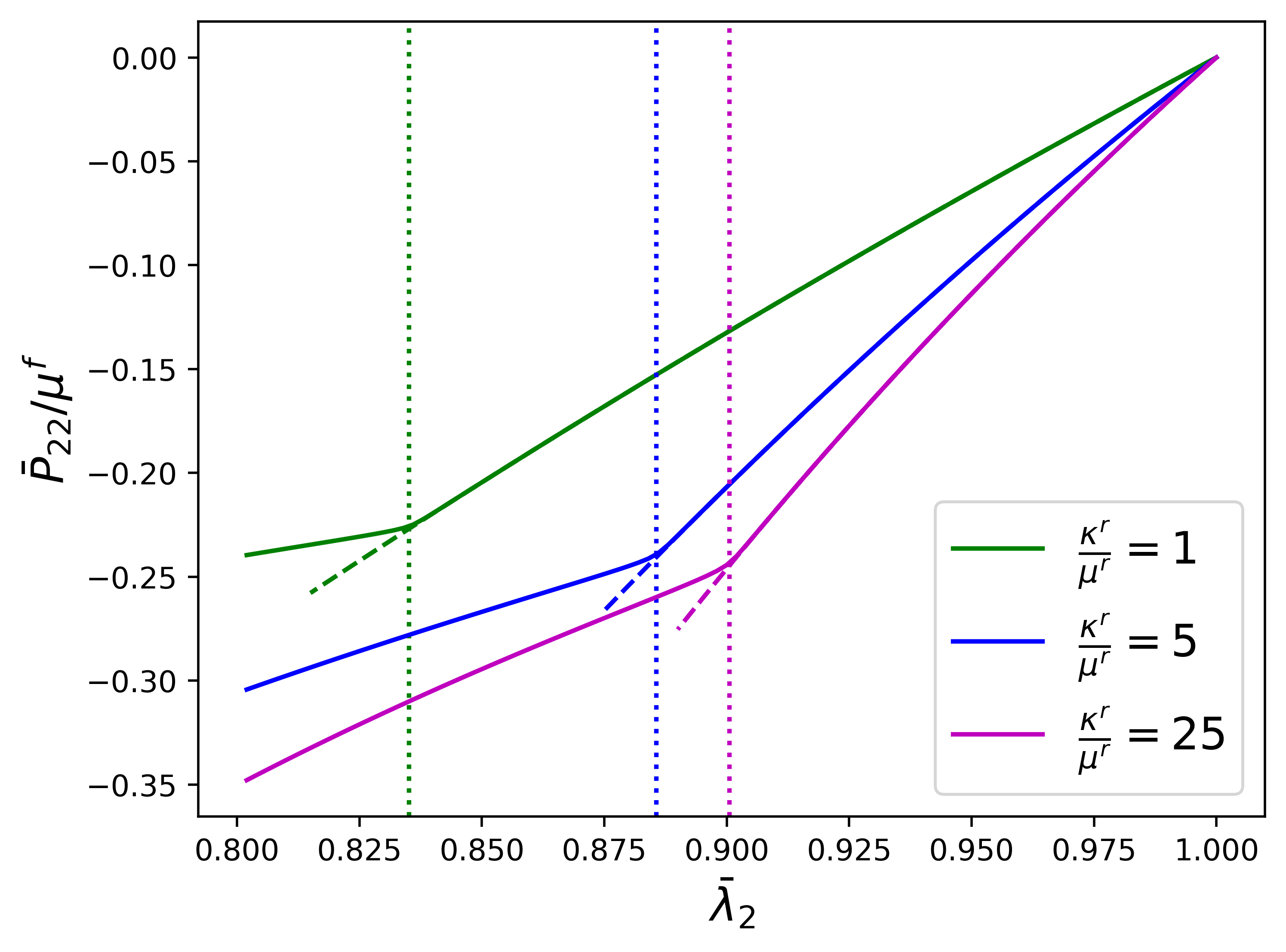}\label{fig:BulkModRatio}}
  \caption{Response of the composite, under monotonic compression, for varying elastic coefficients. The dashed lines represent the analytical principal solution, while the dotted vertical lines indicate the LOE for the respective elastic parameters .\label{fig:ElasticCoef}}
\end{figure}

A parametric study on the effect of elastic coefficients, on hyper elastic laminates, is performed. The sinusoidal perturbation $\Delta X_1^S$ with $\alpha=H/100$ and $w=1$ is used. By first varying the contrast between the shear moduli of the fiber and matrix phase and keeping the ratio $\kappa^{(r)}/\mu^{(r)}$ constant, the results in Fig.\ref{fig:ShearMod} are obtained. On the other hand, when keeping the stiffness contrast between the phase constant and varying $\kappa^{(r)}/\mu^{(r)}$ the plot on Fig. \ref{fig:BulkModRatio} is obtained. This is complementary to the results presented in Part-I. As is expected from that study and the accompanying analytical results for LOE, the higher the heterogeneity between the phases, the lesser the compressive strain required to trigger LOE, which has been found to be an accurate predictor of domain formation for low amplitude perturbations. Additionally, by interpreting approaching incompressibility as the increasing energetic penalization of volumetric deformations, the higher the penalty parameter $\kappa^{r}$ is, the earlier the model leads to domain formation; a deformation mode that favors rotations and shear over compression and increased volumetric changes. In the results of Fig. \ref{fig:ElasticCoef} only the principal path of the analytical solution is included to highlight the softening caused by domain formation.

\subsection*{Bending stresses}
\begin{figure}[h!tbp]
  \centering
  \subfloat[$\bar{\lambda}_2=0.95$]{\includegraphics[width=0.40\textwidth]{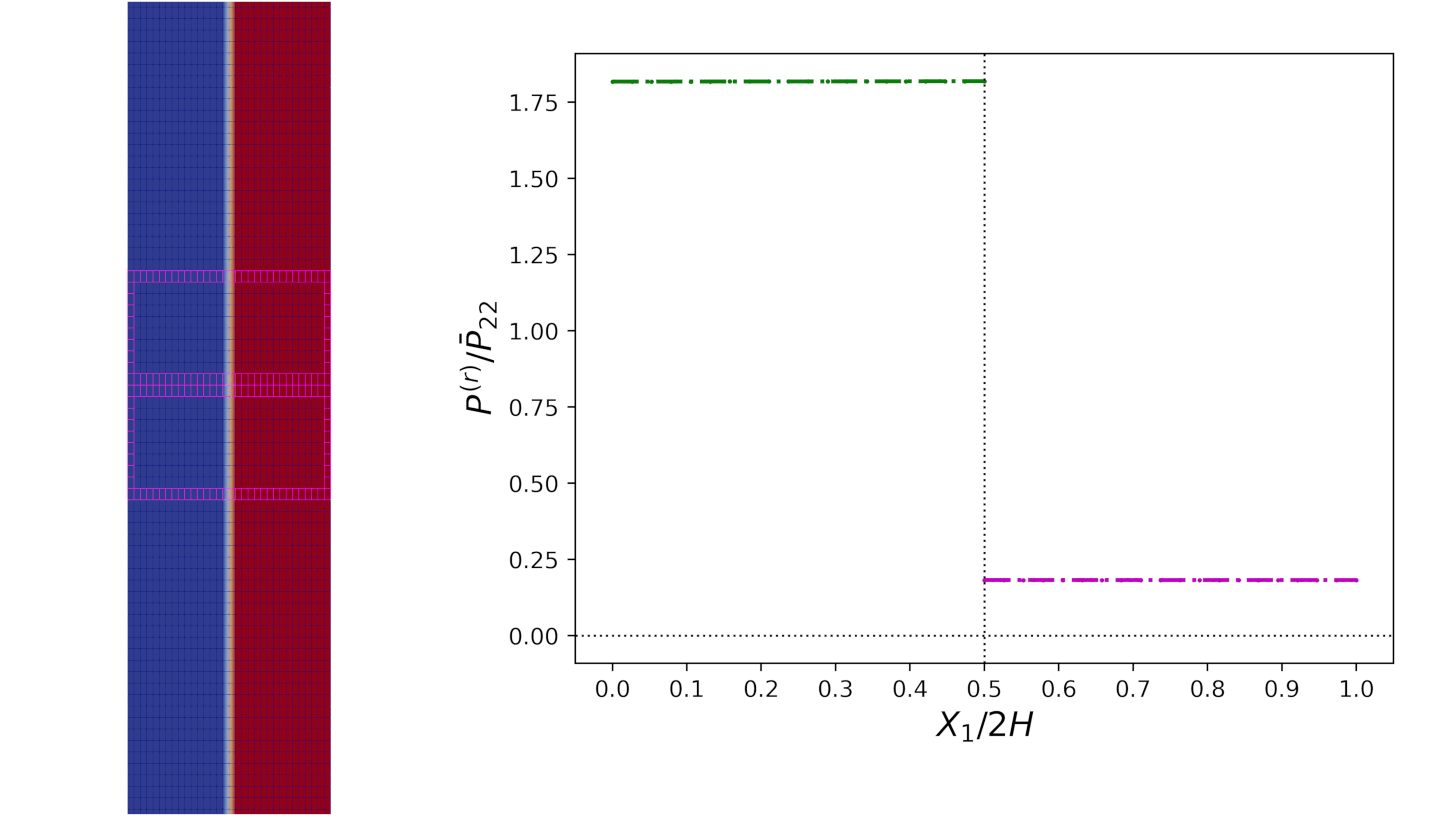}\label{fig:Bstr1}}

 \vspace{3mm}
  
  \subfloat[$\bar{\lambda}_2=0.75$]{\includegraphics[width=0.40\textwidth]{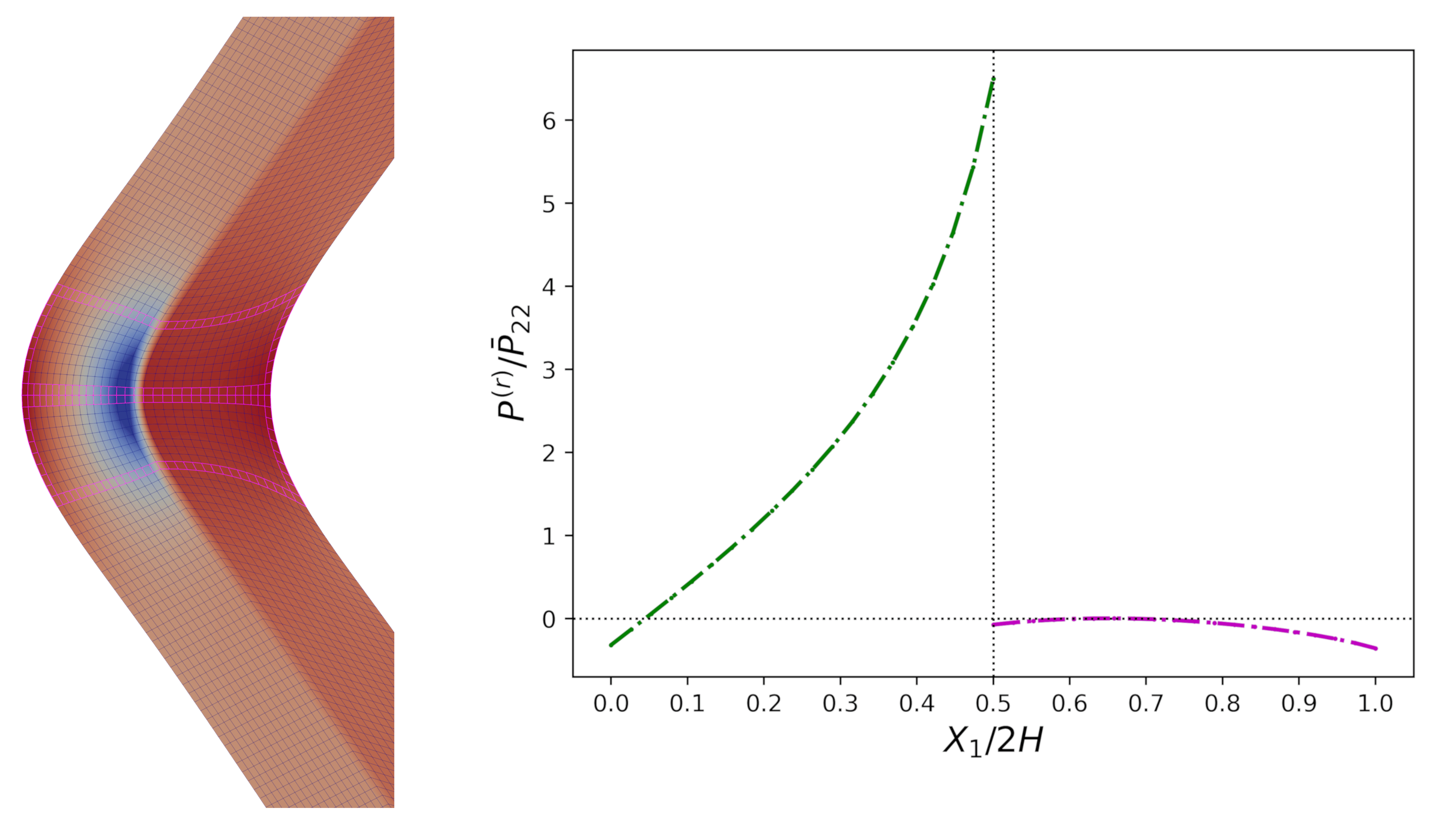}\label{fig:Bstr2}}

   \vspace{3mm}

  \subfloat[$\bar{\lambda}_2=0.55$]{\includegraphics[width=0.40\textwidth]{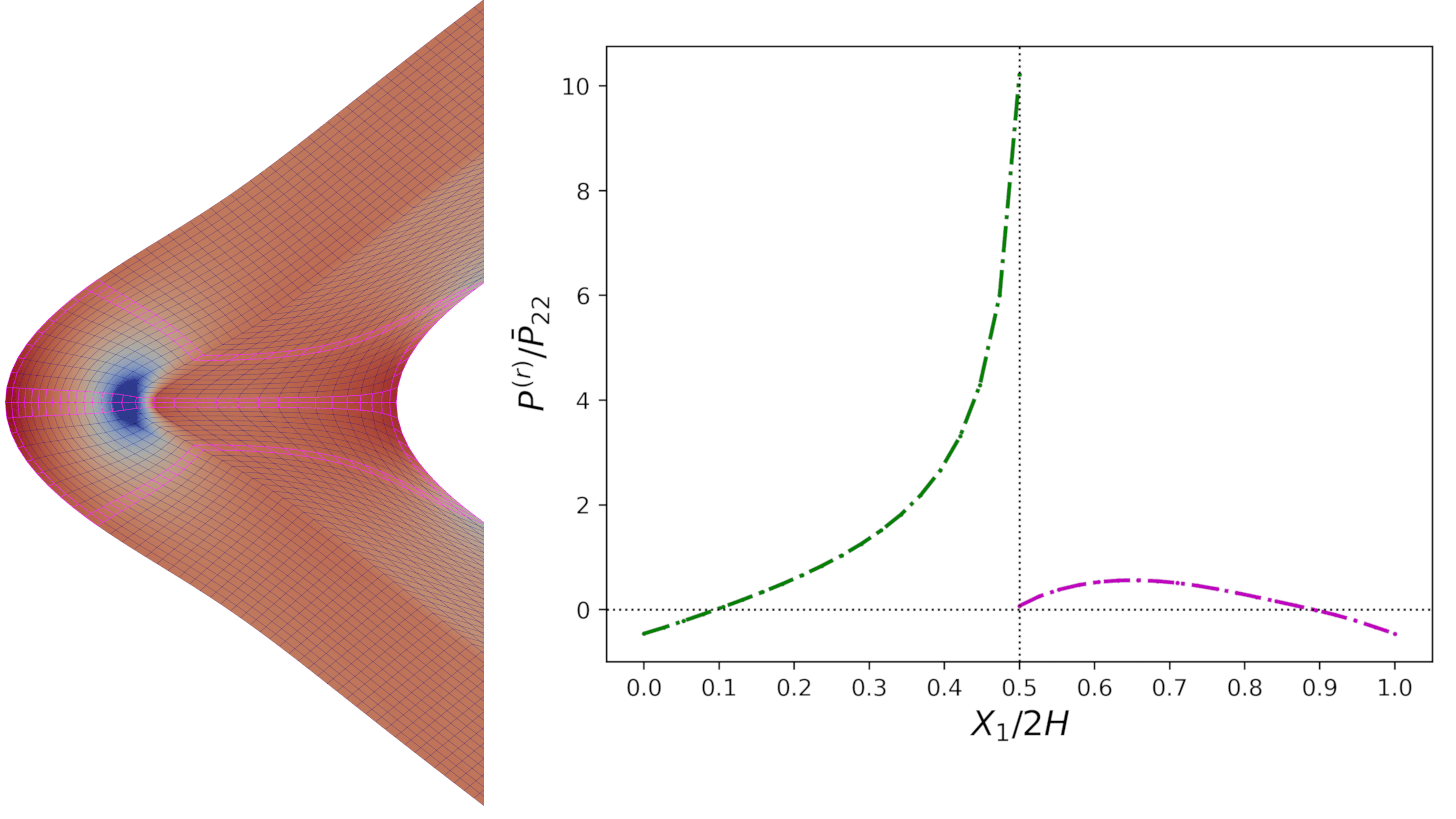}\label{fig:Bstr4}}
  
  \caption{Contour of $P_{22}$ and stress distribution across the thickness of the composite, during monotonic compression for the purely elastic phases of the baseline paradigm with with $\mu^{(f)}/\mu^{(m)}=10$, $\kappa^{(f)}=100\mu^{(f)}$,$\kappa^{(m)}=100\mu^{(m)}$.}\label{fig:BendingStresses}
\end{figure}

Domain formation, as extensively discussed in this manuscript accommodates the formation of phase boundaries for the rank-2 laminate that exhibit highly localized deformation. At these transition regions significant bending of the layers is observed. This occurs both for the purely elastic composite and for the case when elastoplastic fibers are considered. To quantitatively access the bending stresses that arise in these regions we focus on the path that goes through the peak of the sinusoidal perturbation in the undeformed state, around which the domains will form. The stress is non-dimensionalized with respect to the average compressive macroscopic stress, and its distribution along the thickness of the matrix and fiber phases is plotted in Fig. \ref{fig:BendingStresses} where the coordinate system of Fig. \ref{FEMunitCell} placed on node (4) is used. As expected, the initial distribution is uniform everywhere in each phase. After domains have formed, the local response is dominated by bending causing a significant increase on stress on the interface between the two layers. It is noted that the intuition from bending kinematics, which require a higher compressive stress at the lower parts of each phase, is not quite satisfied by the produced stress field on the matrix phase. It is hypothesized that, as the macroscopic compression develops, in order to merge the domains a tensile stress develops on the matrix phase.

\subsection*{Tension to compression: Hardening modulus study}

\begin{figure}[h!tbp]
  \centering
  \subfloat[Composite]{\includegraphics[width=0.46\textwidth]{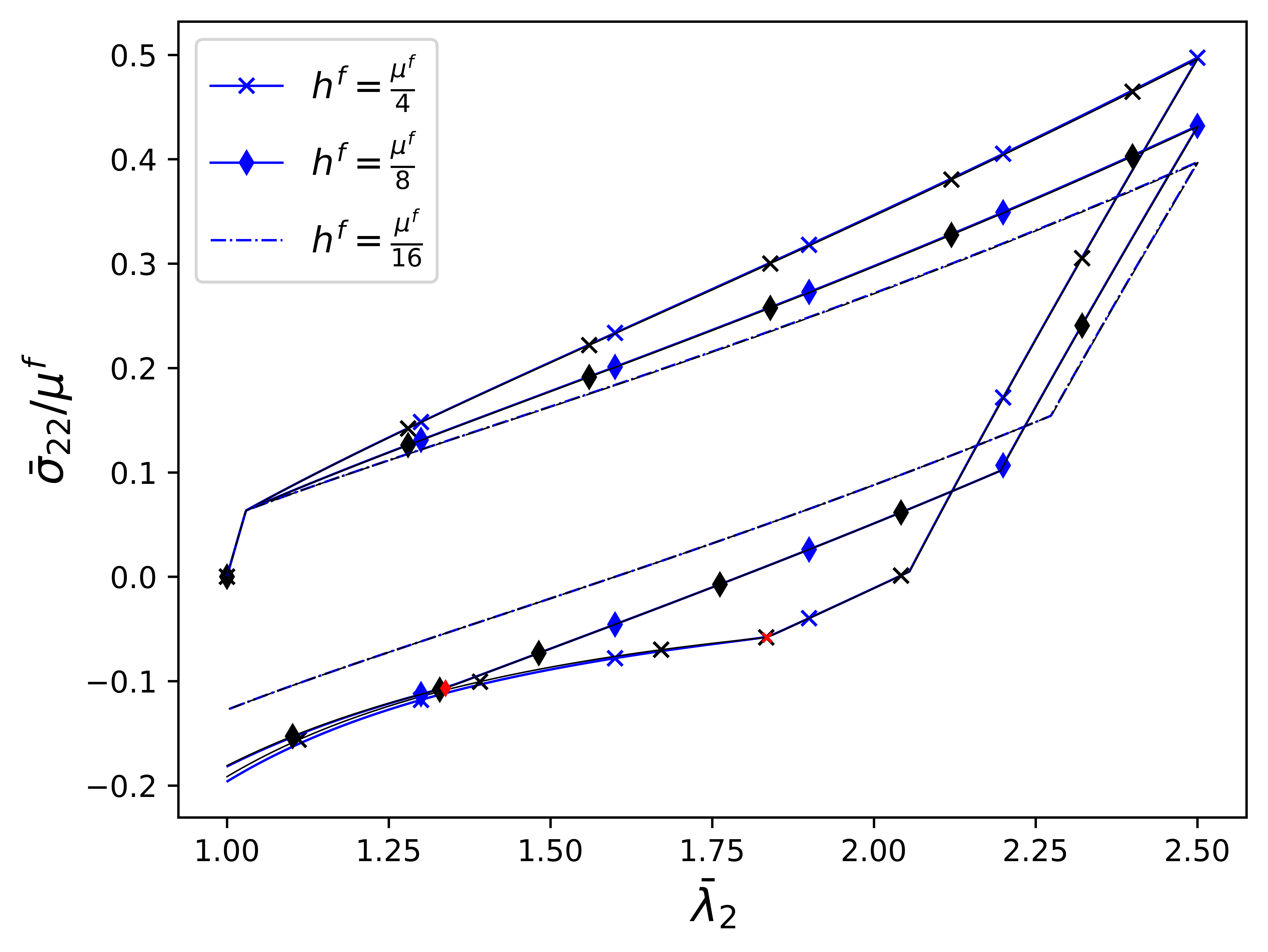}\label{fig:OUF1}}
  \hspace{0mm}
  \subfloat[Fiber phase]{\includegraphics[width=0.46\textwidth]{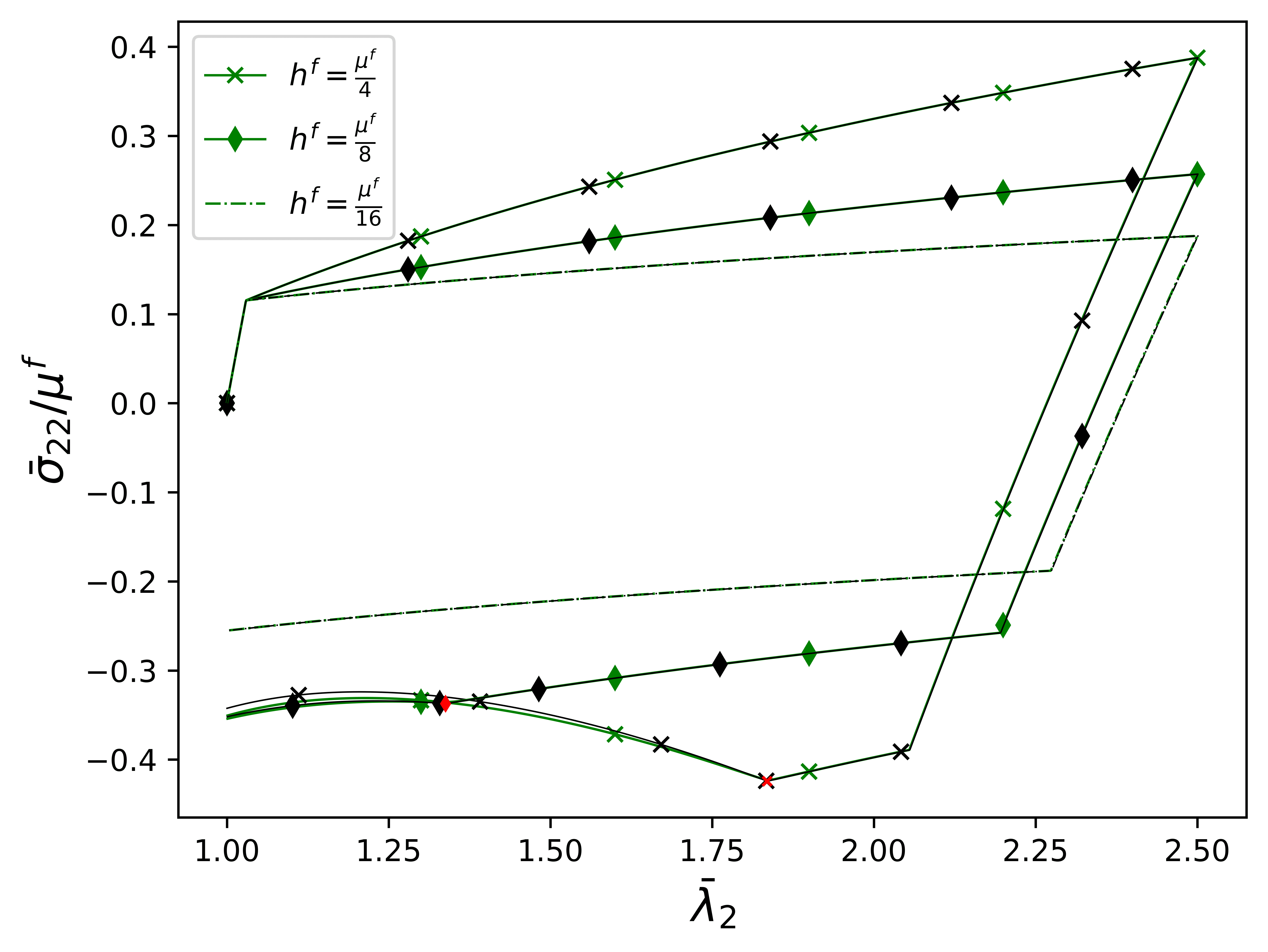}\label{fig:OUF2}}
  
  \vspace{4mm}
  
  \subfloat[Accumulated plastic strain]{\includegraphics[width=0.46\textwidth]{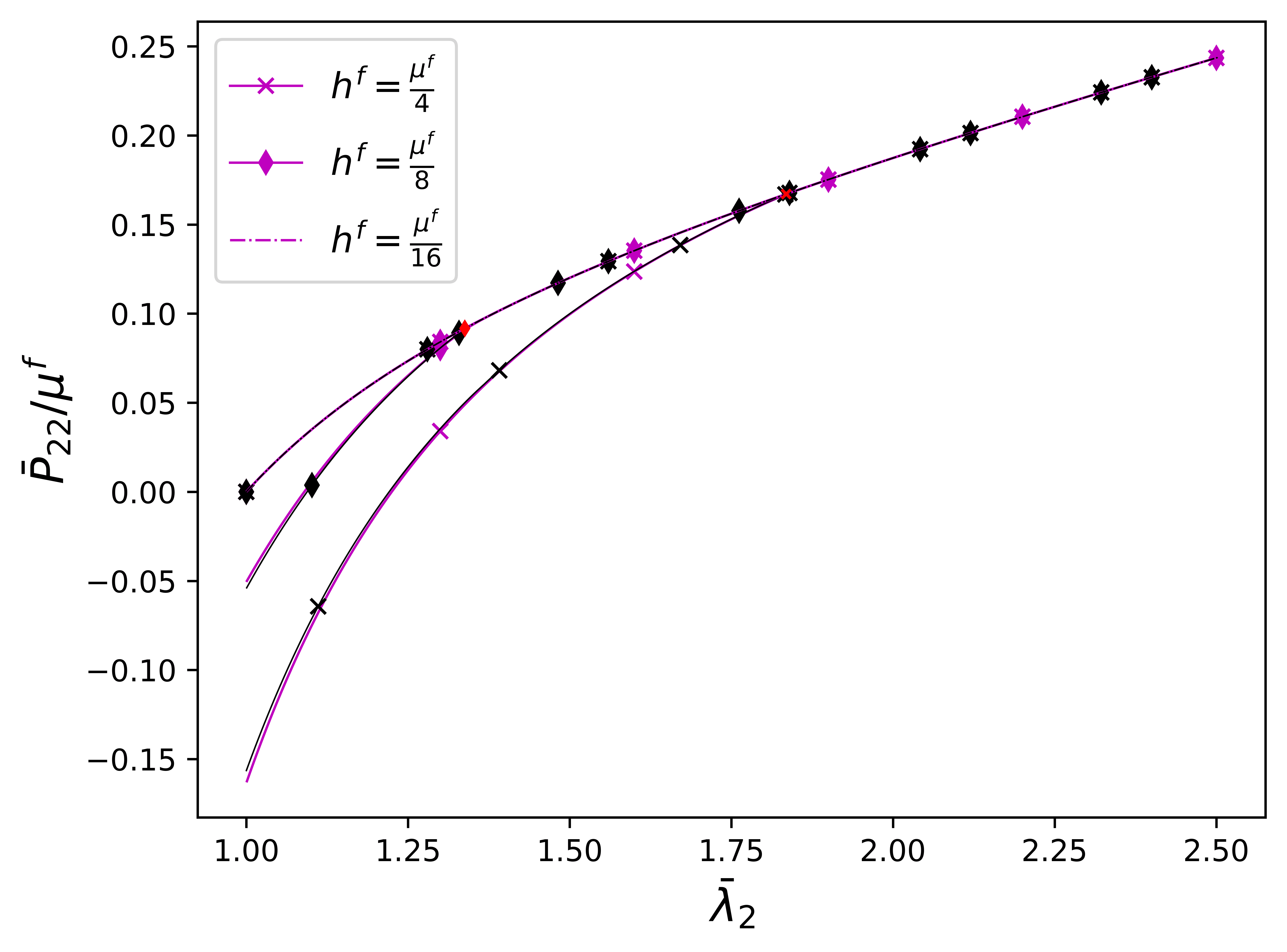}}
  \hspace{0mm}
  \subfloat[Accumulated plastic strain]{\includegraphics[width=0.46\textwidth]{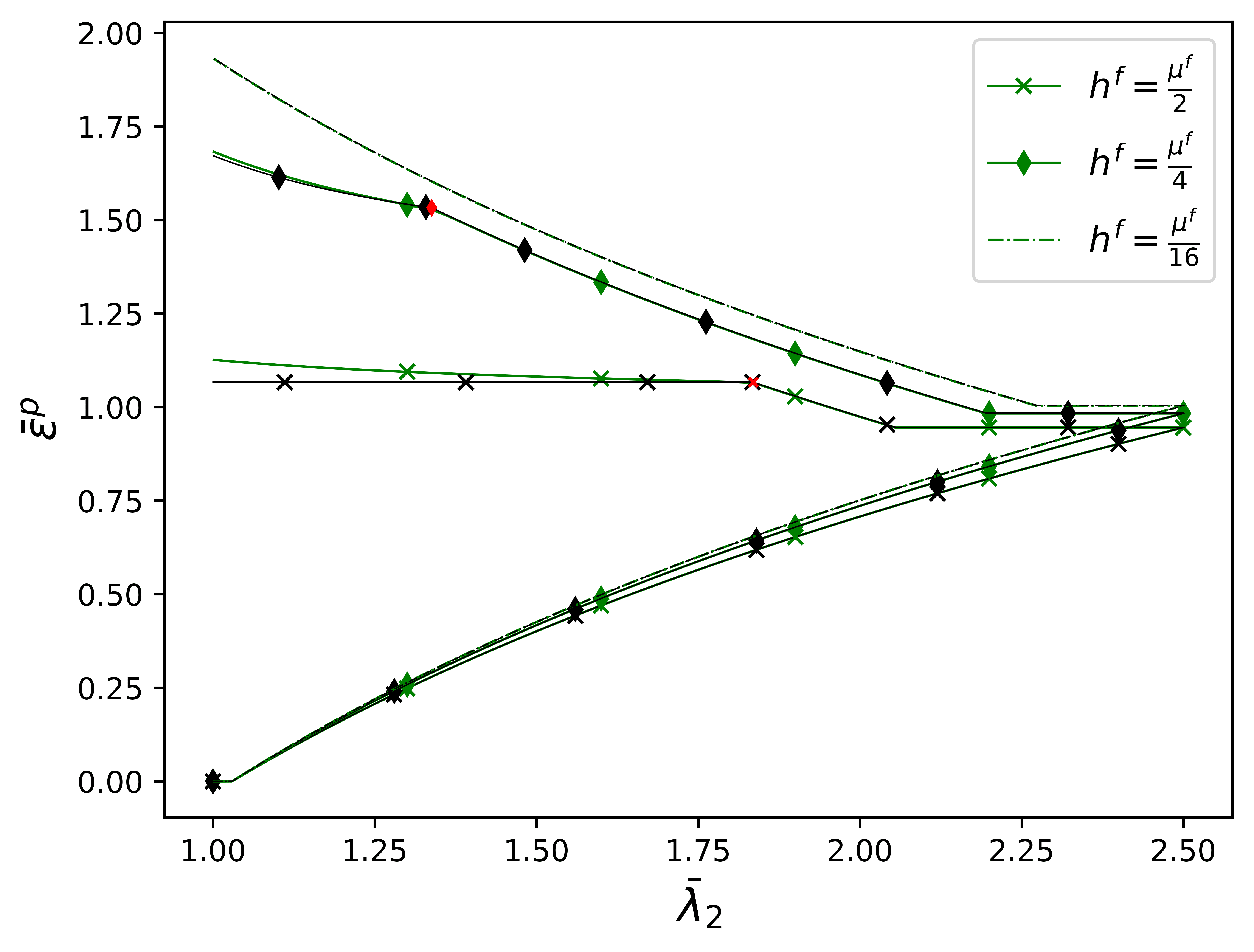}\label{fig:OUF4}}
  \caption{Results for laminates with different hardening, the same initial yield limit $\Sigma_y^{(f)}=\mu^{(f)}/5$ and the same elastic coefficients $\mu^{(f)}=10\mu^{(m)}$, $\kappa^{(f)}=100\mu^{(f)}$, $\kappa^{(m)}=100\mu^{(m)}$. With discontinuous lines are the cases where no domains were observed, while the black lines represent the analytical solution of the perfect laminates. The red dot indicates the loss of ellipticity in the analytical solution. }\label{fig:NonMonH}
\end{figure}

Complementing the results in Fig. \ref{fig:NonMonotLinePlts}, withing the same loading program --consisting of an aligned tensile step until $\lambda^{\text{max}}_2 = 2.5$ followed by elastic unloading and subsequent elastoplastic compression-- the effect of the hardening modulus is examine, under a constant yield limit $\Sigma_y^{(f)}=\mu^{(f)}/10$. The results are presented in Fig. \ref{fig:NonMonH}, from which it is concluded that when plasticity dominates elastic phenomena, as in the case with $h^{(f)}=\mu^{(f)}/16$, no domains emerge.

\subsection*{Tension to compression: Increased tensile stretch under decreasing imperfection amplitude}
\begin{figure}[ht!bp]
  \centering
    \includegraphics[width=0.6\linewidth]{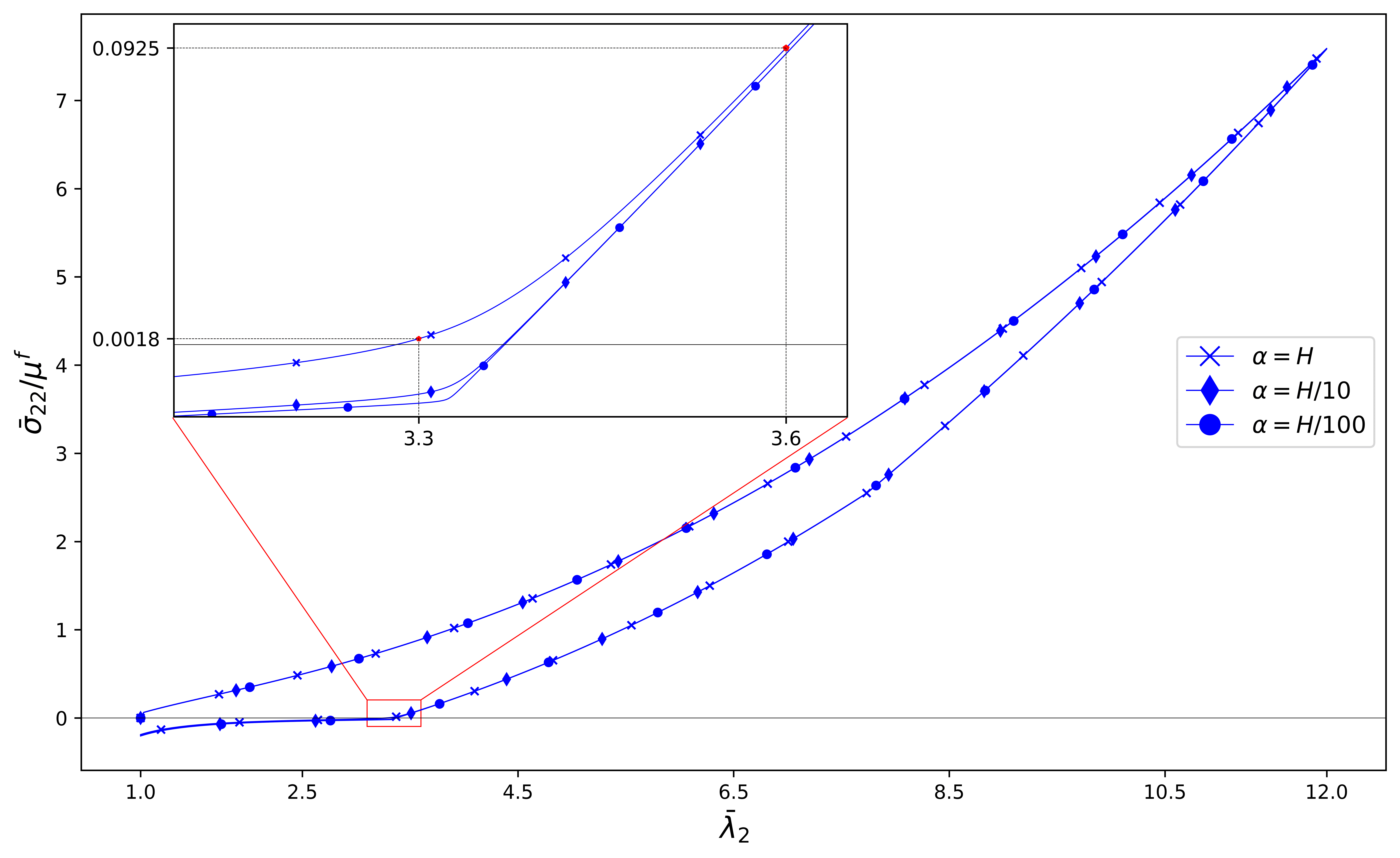}
    \caption{\label{fig:VaringLamdaAlot} Results for laminates with the same plastic parameters $h^{(f)}=\mu^{(f)}/4$, $\Sigma_y^{(f)}=\mu^{(f)}/10$, the same elastic coefficients $\mu^{(f)}=10\mu^{(m)}$, $\kappa^{(f)}=100\mu^{(f)}$, $\kappa^{(m)}=100\mu^{(m)}$ and different amplitudes of the sinusoidal geometric perturbation.} 
\end{figure}

For this study we return to the lower amplitude sinusoidal imperfections and explore, again, the possibility of domain formation under tensile stresses. The results of Fig. \ref{fig:DomainsInTensionALl} made clear that its more unlikely for domains to emerge, under tensile stresses, the less imperfect the body is. In that direction, following the discussion of section \ref{TenToComp}, the maximum applied stretch is increase to $\lambda_2^{\text{max}}=12$ and the results are shown in Fig. \ref{fig:VaringLamdaAlot}.
Now that the perturbations in the undeformed geometry are relatively small, the responses during the elastoplastic tensile loading and elastic unloading parts, of the prescribed loading program, are indistinguishable. There is also great similarity between the numerical solutions for the better part of the reversed elastoplastic loading, until the bodies reaches near the stress free state, for which a close up window is included in Fig. \ref{fig:VaringLamdaAlot}.

\begin{figure}[h!tbp]
  \centering
  \subfloat[$\bar{\lambda}_2=3.6$]{\includegraphics[width=0.44\textwidth]{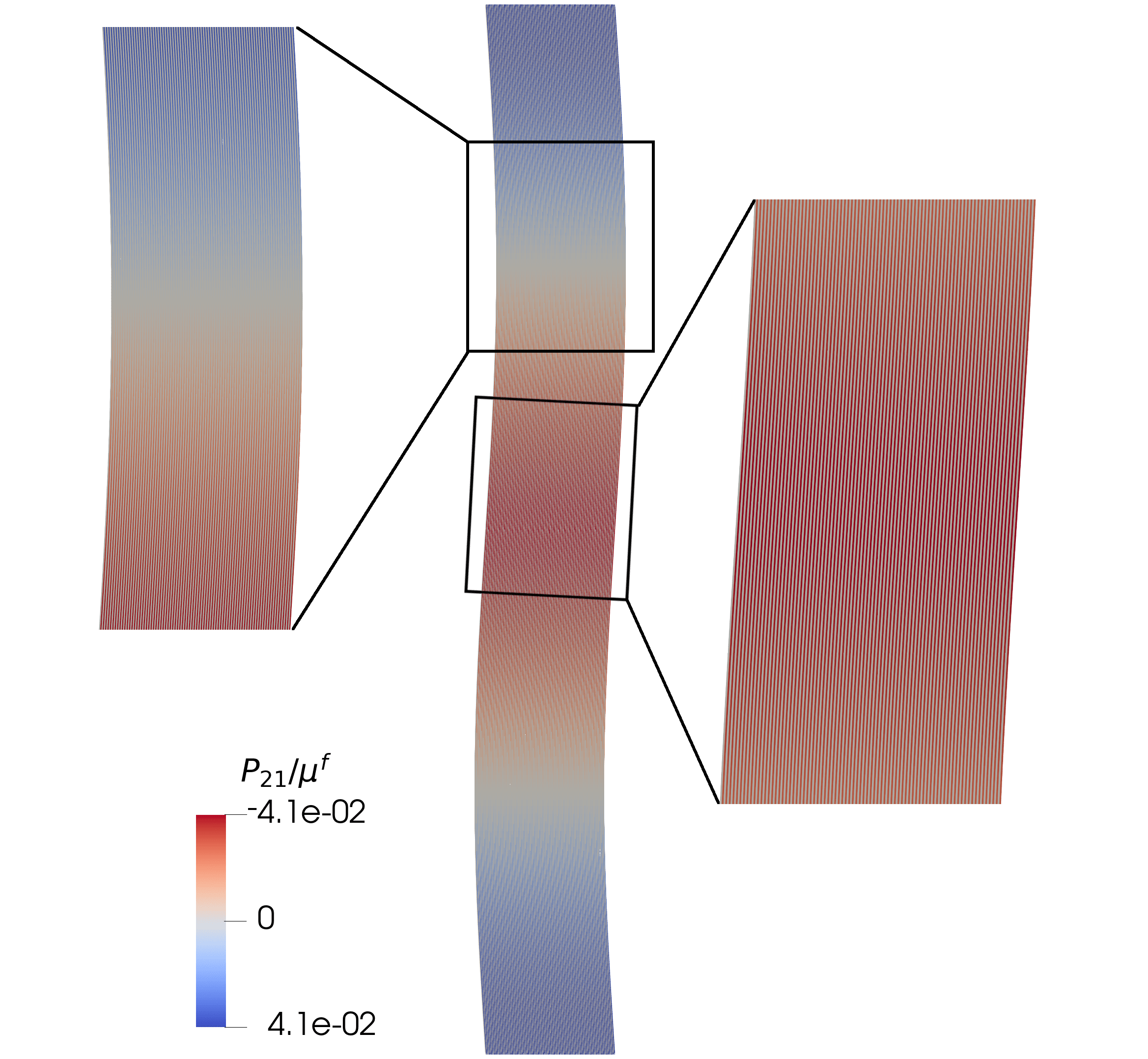}\label{}}
  \hspace{5mm}
   \subfloat[$\bar{\lambda}_2=3.4$]{\includegraphics[width=0.44\textwidth]{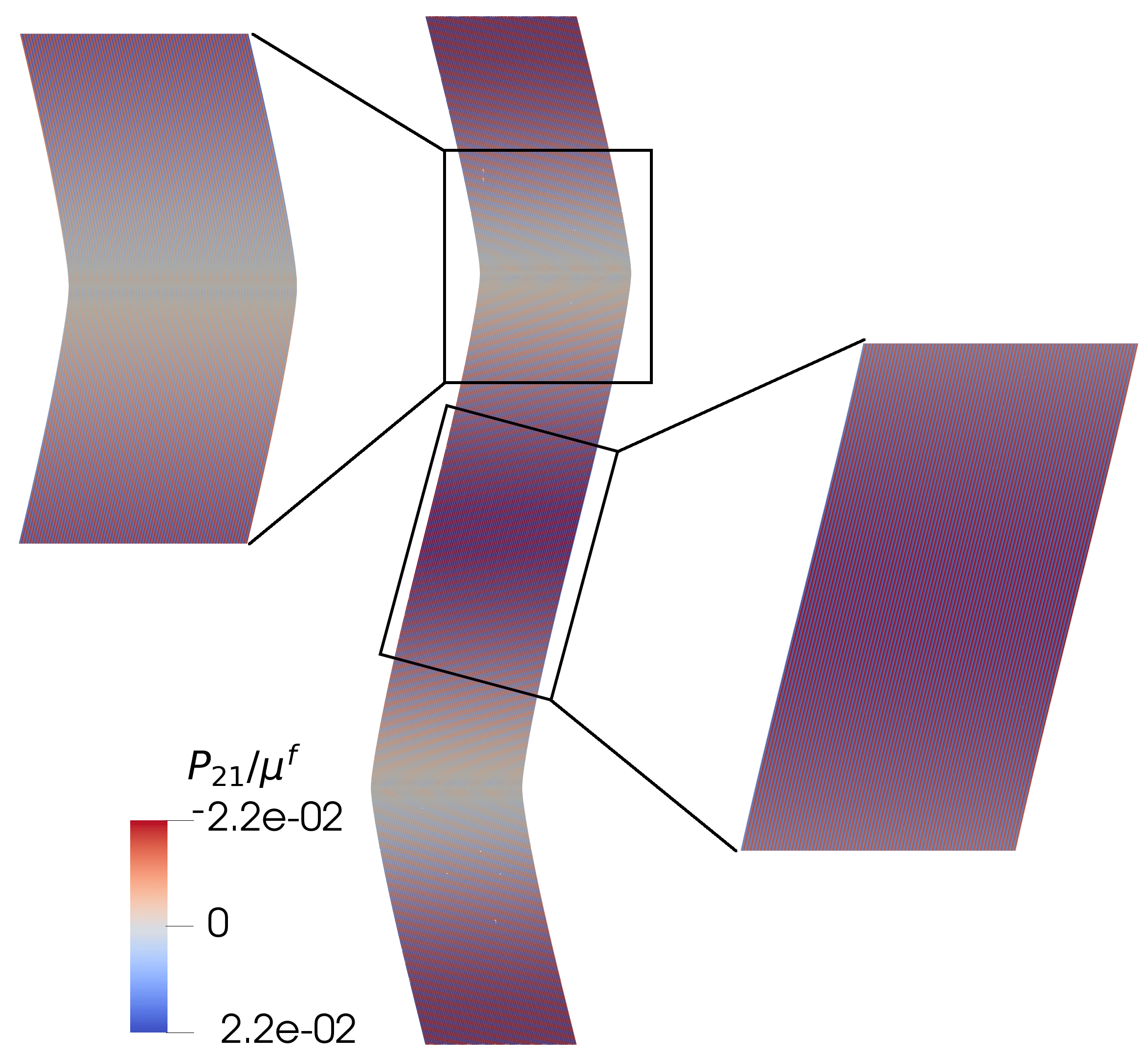}\label{}}
 \caption{Domains in tension for the elastoplastic laminate of figure \ref{fig:VaringLamdaAlot} with the highest imperfection ($\alpha = H$).\label{fig:DomainsInTensionALl_Alot}}
\end{figure}

As the imperfection gets smaller, this transition from the principal to the bifurcated solution becomes sharper and it is in these cases where LOE was found to accurately indicate domain formation. In the current study, only the laminate with the highest perturbation was observed to form domains under tensile normal stresses (and stretch), due to the earlier softening caused by the imperfection. The development of domains in tension is presented in Fig. \ref{fig:DomainsInTensionALl_Alot} --where 80 unit cells have been used for the visualization of the final RVE-- which correspond to the red dots in the window of Fig. \ref{fig:VaringLamdaAlot}.

\bibliographystyle{unsrt}  
\bibliography{references}  %%% Remove comment to use the external .bib file (using bibtex).
%%% and comment out the ``thebibliography'' section.

%%% Comment out this section when you \bibliography{references} is enabled.
% \begin{thebibliography}{1}

% \bibitem{kour2014real}
% George Kour and Raid Saabne.
% \newblock Real-time segmentation of on-line handwritten arabic script.
% \newblock In {\em Frontiers in Handwriting Recognition (ICFHR), 2014 14th
%   International Conference on}, pages 417--422. IEEE, 2014.

% \bibitem{kour2014fast}
% George Kour and Raid Saabne.
% \newblock Fast classification of handwritten on-line arabic characters.
% \newblock In {\em Soft Computing and Pattern Recognition (SoCPaR), 2014 6th
%   International Conference of}, pages 312--318. IEEE, 2014.

% \bibitem{hadash2018estimate}
% Guy Hadash, Einat Kermany, Boaz Carmeli, Ofer Lavi, George Kour, and Alon
%   Jacovi.
% \newblock Estimate and replace: A novel approach to integrating deep neural
%   networks with existing applications.
% \newblock {\em arXiv preprint arXiv:1804.09028}, 2018.

% \end{thebibliography}

\end{document}